\documentclass[]{JFM-FLM_Au}

\usepackage{graphicx}
\usepackage{newtxtext}
\usepackage{newtxmath}
\usepackage{natbib}
\usepackage{hyperref}
\hypersetup{
    colorlinks = true,
    urlcolor   = blue,
    citecolor  = black,
}

\newcommand{\RomanNumeralCaps}[1]
\linenumbers

\newcommand{\ue}{\mathrm{e}}
\newcommand{\ui}{\mathrm{i}}
\newcommand{\ud}{\mathrm{d}}

\lefttitle{Athanasios Dermatis et al.}
\righttitle{A time-domain approach for motion-explicit evaluation of loads on floating structures}

\title{A time-domain approach for motion-explicit evaluation of loads on floating structures in fully nonlinear waves}

\author{Athanasios Dermatis\aff{1}, Henrik Bredmose\aff{2}, Harry B. Bingham\aff{3}, Benjamin Bouscasse\aff{1} \and Guillaume Ducrozet\aff{1}}

\affiliation{\aff{1}Nantes Université, École Centrale Nantes, CNRS, LHEEA, UMR 6598, F-44000, Nantes, France
\aff{2}Department of Wind and Energy Systems, Technical University of Denmark, Kgs. Lyngby, Denmark
\aff{3}Department of Civil and Mechanical Engineering, Technical University of Denmark, Kgs. Lyngby, Denmark}

\corresau{Henrik Bredmose, hbre@dtu.dk}

\begin{document}
\maketitle

\begin{abstract}
This paper presents a novel method for evaluating second-order consistent hydrodynamic loads, which employs nonlinear wave and body kinematics. The pseudo-spectral formulation of nonlinear potential flow wave solvers is exploited, permitting the application of transfer functions on the nonlinear incident wave field. A closed-form expression is accordingly derived for the potential force component, which constitutes a generalisation of the Pinkster approximation to fully nonlinear waves. Moreover, the quadratic force component is reformulated to account for the total nonlinear body motion and velocity rather than their first-order counterparts. Hence, the traditional assumption that first-order body motions are significantly larger than the second-order components, which is violated in the case of moored floating structures, is circumvented. To this end, the radiation potential is treated in the time domain and is distinguished from the incident and scattering wave contributions, which are considered through wavenumber-domain transfer functions. An important advantage of the proposed approach is that it is established on the output of radiation-diffraction analysis in the frequency domain, and therefore is highly practical and efficient. Finally, the derived force model is coupled with a time-domain motion solver, which permits the consideration of the instantaneous body motion and velocity in the force calculation. The solver is employed to investigate the motions of a moored container ship, and the results demonstrate significant improvements over standard second-order radiation-diffraction theory. 
\end{abstract}

\begin{keywords} 

\end{keywords}


\section{Introduction}
\label{sec:introduction}

The offshore industry has shown increasing interest lately in using floating structures as an alternative to bottom-fixed structures, attempting to harness wind and wave energy resources in regions of large depth. The majority of these structures are anchored at the designed location using a mooring system that consists of several lines or risers. To prevent the occurrence of resonant motions in the horizontal plane, the natural frequency of the coupled system is designed not to coincide with the wave frequency range. However, second-order difference- or sum-frequency wave loads can potentially excite these resonant motions, significantly stressing the mooring system. Therefore, accurate modelling of these extreme responses and associated mooring loads is important for the safe and efficient design of moored floating structures.

Considering that the driving mechanisms of the problem are second-order, the use of linear theory is expected to fall short in modelling these responses. The common approach is to solve the first- and second-order radiation-diffraction problem using frequency-domain tools \citep{lee1995wamit, chen_middle-field_2007, kurnia2023}. The results can then be delivered in a parametric form, referred to as the response amplitude operator (RAO) for the first-order loads and motions, and the quadratic transfer function (QTF) for the second-order quantities. The solution of the full second-order problem entails considerable computational burden, and thus, some approximations have been developed in the literature \citep{de_hauteclocque_review_2012}. For instance, the Newman approximation \citep{newman1974second} considers only the diagonal elements of the QTF matrix, corresponding to zero-difference-frequency pairs. Moreover, \cite{pinkster1980} approximated the second-order potential force by assuming that the so-called scattering second-order effects are secondary and approximating them based on the undisturbed incident wave field. The common feature of these approaches is the approximation of second-order loads using exclusively first-order quantities. 

Upon solution of the radiation-diffraction problem and the evaluation of the hydrodynamic loads in the frequency domain, a second-order wavefield can be considered, and the first- and second-order responses can be obtained. The reconstruction of the linear responses in the time domain is straightforward by multiplying the RAO by the complex wave amplitudes and then applying an inverse fast Fourier transform (IFFT). Similarly, a double-summation scheme is employed for the time-domain reconstruction of second-order responses. This procedure constitutes the pure frequency domain approach and does not permit the inclusion of any additional nonlinearities in the mechanical system. Alternatively, the first- and second-order loads can be transformed into the time domain, and they can be used to solve the Cummins equation \citep{cummins1962impulse} at a post-processing stage. This approach is more refined, since it permits the consideration of nonlinear forces, and is referred to as the hybrid frequency-time domain approach. Finally, the overall problem can also be solved entirely in the time domain at an additional computational expense, through boundary element methods \citep{bai2006higher, ferrant2003nonlinear} or field methods that discretise the fluid domain using finite differences \citep{ducrozet2010highorder} or finite elements \citep{wu2003coupled}.


Nevertheless, the importance of including nonlinear wave kinematics in the modelling of the low- or high-frequency responses has been highlighted in numerous recent studies.  According to the European standard IEC 61400-3 \citep{IEC2020} for floating wind turbines, considering the nonlinearity of the waves is necessary, especially for extreme load scenarios. \cite{robertson_2017} showed that the majority of the state-of-the-art coupled hydrodynamic tools for FOWT underestimated the ultimate and fatigue limit state loads, because of the low-frequency responses. It was later concluded by \cite{robertson2020} that nonlinear wave kinematics and wave stretching might be significant for the accurate estimation of the low-frequency loads. In addition, \cite{agarwal2011nonlinear} employed a second-order irregular wave model to investigate the longitudinal bending moment of a monopile structure in shallow water depths. It was shown that for the same sea state, the structural loads due to nonlinear waves followed a non-Gaussian distribution and were around 10\% higher compared to the loads due to linear waves. It was concluded that considering linear waves might provide unconservative estimates of the long-term structural loads on offshore wind turbines. In \cite{schloer2016}, the hydrodynamic loads on a fixed monopile were investigated using both linear and fully nonlinear wave kinematics for different depths.  The inline force and tower bending moment were evaluated using the Rainey model \citep{rainey1995}, and a considerable underestimation of the load statistics was observed using linear waves. Moreover, the inclusion of wave nonlinearity resulted in an energy redistribution between free and bound wave components, which had a notable impact on the fatigue analysis.

Today, several approaches exist that deliver nonlinear wave kinematics at a quite reasonable computational cost. For regular waves, Stokes wave theory \citep{stokes1847theory} and stream function theory \citep{rienecker1981fourier} are widely employed approaches. For irregular waves, notable fully nonlinear wave models exist such as \texttt{OceanWave3D} \citep{engsigkarup2009efficient}, \texttt{HOS-NWT }\citep{ducrozet2012modified} and \texttt{HOS-ocean} \citep{ducrozet2026hosocean}, among others. In the case of slender structures, the scattered and radiated wavefields can be neglected, and the forces can be evaluated exclusively based on the incident waves. Under these conditions, semi-empirical force models can be employed \citep{morison1950force, rainey1995} to evaluate the wave-induced loads using fully nonlinear wave kinematics. Towards the inclusion of some scattering effects, the FNV model was proposed by \cite{Faltinsen_Newman_Vinje_1995} for third-order regular waves, and was later extended to include fully nonlinear irregular wave kinematics by \cite{Kristiansen_Faltinsen_2017}. For the case of non-slender structures, a novel force model was introduced by \cite{bredmose2024forcemodel} that incorporates fully nonlinear wave kinematics in the calculation of second-order forces on fixed structures. In the same study, it was shown that, when using second-order waves, the proposed method is equivalent to the Pinkster approximation. 

Moreover, another limitation arises from the expression of the second-order loads in an inertial reference frame. More precisely, the formulation is based on the assumption that second-order motions are significantly smaller than first-order motions. In practical applications such as the resonant low-frequency responses of moored floating bodies, this assumption can be largely violated. To address this issue, building on the earlier work of \cite{shao_bodyfixed_2010}, \cite{shao_consistent_2022} reformulated the second-order boundary value problem in a body-fixed reference frame. This formulation eliminates the need for Taylor series expansions to approximate the pressure on the instantaneous body surface relative to the mean body surface. Consequently, large second-order motions in the inertial reference frame do not affect the accuracy of the pressure integration. An alternative approach was introduced by \cite{teng2016twice}, who addressed the problem using a non-inertial reference frame that oscillates horizontally with the slow-drift velocity of the body. By formulating the problem relative to this frame, the motion components could be consistently expanded in a power series, with second-order contributions remaining smaller than the first-order terms. This separation allowed the low-frequency horizontal motions to be decoupled from the wave-frequency motions, leading to a more accurate evaluation of the wave-induced loads. However, both these approaches require the solution of the boundary value problem in the time domain, which can be computationally demanding.

These two principal limitations in the modelling of the second-order wave-induced loads and responses of large-volume structures are the motivation for the present work, which extends the formulation of \cite{bredmose2024forcemodel} to the case of moving bodies. The proposed approach is based on the output of a radiation-diffraction analysis in the frequency domain, and therefore is particularly efficient. In addition, the hydrodynamic loads are formulated employing both nonlinear wave and body kinematics, and regarding the latter, two advantages are presented. First, the distinction between first- and second-order body motions and assumptions about their respective order of magnitude is circumvented. As discussed in the introduction, this is particularly advantageous in the case of moored floating structures, where second-order motions become important due to resonant effects and therefore introduce inaccuracies in the pressure integration method. Moreover, the radiation potential is developed in the time domain, with respect to the nonlinear body kinematics, which partially decouples the resulting wavefield from the wave-induced motions. This feature permits the explicit consideration of other contributions, such as aerodynamic loads, in an integrated coupled load analysis, which is useful for application such as floating wind turbines. The new model is validated for the wave-induced surge and pitch motion of a moored container ship in unidirectional waves. The results are compared against classical second-order theory and experimental results. 

The remainder of this paper is structured as follows. Section \ref{sec:nonlinear_problem} describes the fully nonlinear potential flow boundary value problem and the associated hydrodynamic loads. Section \ref{sec:second-order_problem} describes the solution of the problem in orders and the hydrodynamic loads up to second order. Section \ref{sec:proposed_force_model} introduces the proposed novel approach for the evaluation of the hydrodynamic loads and the solution of the motion equations. Section \ref{sec:experiments} outlines the experimental setup of the container ship vessel. Finally, Section \ref{sec:results} includes a verification of the proposed method and a comparison of the results obtained against standard second-order theory and experiments, while Section \ref{sec:conclusions} summarises the conclusions of this work.

\section{Fully nonlinear problem} \label{sec:nonlinear_problem}
\subsection{Frames of reference} \label{sec:reference_frames}
\begin{figure}[ht!]
    \centering
    \includegraphics[width=0.8\linewidth]{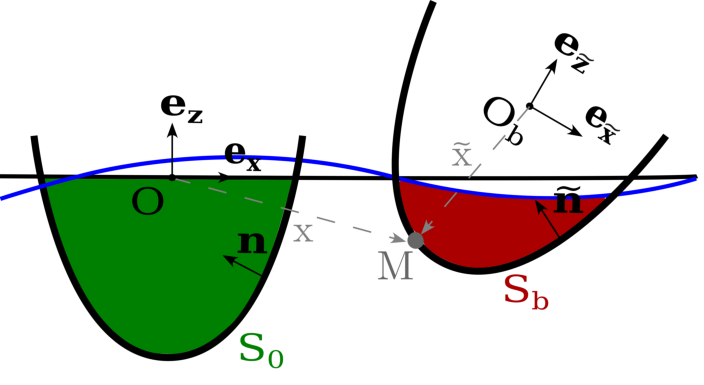}
    \caption{Reference frame definition}
    \label{fig:reference-frame}
\end{figure}

Two reference frames are considered for the definition of the rigid body motions, which are depicted in Figure \ref{fig:reference-frame} and more precisely,
\begin{itemize}
    \item A fixed inertial reference frame $R_O=(\mathbf{O}, \mathbf{e}_x, \mathbf{e}_y, \mathbf{e}_z )$, the origin of which is located on the still waterline and the horizontal centre of gravity (COG) of the body at rest. The position vector is $\mathbf{x} =[x,y,z]$ and the unit normal vector, pointing out of the fluid domain, is defined as $\mathbf{n}=[n_x,n_y,n_z]$. 
    \item A body-fixed reference frame $R_b = (\mathbf{O}_b, \mathbf{e}_{\tilde{x}}, \mathbf{e}_{\tilde{y}}, \mathbf{e}_{\tilde{z}}) $, with position and normal vectors $\mathbf{\tilde{x}}=(\tilde{x}, \tilde{y},\tilde{z} )$ and $\mathbf{\tilde{n}} = [\tilde{n}_x,\tilde{n}_y,\tilde{n}_z]$. 
\end{itemize}

The instantaneous submerged body surface is denoted as $S_b$, while the time-invariant mean body surface is $S_0$. Moreover, following \cite{lee1995wamit}, a constant distance $\tilde{z}_0$ can be introduced, allowing for a vertical displacement of $\mathbf{O}$ from the free-surface, and of $\mathbf{O}_b$ along the body axis $\mathbf{e}_{\tilde{z}}$.

\subsection{Rigid body kinematics} \label{sec:rigid_body_kinematics}
The body is considered free to move in all $6$ DoFs, with $\boldsymbol{\xi}=[\xi_1,\xi_2,\xi_3]=\mathbf{OO}_b$, denoting the vector of translational body motions. The vector of rotational motions $\boldsymbol{\alpha}=[\alpha_1,\alpha_2,\alpha_3]$ is defined through the relative rotation between the axes of the two reference frames. Further assuming $\boldsymbol{\alpha}$ consists of Tait-Bryan angles, and adopting the "\textit{xyz}" convention of elementary rotations, the transformation matrix $\boldsymbol{\mathcal{R}}$ is introduced,

\begin{equation} \small
    \begin{aligned} 
    \boldsymbol{\mathcal{R}} =\begin{bmatrix}
        \cos\alpha_3 \cos\alpha_2     & \sin\alpha_3 \cos\alpha_2     & -\sin\alpha_2 \\
        -\sin\alpha_3\cos\alpha_1 + \cos \alpha_3 \sin\alpha_2 \sin\alpha_1     & \cos\alpha_3\cos\alpha_1 + \sin\alpha_3 \sin\alpha_2 \sin\alpha_1     & \cos\alpha_2 \sin\alpha_1 \\
        \sin\alpha_3 \sin\alpha_1 + \cos\alpha_3 \sin\alpha_2 \cos\alpha_1     & -\cos\alpha_3 \sin\alpha_1 + \sin\alpha_3 \sin\alpha_2 \cos\alpha_1     & \cos\alpha_2 \cos\alpha_1
    \end{bmatrix}
    \end{aligned}
\end{equation}

The definition of the position vector for the global reference frame is defined for an arbitrary point $\mathbf{M}$ as $\mathbf{OM} = \mathbf{x}$, shown in Figure \ref{fig:reference-frame} and $\mathbf{O}_b \mathbf{M} = \boldsymbol{\mathcal{R}}\mathbf{\tilde{x}}$. The relationship between the position and normal vectors in $R_O$ and the respective vectors in $R_b$ is,

\begin{equation} \label{eq:point_definition}
    \mathbf{x}=\mathbf{OM} = \mathbf{OO}_b + \mathbf{O M}_b  = \boldsymbol{\xi} + \boldsymbol{\mathcal{R}}\tilde{\mathbf{x}}
\end{equation}
\begin{equation} \label{eq:normal_definition}
    \mathbf{n} = \boldsymbol{\mathcal{R}} \tilde{\mathbf{n}}
\end{equation} 
It is convenient to also define the generalised motion vector $\underline{\boldsymbol{\xi}}$ and normal vectors $\underline{\mathbf{n}}, \tilde{\underline{\mathbf{n}}}$ in all 6 DoFs as,

\begin{equation}
    \underline{\boldsymbol{\xi}} = \begin{bmatrix}
    \boldsymbol{\xi}  \\ \boldsymbol{\alpha}
        \end{bmatrix}
    \qquad \qquad
    \underline{\mathbf{n}}= \begin{bmatrix}
    \mathbf{n} \\\mathbf{x} \times \mathbf{n}
        \end{bmatrix}
    \qquad \qquad
    \tilde{\underline{\mathbf{n}}}= \begin{bmatrix}
    \tilde{\mathbf{n}} \\ \tilde{\mathbf{x}} \times \tilde{\mathbf{n}}
        \end{bmatrix}
\end{equation}

\subsection{Exact boundary value problem} \label{sec:pf_bvp}
Assuming a homogeneous, incompressible and inviscid fluid and an irrotational flow, the latter can be fully described through a scalar velocity potential field $\phi=\phi(\mathbf{x},t)$, so that the fluid velocity is given as $\textbf{u}=\nabla\phi$. A fluid domain $\mathbb{D}$ is considered, bounded by a free surface at $z=\eta(x,y,t)$, by a seabed of finite and constant depth $h$ at $z=-h$ and by the body's surface $S_b$. Under these assumptions, the mass conservation law is replaced by the Laplace equation throughout the fluid domain $\mathbb{D}$. Furthermore, an impermeability boundary condition is considered on the seabed and the body boundary, along with the dynamic and kinematic free-surface boundary conditions. The exact boundary value problem is formulated as, 

\begin{align} 
        \nabla^2\phi =0 &\qquad \mathbf{x} \in \mathbb{D}  \label{eq:nonlinear_laplace} \\
        \partial_z\phi=0 &\qquad z=-h  \label{eq:nonlinear_bottom} \\ 
        \partial_{t} \eta + \nabla_h\phi \cdot \nabla_h\eta - \partial_z\phi =0 &\qquad z=\eta(x,y,t) \label{eq:nonlinear_kfsbc} \\
        \partial_t\phi + g\eta + \frac{1}{2} \nabla \phi \cdot \nabla \phi = 0 &\qquad z=\eta(x,y,t) \label{eq:nonlinear_dfsbc} \\ 
        \nabla \phi \cdot \mathbf{n} =  \partial_t\mathbf{x} \cdot \mathbf{n}  & \qquad \mathbf{x} \in S_b  \label{eq:nonlinear_bbc} 
    \end{align}
where $\partial_q$ denotes the derivative with respect to $q$ and $\nabla_h=(\partial_x,\partial_y)$ denotes the gradient in the horizontal plane. 

\subsection{Hydrodynamic loads}
Solution of the exact boundary value problem would provide the nonlinear velocity potential, which is related to the pressure through the Bernoulli equation

\begin{equation} \label{eq:bernoulli_x}
    p(\mathbf{x},t)=-\rho \left ( \partial_t\phi(\mathbf{x},t)+\frac{1}{2} \nabla \phi(\mathbf{x},t) \cdot \nabla \phi(\mathbf{x},t) + g(z + \tilde{z}_0)    \right)
\end{equation}
where, without loss of generality, the Bernoulli constant is considered zero. The fully nonlinear wave loads are then obtained by integration of the pressure on the exact body surface $S_b(t)$,

\begin{equation} \label{eq:force_exact}
    \underline{\mathbf{F}}(t) = \begin{bmatrix}
        \mathbf{F}(t) \\
        \mathbf{Q}(t) 
    \end{bmatrix} = \int_{S_b(t)} p(\mathbf{x},t) \underline{\mathbf{n}} ~\ud S
\end{equation}
where the generalised 6DoF vector $\underline{\mathbf{F}}$ includes the forces $\mathbf{F}$ and the moments $\mathbf{Q}$.

\section{Second-order radiation-diffraction theory} \label{sec:second-order_problem}
The boundary value problem described by \eqref{eq:nonlinear_laplace}-\eqref{eq:nonlinear_bbc} is highly nonlinear. More precisely, apart from the nonlinear terms in \eqref{eq:nonlinear_kfsbc}-\eqref{eq:nonlinear_bbc}, the free-surface conditions are also expressed on the unknown free-surface elevation. In addition, the body boundary condition is applied on the time-variant instantaneous wetted surface $S_b$, while the same surface is used for the pressure integration in \eqref{eq:force_exact}. Most state-of-the-art radiation-diffraction tools solve the problem using the perturbation approach \citep{lee1995wamit, newman2018marine}, which is outlined in the present section.

\subsection{Body kinematics in orders} \label{sec:ordered_body_kinematics}
The problem can be simplified through perturbation of the wave and body motion variables in a power series with respect to a small parameter $\epsilon$. Traditionally, this perturbation analysis is performed in nondimensional variables, in which case $\epsilon$ corresponds to the wave steepness. However, dimensional variables are used throughout the present analysis, so that $\epsilon$ has no direct physical meaning and is only used to indicate the order of magnitude. The body motions are accordingly developed in perturbation series as,

\begin{equation} \label{eq:series_motion}
\begin{aligned}
    \boldsymbol{\xi} = \boldsymbol{\xi}^{(1)} + \boldsymbol{\xi}^{(2)} + O(\epsilon^3) \qquad \qquad
    \boldsymbol{\alpha} = \boldsymbol{\alpha}^{(1)} + \boldsymbol{\alpha}^{(2)} + O(\epsilon^3)
\end{aligned}
\end{equation}
where $q^{(m)}$ denotes a quantity $q$ of order $m$ and $O(\epsilon^m)$ includes terms of order $m$ and higher. The transformation matrix $R$ can also be developed as,

\begin{equation} \label{eq:series_transform_matrix}
        \boldsymbol{\mathcal{R}} =  \boldsymbol{\mathcal{R}}^{(0)} + \boldsymbol{\mathcal{R}}^{(1)} + \boldsymbol{\mathcal{R}}^{(2)} + O(\epsilon^3) = \boldsymbol{\mathbb{I}}_3 + \boldsymbol{\mathcal{A}}^{(1)}_{\times} + \boldsymbol{\mathcal{A}}^{(2)}_{\times} + \boldsymbol{\mathcal{H}}^{(2)} + O(\epsilon^3)
\end{equation}
where $\boldsymbol{\mathbb{I}}_3$ denotes the $3\times 3$ identity matrix. Moreover, $\boldsymbol{\mathcal{A}}^{(1)}_{\times}$ and $\boldsymbol{\mathcal{A}}^{(2)}_{\times}$ are the skew-symmetric matrices of the body rotation and are defined as,

\begin{equation} \label{eq:rot_skew_symmetric}
    \boldsymbol{\mathcal{A}}^{(m)}_{\times} = \begin{bmatrix}
        0 & - \alpha^{(m)}_3 &  \alpha^{(m)}_2 \\
        \alpha^{(m)}_3 & 0 & -\alpha^{(m)}_1 \\
        - \alpha^{(m)}_2 & \alpha^{(m)}_1 & 0
    \end{bmatrix} \qquad m = 1,2
\end{equation}
and for an arbitrary three-dimensional vector $\mathbf{q}$ they follow $\boldsymbol{\mathcal{A}}^{(m)}_{\times}\mathbf{q}=\boldsymbol{\alpha}^{(m)} \times \mathbf{q}$. The matrix $\boldsymbol{\mathcal{H}}^{(2)}$ is defined in accordance with the "$xyz$" convention as, 

\begin{equation}
    \boldsymbol{\mathcal{H}}^{(2)} = \begin{bmatrix}
-\frac{1}{2}\left((\alpha_2^{(1)})^2 + (\alpha_3^{(1)})^2\right) & 0 & 0\\
\alpha^{(1)}_1 \alpha^{(1)}_2 & -\frac{1}{2}\left((\alpha^{(1)}_1)^2 + (\alpha_3^{(1)})^2\right) & 0 \\
\alpha^{(1)}_1 \alpha^{(1)}_3 & \alpha^{(1)}_2 \alpha^{(1)}_3 & -\frac{1}{2}\left((\alpha_1^{(1)})^2 + (\alpha_2^{(1)})^2\right) 
\end{bmatrix}
\end{equation}
Therefore, \eqref{eq:point_definition} and \eqref{eq:normal_definition} can be written in an order-consistent form as,

\begin{equation} \label{eq:posvec_expansion}
    \begin{aligned}
        \mathbf{x} &= \tilde{\mathbf{x}} + \boldsymbol{\xi}^{(1)} + \boldsymbol{\alpha}^{(1)} \times \tilde{\mathbf{x}}  + \boldsymbol{\mathcal{H}}^{(2)} \tilde{\mathbf{x}}  + \boldsymbol{\xi}^{(2)} + \boldsymbol{\alpha}^{(2)} \times \tilde{\mathbf{x}}  + O(\epsilon^3) \\ 
        & = \mathbf{x}^{(0)} + \mathbf{x}^{(1)} + \mathbf{x}^{(2)} + O(\epsilon^3) 
    \end{aligned}
\end{equation}

\begin{equation} \label{eq:normvec_expansion}
    \mathbf{n} = \mathbf{\tilde{n}} +  \boldsymbol{\alpha}^{(1)} \times \mathbf{\tilde{n}}  + \boldsymbol{\mathcal{H}}^{(2)} \mathbf{\tilde{n}}+ O(\epsilon^3)=\mathbf{n}^{(0)} + \mathbf{n}^{(1)} + \mathbf{n}^{(2)} + O(\epsilon^3) \\ \\
\end{equation}
and similarly for the generalised normal vector,

\begin{equation}
    \underline{\mathbf{n}} = \underline{\mathbf{\tilde{n}}} +  \tilde{\mathbf{N}}_{\underline{\boldsymbol{\xi}}^{(1)}}  + \tilde{\mathbf{N}}_{\underline{\boldsymbol{\xi}}^{(2)}} + \tilde{\mathbf{N}}_{\underline{\boldsymbol{\xi}}^{(1)}\underline{\boldsymbol{\xi}}^{(1)}} + O(\epsilon^3) =\underline{\mathbf{n}}^{(0)} + \underline{\mathbf{n}}^{(1)} + \underline{\mathbf{n}}^{(2)} + O(\epsilon^3)
\end{equation}
where the operators $\tilde{\mathbf{N}}_{\underline{\boldsymbol{\xi}}^{(1)}}$, $\tilde{\mathbf{N}}_{\underline{\boldsymbol{\xi}}^{(2)}}$ and $\tilde{\mathbf{N}}_{\underline{\boldsymbol{\xi}}^{(1)}\underline{\boldsymbol{\xi}}^{(1)}}$ are defined as,

\begin{equation}
    \begin{aligned}
        \tilde{\mathbf{N}}_{\underline{\boldsymbol{\xi}}^{(m)}} &= \begin{bmatrix}
           \boldsymbol{\alpha}^{(m)} \times \mathbf{\tilde{n}} \\
            \boldsymbol{\xi}^{(m)} \times \mathbf{\tilde{n}} + \boldsymbol{\alpha}^{(m)} \times (\tilde{\mathbf{x}}
            \times \mathbf{\tilde{n}})
        \end{bmatrix}  \qquad \qquad m =1,2
        \\ \\
       \tilde{\mathbf{N}}_{\underline{\boldsymbol{\xi}}^{(1)}\underline{\boldsymbol{\xi}}^{(1)}} &= \begin{bmatrix}
            \boldsymbol{\mathcal{H}}^{(2)} \mathbf{\tilde{n}}\\
            \boldsymbol{\xi}^{(1)} \times ( \boldsymbol{\alpha}^{(1)}\times \mathbf{\tilde{n}}) + \boldsymbol{\mathcal{H}}^{(2)}(\tilde{\mathbf{x}} \times \mathbf{\tilde{n}}) 
        \end{bmatrix}
    \end{aligned}
\end{equation}

\subsection{Boundary value problem in orders} \label{sec:bvp_ordered}
The velocity potential can also be developed in a perturbation series as,

\begin{equation} \label{eq:series_potential}
    \phi = \phi^{(1)} + \phi^{(2)} + O(\epsilon^3)
\end{equation}
and the boundary value problem can be formulated at each order as,

\begin{equation} \label{eq:bvp_ordered}
    \begin{aligned}
        \nabla^2 \phi^{(m)} = 0 &\qquad \mathbf{x} \in \mathbb{D}\\ 
        \partial_z \phi^{(m)} =0 &\qquad z=-h  \\ 
        \partial_{tt}\phi^{(m)} + g \partial_z\phi^{(m)} = \Theta^{(m)}_F &\qquad z =0 \\ 
        \nabla \phi^{(m)} \cdot \mathbf{\tilde{n}} =  \partial_t\underline{\boldsymbol{\xi}}{}^{(m)} \cdot \underline{\mathbf{\tilde{n}}} + \Theta_B^{(m)} &\qquad \mathbf{x}=\tilde{\mathbf{x}} \in S_0 \\
    \end{aligned}
    \qquad m = 1,2
\end{equation}
where the kinematic and dynamic free-surface conditions were combined and expressed around $z=0$ through Taylor expansion \citep{lee1995wamit}. The free-surface forcing functions are defined as,

\begin{equation} \label{eq:fs_forcing_functions}
    \begin{aligned}
    \Theta_F^{(1)} &= 0 \\
        \Theta_F^{(2)} &=\frac{1}{g}\partial_t\phi^{(1)} \partial_z\left ( \partial_{tt}\phi^{(1)} + g \partial_{z}\phi^{(1)}    \right) - \partial_t  \left ( \nabla \phi^{(1)} \cdot \nabla\phi^{(1)} \right ) \\
        \end{aligned}
\end{equation}
and the body boundary forcing functions are,

\begin{equation} \label{eq:body_forcing_functions}
    \begin{aligned}
        \Theta_B^{(1)} &= 0 \\
        \Theta_B^{(2)} &=  \mathbf{\tilde{n}} \cdot  \partial_t\boldsymbol{\mathcal{H}}^{(2)} \mathbf{\tilde{x}} + (\boldsymbol{\alpha}^{(1)} \times \mathbf{\tilde{n}}) \cdot \left (  \partial_t\boldsymbol{\xi}{}^{(1)} + \partial_t\boldsymbol{\alpha}{}^{(1)} \times 
    \mathbf{\tilde{x}}   - \nabla \phi^{(1)}   \right) \\ & \qquad - \mathbf{\tilde{n}} \left [ \left ( \boldsymbol{\xi}^{(1)} + \boldsymbol{\alpha}^{(1)} \times 
    \mathbf{\tilde{x}}  \right) \cdot \nabla \right ] \nabla \phi^{(1)} 
    \end{aligned}
\end{equation}

\subsection{Solution of the problem in orders}
The total velocity potential that satisfies \eqref{eq:bvp_ordered} is decomposed into three components as,

\begin{equation}
     \phi^{(m)} = \phi_I^{(m)} + \phi_R^{(m)} + \phi_S^{(m)}
\end{equation}
where $\phi^{(m)}_I$ is the incident wave potential and corresponds to the undisturbed wavefield. Moreover, $\phi_R^{(m)}$ is the radiation potential and is related to the wavefield generated by the body motion. The last component $\phi_S^{(m)}$ at first order corresponds to the wavefield scattered by the body. As shown below, its physical interpretation is not straightforward at second order \citep{pinkster1980}, since it also includes all quadratic interactions between $\phi^{(1)}_S$ and $\phi^{(1)}_R$.

Assuming the incident wave solution is an irregular plane wave propagating towards the positive $\mathbf{e}_x$-direction, the first- and second-order incident potentials are given as,

\begin{equation} 
    \begin{aligned} \label{eq:linear_waves}
        \phi_I^{(1)}(\mathbf{x},t) &= \sum_{n=-N}^N B_n  \frac{\cosh[k_n(z + h)]}{\cosh(k_nh)} \ue^{\ui \left ( \omega_n t -k_nx \right )}\\
    \end{aligned}
\end{equation} 
\begin{equation} \label{eq:quadratic_incident_potential}
\begin{aligned}
       \phi_I^{(2)}(\mathbf{x},t) &= \sum_{m=-N}^N\sum_{n=-N}^N B_m B_n 
    \hat{\mathbb{T}}{}^{(2)}_{\phi}(\omega_m, \omega_n)\frac{\cosh  \left [ |k_m+k_n| (z+h) \right ]}{\cosh\left ( |k_m+k_n|h \right )}
    \ue^{\ui\left[(\omega_m+\omega_n)t-(k_m+k_n) x\right]} \\
\end{aligned}
\end{equation} 
where $\omega_n$ is the wave frequency, $k_n$ is the wavenumber, $A_n$ is the complex wave amplitude, and $B_n=\ui g A_n/\omega_n$ is the respective velocity potential amplitude. Moreover, the following symmetry conditions are considered, $A_{-n}=A^\ast_{n}$, $B_{-n}=B^\ast_{n}$, $k_{-n}=-k_n$ and $\omega_{-n}=-\omega_n$, with $[\cdot]^\ast$ denoting complex conjugate. Such a definition in the negative frequency range yields real outputs for the velocity potential, and also avoids distinguishing between sum- and difference-frequency pairs at second-order \citep{bredmose_second-order_2021}. Finally, following \cite{sharma1981}, the quadratic transfer function defined as,

\begin{equation} \label{eq:sharma_dean_tf}
    \begin{aligned}
       \hat{\mathbb{T}}{}^{(2)}_{\phi}(\omega_m, \omega_n) = & -\ui\frac{2(\omega_m + \omega_n) \left ( k_m k_n - \omega_m^2 \omega_n^2/g^2 \right ) + \omega_m (k_n^2-\omega_n^4/g)+\omega_n(k_m^2-\omega_m^2/g)}{(\omega_m+\omega_n)^2-g(k_m+k_n) \tanh [(k_m+k_n)h]} \\
    \end{aligned}
\end{equation}
Replacing \eqref{eq:linear_waves} into the free-surface boundary condition of \eqref{eq:bvp_ordered} provides the well-known linear dispersion relation, which associates the wavenumber and frequency as,

\begin{equation} \label{eq:linear_dispersion_relation}
    \omega^2_n = gk_n ~\text{tanh} (k_nh)
\end{equation}

Regarding the radiation potential, it can be defined at each order as,

\begin{equation} 
    \phi_R^{(m)}(\mathbf{x},t) =  \underline{\boldsymbol{\phi}}{}_r(\mathbf{x},t) \circledast \partial_t\underline{\boldsymbol{\xi}}{}^{(m)}(t)
\end{equation}
where $\circledast$ denotes convolution and summation over all 6 DoFs, while $\underline{\boldsymbol{\phi}}{}_r$ is a 6DoF vector containing the elementary radiation potentials. 

Therefore, the solution of the simplified flow problem consists of determining the velocity potentials $\underline{\boldsymbol{\phi}}{}_r$ and $\phi_S^{(m)}$. All potential components satisfy the Laplace equation and the bottom boundary condition. Regarding the free-surface condition, its second-order forcing function can be further decomposed as,

\begin{equation} 
    \Theta_F^{(2)} = \Theta_{F,II}^{(2)} + \Theta_{F,IB}^{(2)} + \Theta_{F,BB}^{(2)}
\end{equation}
where, each subcomponent of $\Theta_F^{(2)}$ involves different combinations of quadratic products between the potentials $\phi_I^{(1)}$ and $\phi_B^{(1)}= \phi_S^{(1)}+\phi_R^{(1)}$ on the right-hand side of \eqref{eq:fs_forcing_functions}. Thus, due to the independence of the incident potential from the wave-structure interaction, it satisfies the free-surface condition that involves $\Theta_{F,~II}^{(2)}$. Furthermore, the radiation potential satisfies the homogeneous free-surface boundary condition. Therefore, the scattering potential is defined to account for the terms involving $\phi_B^{(1)}$ self-interactions. Letting $\Theta_{F,II}^{(1)}=\Theta_{F,IB}^{(1)}=\Theta_{F,BB}^{(1)}=0$, the condition is formulated at each order $m=1,2$ as,
\begin{equation} \label{eq:fs_forcing_function_decomposition}
    \begin{aligned}
       \partial_{tt}\phi^{(m)}_I + g \partial_z\phi^{(m)}_I = \Theta_{F,II}^{(m)}  \qquad z =0 \\
       \partial_{tt}\phi^{(m)}_S + g \partial_z\phi^{(m)}_S =  \Theta_{F,IB}^{(m)} + \Theta_{F,BB}^{(m)}  \qquad z =0 \\
       \partial_{tt}\phi^{(m)}_R + g \partial_z\phi^{(m)}_R = 0 \qquad z=0
    \end{aligned}
\end{equation}
The boundary value problem is completed by the following body boundary conditions

\begin{equation} \label{eq:bc-phi}
\begin{aligned}
      \nabla \phi_I^{(m)} \cdot \mathbf{\tilde{n}} + \nabla \phi_S^{(m)} \cdot \mathbf{\tilde{n}} = \Theta_B^{(m)} \qquad & \mathbf{x}=\tilde{\mathbf{x}} \in S_0 \\
     \nabla \phi_R^{(m)} \cdot\mathbf{\tilde{n}} = \partial_t\underline{\boldsymbol{\xi}}{}^{(m)} \cdot \underline{\mathbf{\tilde{n}}} \qquad & \mathbf{x}=\tilde{\mathbf{x}} \in S_0
\end{aligned}
\end{equation}
where $ \mathbf{x}=\tilde{\mathbf{x}} \in S_0$ means that the boundary condition is expressed on the time-invariant mean body surface $S_0$. Finally, the boundary conditions for the radiation potential can be written directly for the elementary radiation vector as,

\begin{equation} \label{eq:phiR_fsbc}
    \partial_{tt}\underline{\boldsymbol{\phi}}{}_r + g \partial_z \underline{\boldsymbol{\phi}}{}_r = 0  \qquad z =0
\end{equation}

\begin{equation} \label{eq:bc-phirj}
     \nabla \underline{\boldsymbol{\phi}}{}_r \cdot \mathbf{\tilde{n}}= \underline{\mathbf{\tilde{n}}}~\delta(t)  \qquad \mathbf{x}=\tilde{\mathbf{x}} \in S_0
\end{equation}
where $\delta(t)$ denotes the Dirac delta function.

\subsection{First- and second-order hydrodynamic loads} \label{sec:ordered_wave_loads}
The pressure on the instantaneous body surface $S_b$ is approximated by a Taylor expansion of the pressure around the position $\tilde{\mathbf{x}}$ with respect to the mean body surface $S_0$ as, 

\begin{equation} \label{eq:pressure_taylor}
    \begin{aligned}
        p(\mathbf{x},t) &= p(\mathbf{\tilde{x}},t) + \left ( \mathbf{x} -\mathbf{\tilde{x}}     \right ) \cdot \nabla p(\mathbf{\tilde{x}},t) + O(\epsilon^3) \\&  = p(\mathbf{\tilde{x}},t) + \left [ \boldsymbol{\xi}(t) +(\boldsymbol{\mathcal{R}}(t)-\mathbb{I}_3)\mathbf{\tilde{x}}    \right ] \cdot \nabla p(\mathbf{\tilde{x}},t) + O(\epsilon^3) \\
        &= p^{(0)}(\tilde{\mathbf{x}}) + p^{(1)}(\tilde{\mathbf{x}},t)  + p^{(2)}(\tilde{\mathbf{x}},t)  + O(\epsilon^3)
    \end{aligned}
\end{equation}
where the pressure at $\mathbf{\tilde{x}} \in S_0$ follows Bernoulli's equation in \eqref{eq:bernoulli_x}, and thus the pressure components follows,

\begin{equation}
    \begin{aligned}
        p^{(0)}(\tilde{\mathbf{x}}) &= - \rho   g(\tilde{z} + \tilde{z}_0) \\
        p^{(1)}(\tilde{\mathbf{x}},t) &= - \rho \left [ \partial_t \phi^{(1)}(\tilde{\mathbf{x}},t)  + g\xi^{(1)}_r(\tilde{x}, \tilde{y}, t) \right ] \\
        p^{(2)}(\tilde{\mathbf{x}},t) &= - \rho  \Big [\partial_t \phi^{(2)}(\tilde{\mathbf{x}},t) + g \xi^{(2)}_r(\tilde{x}, \tilde{y}, t) +\frac{1}{2} \nabla \phi^{(1)}(\tilde{\mathbf{x}},t) \cdot \nabla \phi^{(1)}(\tilde{\mathbf{x}},t)    + g \boldsymbol{\mathcal{H}}^{(2)}(t)  \tilde{\mathbf{x}} \cdot \nabla \tilde{z} \\&  \qquad \qquad+  (\boldsymbol{\xi}^{(1)}(t)  + \boldsymbol{\alpha}^{(1)}(t) \times \tilde{\mathbf{x}}) \cdot \nabla \partial_t \phi^{(1)}(\tilde{\mathbf{x}},t) \Big ]
    \end{aligned}
\end{equation}
with $\xi^{(m)}_{r} = \xi_3^{(m)} + \alpha_1^{(m)} \tilde{y}- \alpha^{(m)}_2 \tilde{x}$ for each order $m=1,2$. The total force is obtained from integrating the pressure terms on the approximated body surface as,

\begin{equation}
    \begin{aligned}
     \underline{\mathbf{F}}(t)& =  \int_{S_0+S_w(t)} \left ( p^{(0)}(\tilde{\mathbf{x}}) + p^{(1)}(\tilde{\mathbf{x}},t)  + p^{(2)}(\tilde{\mathbf{x}},t)   \right ) \left ( \underline{\mathbf{n}}^{(0)} + \underline{\mathbf{n}}^{(1)}(t) + \underline{\mathbf{n}}^{(2)}(t) \right ) ~\ud S + O(\epsilon^3)\\
      &=  \underline{\mathbf{F}}^{(0)} + \underline{\mathbf{F}}^{(1)}(t) + \underline{\mathbf{F}}^{(2)}(t) + O(\epsilon^3)
    \end{aligned}
\end{equation}
where the instantaneous body surface $S_b$ is decomposed into $S_0$ and $S_w=S^{(1)}_w+O(\epsilon^2)$. The latter accounts for the oscillations of the wetted surface, attributed to the variations of the waterline by the free-surface elevation and associated body motions at each order. 

Starting with the zero-order hydrostatic forces, they follow, 

\begin{equation} \label{eq:force_buoyancy}
    \underline{\mathbf{F}}^{(0)} = -\rho g  \int_{S_0} (\tilde{z} + \tilde{z}_0) ~ \underline{\tilde{\mathbf{n}}} ~\ud S = -\rho g V
\end{equation}
where $V$ denotes the wetted volume of the body at rest. The first-order force is composed of the following terms,

\begin{equation}
   \underline{\mathbf{F}}^{(1)}(t) = \underline{\mathbf{F}}^{(1)}_P(t) + \underline{\mathbf{F}}^{(1)}_H(t)
\end{equation}
where, 

\begin{equation}
    \begin{aligned}
        \underline{\mathbf{F}}^{(1)}_P(t) &= - \rho \int_{S_0} \partial_t\phi^{(1)}(\tilde{\mathbf{x}},t) \underline{\tilde{\mathbf{n}}} ~\ud S \\
        \underline{\mathbf{F}}^{(1)}_H(t) &= 
         -\rho g \int_{S_0}  \Big ( \xi^{(1)}_r(\tilde{x}, \tilde{y}, t) \underline{\tilde{\mathbf{n}}} +  (\tilde{z} + \tilde{z}_0)\tilde{\mathbf{N}}_{\underline{\boldsymbol{\xi}}^{(1)}}(t) \Big ) ~\ud S =\boldsymbol{\mathcal{C}}_H ~ \underline{\boldsymbol{\xi}}{}^{(1)}(t)
    \end{aligned}
\end{equation}
where $\boldsymbol{\mathcal{C}}_H$ denotes the hydrostatic stiffness matrix \citep{newman2018marine}. Finally, the total second-order force will consist of three components,

\begin{equation}
    \underline{\mathbf{F}}^{(2)}(t) = \underline{\mathbf{F}}_P^{(2)}(t) + \underline{\mathbf{F}}^{(2)}_H(t) + \underline{\mathbf{F}}_Q^{(2)}(t)
\end{equation}
where $\underline{\mathbf{F}}^{(2)}_P$ is related to the second-order potential time derivative, and the restoring force $\underline{\mathbf{F}}^{(2)}_H$ is proportional to the second-order body motion, defined respectively as,

\begin{equation}
    \underline{\mathbf{F}}_P^{(2)}(t) = -\rho \int_{S_0} \partial_t\phi^{(2)}(\tilde{\mathbf{x}},t) \underline{\tilde{\mathbf{n}}}~\ud S
\end{equation}

\begin{equation}
    \begin{aligned}
        \underline{\mathbf{F}}^{(2)}_H(t) &= -\rho g \int_{S_0}  \xi_r^{(2)}(\tilde{x}, \tilde{y}, t)~ \underline{\tilde{\mathbf{n}}}~\ud S  -\rho g \int_{S_0}  (\tilde{z} + \tilde{z}_0)\tilde{\mathbf{N}}_{\underline{\boldsymbol{\xi}}^{(2)}}(t) ~\ud S = \boldsymbol{\mathcal{C}}_H ~\underline{\boldsymbol{\xi}}{}^{(2)}(t)
    \end{aligned}
\end{equation}
In addition, $\underline{\mathbf{F}}^{(2)}_Q$ includes the quadratic interaction of first-order wave and motion quantities

\begin{equation} \label{eq:quadratic_force_wamit}
    \begin{aligned}
        &\underline{\mathbf{F}}_Q^{(2)}(t) = -\rho \int_{S_0} \left [ \frac{1}{2} \nabla \phi^{(1)}(\tilde{\mathbf{x}},t) \cdot \nabla \phi^{(1)}(\tilde{\mathbf{x}},t) + \Big (\boldsymbol{\xi}^{(1)}(t) + \boldsymbol{\alpha}^{(1)}(t) \times \tilde{\mathbf{x}} \Big) \cdot \nabla \partial_t\phi^{(1)}(\tilde{\mathbf{x}},t) \right ] \underline{\tilde{\mathbf{n}}}~\ud S  \\  
        &  -\rho \int_{S_0}\tilde{\mathbf{N}}_{\underline{\boldsymbol{\xi}}^{(1)}}(t) \left ( \partial_t \phi^{(1)}(\tilde{\mathbf{x}},t) + g\xi_r^{(1)}(\tilde{x}, \tilde{y}, t) \right ) \ud S    -\rho g \int_{S_0}  (\tilde{z} + \tilde{z}_0) \tilde{\mathbf{N}}_{ \underline{\boldsymbol{\xi}}^{(1)}\underline{\boldsymbol{\xi}}^{(1)} }(t) ~\ud S \\
         &  -\rho g\int_{S_0}  (\boldsymbol{\mathcal{H}}^{(2)}(t) \tilde{\mathbf{x}}\cdot \nabla \tilde{z})~\underline{\tilde{\mathbf{n}}}~\ud S  +\frac{1}{2} \rho g \int_{WL} \left (\eta^{(1)}(\tilde{x}, \tilde{y},t) - \xi^{(1)}_r(\tilde{x}, \tilde{y}, t) \right )^2 (1-\tilde{n}_z)^{-1/2} \underline{\tilde{\mathbf{n}}}~\ud l\\ 
    \end{aligned}
\end{equation}
where the integral over $S_w^{(1)}(t)$ is transformed into a line integral over the waterline \citep{pinkster1980,lee1995wamit}.

\subsection{Solution in the frequency domain} \label{sec:freqdom_solution}
The principal advantage of the boundary value problem in orders is that it can be solved efficiently in the frequency domain, through the so-called radiation-diffraction solvers \citep{lee1995wamit, kurnia2023}, which are standard tools in engineering practice. The incident potential field is expressed in the frequency domain as transfer functions with respect to the potential amplitudes $B_n$,

\begin{equation} \label{eq:linear_inc_tf}
    \hat{\phi}_{I}^{(1)}(\omega_n,\mathbf{x})  =\frac{\cosh[k_n(z + h)]}{\cosh(k_nh)} \ue^{-\ui k_n x }
\end{equation}

\begin{equation}
    \hat{\phi}_{I}^{(2)}(\omega_m,\omega_n,\mathbf{x}) =\hat{\mathbb{T}}{}^{(2)}_{\phi}(\omega_m, \omega_n)\frac{\cosh  \left [ |k_m+k_n| (z+h) \right ]}{\cosh\left ( |k_m+k_n|h \right )}
    \ue^{-\ui(k_m+k_n) x}
\end{equation}
Therefore, the problem degenerates to finding the associated elementary radiation potentials $\underline{\boldsymbol{\phi}}{}_r(\omega_n, \mathbf{x})$ on the body and the first- and second-order scattering potentials $\phi^{(1)}_S(\omega_n, \mathbf{x})$ and $\phi^{(2)}_S(\omega_m, \omega_n, \mathbf{x})$. Then, letting $\hat{\phi}_{IS}^{(m)}=\hat{\phi}_{I}^{(m)}+\hat{\phi}_{S}^{(m)}$ for $m=1,2$, the potential force transfer functions from the incident and scattering wavefields are obtained as,

\begin{equation} \label{eq:linear_exc_force_tf}
    \hat{\underline{\mathbf{F}}}_{P_{IS}}^{(1)}(\omega_n) = \rho g \int_{S_0}  \hat{\phi}^{(1)}_{IS}(\omega_n, \tilde{\mathbf{x}})~\underline{\tilde{\mathbf{n}}} ~\ud S
\end{equation}

\begin{equation}
    \hat{\underline{\mathbf{F}}}_{P_{IS}}^{(2)}(\omega_m,\omega_n) = \ui \rho g \frac{\omega_m+\omega_n}{\omega_m \omega_n} \int_{S_0} \hat{\phi}^{(2)}_{IS}(\omega_m,\omega_n, \tilde{\mathbf{x}}) ~ \underline{\tilde{\mathbf{n}}} ~\ud S
\end{equation}
Moreover, letting $\hat{\underline{\boldsymbol{\xi}}}{}^{(1)}(\omega_n)$ be the RAOs of the body and $\hat{\underline{\boldsymbol{\xi}}}{}^{(2)}(\omega_m,\omega_n)$ the motion QTFs, the radiation potential is obtained as,

\begin{equation} \label{eq:rad_pot_freq}
     \hat{\phi}_R^{(1)}(\omega, \mathbf{x}) =  \ui \omega ~\hat{\underline{\boldsymbol{\phi}}}{}_r(\omega, \mathbf{x})  \cdot \hat{\underline{\boldsymbol{\xi}}}{}^{(1)}(\omega)
\end{equation}

\begin{equation}
     \hat{\phi}_R^{(2)}(\omega_m, \omega_n,\mathbf{x}) =  \ui (\omega_m + \omega_n)~ \hat{\underline{\boldsymbol{\phi}}}{}_r(\omega_m+\omega_n, \mathbf{x})  \cdot \hat{\underline{\boldsymbol{\xi}}}{}^{(2)}(\omega_m,\omega_n)
\end{equation}
Finally, inserting pairs of $\hat{\phi}^{(1)}(\omega,\mathbf{x})$ and $\hat{\underline{\boldsymbol{\xi}}}{}^{(1)}(\omega)$ in the frequency domain expression of \eqref{eq:quadratic_force_wamit}, provides the QTF of the quadratic force $\hat{\underline{\mathbf{F}}}_Q^{(2)}(\omega_m,\omega_n)$. It is noted that if the transfer functions are provided with respect to the wave elevation amplitudes $A_n$, they can be transformed to be applicable for the velocity potential amplitudes $B_n$ by scaling them with a factor of $\ui g/ \omega_n $.

\section{Quadratic Motion-Explicit (QME) approach} \label{sec:proposed_force_model}
In this section, a novel approach for evaluating hydrodynamic loads, termed the Quadratic Motion-Explicit (QME) approach, is introduced, extending the method of  \citet{bredmose2024forcemodel} to moving bodies. The same reference frames defined in Section \ref{sec:reference_frames} and the second-order consistent boundary value problem of Section \ref{sec:bvp_ordered} are considered. As shown in this section, the inclusion of higher-order contributions in the force calculation is possible in the time domain by employing the total nonlinear wave and motion variables, without decomposition into first- and second-order terms.

\subsection{Redefined body kinematics} \label{sec:redefined_body_kinematics}
A small parameter $\zeta$ is introduced, representative of the body motion amplitude. Similar to $\epsilon$, it is only used for tracking the leading-order terms in the force description that are related to the body motion, and no direct physical meaning is assigned. Moreover, instead of expanding the wave and motion variables into perturbation series, as \eqref{eq:series_potential} and \eqref{eq:series_motion}, the total nonlinear variables are used. Therefore, the concept of orders here is only introduced through the products between the individual wave and motion terms. The position and normal vectors in the global reference frame can be written as, 

\begin{equation} \label{eq:motion_vector_expansion}
    \begin{aligned}
        \mathbf{x} &= \tilde{\mathbf{x}} + \boldsymbol{\xi} + \boldsymbol{\alpha} \times \tilde{\mathbf{x}} + \boldsymbol{\mathcal{H}}\tilde{\mathbf{x}} + O(\zeta^3)  \\
        \underline{\mathbf{n}} &= \underline{\mathbf{\tilde{n}}} +  \tilde{\mathbf{N}}_{\underline{\boldsymbol{\xi}}}  + \tilde{\mathbf{N}}_{\underline{\boldsymbol{\xi}}\underline{\boldsymbol{\xi}}} + O(\zeta^3) 
    \end{aligned}
\end{equation}
where, as explained above, $O(\zeta^3)$ denotes terms that include tertiary and higher products between elements of the motion vector $\underline{\boldsymbol{\xi}}$. In addition, the matrix $\boldsymbol{\mathcal{H}}$ and operators $\tilde{\mathbf{N}}_{\underline{\boldsymbol{\xi}}}$, $\tilde{\mathbf{N}}_{\underline{\boldsymbol{\xi}}\underline{\boldsymbol{\xi}}}$ are defined equivalently as in Section \ref{sec:ordered_body_kinematics} as, 

\begin{equation}
    \boldsymbol{\mathcal{H}} = \begin{bmatrix}
    -\tfrac{1}{2}\left(\alpha_2^2 + \alpha_3^2\right) & 0 & 0\\
    \alpha_1 \alpha_2 & -\tfrac{1}{2}\left(\alpha_1^2 + \alpha_3^2\right) & 0 \\
    \alpha_1 \alpha_3 & \alpha_2 \alpha_3 & -\tfrac{1}{2}\left(\alpha_1^2 + \alpha_2^2\right) 
    \end{bmatrix} 
\end{equation}

\begin{equation} \label{eq:transform_operators}
    \begin{aligned}
        \tilde{\mathbf{N}}_{\underline{\boldsymbol{\xi}}} &= \begin{bmatrix}
           \boldsymbol{\alpha} \times \mathbf{\tilde{n}} \\
            \boldsymbol{\xi} \times \mathbf{\tilde{n}} + \boldsymbol{\alpha} \times (\tilde{\mathbf{x}} \times \mathbf{\tilde{n}})
        \end{bmatrix}  \qquad \qquad
                \tilde{\mathbf{N}}_{\underline{\boldsymbol{\xi}}\underline{\boldsymbol{\xi}}} = \begin{bmatrix}
           \boldsymbol{\mathcal{H}}\mathbf{\tilde{n}}  \\
            \boldsymbol{\xi} \times ( \boldsymbol{\alpha}\times \mathbf{\tilde{n}}) + \boldsymbol{\mathcal{H}}(\tilde{\mathbf{x}} \times \mathbf{\tilde{n}})
        \end{bmatrix}
    \end{aligned}
\end{equation}
It is worth noting that insertion of \eqref{eq:series_motion} into the above equations, and further neglecting terms of $O(\epsilon^3)$, provides a consistent result with \eqref{eq:posvec_expansion} and \eqref{eq:normvec_expansion}. In addition, the accuracy of \eqref{eq:motion_vector_expansion} compared to the fully nonlinear \eqref{eq:point_definition} does not depend on the amplitude of the translational motions, but rather the expansion of the rotational transformation matrix $\boldsymbol{\mathcal{R}}$.

\subsection{Boundary value problem}
Recalling the boundary value problem in orders from \eqref{eq:bvp_ordered}, it can be reformulated for the combined first- and second-order potential as,

\begin{equation} \label{eq:bvp_nonlin}
    \begin{aligned}
        \nabla^2 \phi= 0 &\qquad \mathbf{x} \in \mathbb{D}\\ 
        \partial_z \phi  =0 &\qquad z=-h \\ 
        \partial_{tt}\phi + g \partial_z\phi = \Theta_F &\qquad z=0 \\ 
        \nabla \phi \cdot \tilde{\mathbf{n}} = \partial_t \underline{\boldsymbol{\xi}}  \cdot \underline{\tilde{\mathbf{n}}} + \Theta_B &\qquad \mathbf{x}=\tilde{\mathbf{x}} \in S_0 \\
    \end{aligned}
\end{equation}
where $\Theta_F$ and $\Theta_B$, according to \eqref{eq:fs_forcing_functions} and \eqref{eq:body_forcing_functions}, are defined as,

\begin{equation} \label{eq:forcing_functions_nonlin}
    \begin{aligned}
        \Theta_F &=\frac{1}{g}\partial_t\phi~ \partial_z \left(\partial_{tt}\phi + g \partial_{z}\phi\right) - \partial_t\left ( \nabla \phi \cdot \nabla\phi \right ) \\
        \Theta_B &=  \tilde{\mathbf{n}} \cdot \partial_{t}\mathbf{H} \tilde{\mathbf{x}} + (\boldsymbol{\alpha} \times \tilde{\mathbf{n}}) \cdot \left ( \partial_{t}\boldsymbol{\xi} + \partial_{t}\boldsymbol{\alpha} \times 
    \tilde{\mathbf{x}}   - \nabla \phi   \right)  - \tilde{\mathbf{n}} \left [ \left ( \boldsymbol{\xi} + \boldsymbol{\alpha} \times 
    \tilde{\mathbf{x}}  \right) \cdot \nabla \right ] \nabla \phi 
    \end{aligned}
\end{equation}
Once again, equivalence with the classical boundary value problem in orders is attained by replacing \eqref{eq:series_motion} and \eqref{eq:series_potential} and also neglecting terms of $O(\epsilon^3)$ in the forcing functions $\Theta_F$ and $\Theta_B$. The same decomposition of the velocity potential is considered into the total incident, scattering and radiation components,

\begin{equation}
    \phi = \phi_I + \phi_S + \phi_R 
\end{equation}
In the rest of this section, the detailed expression and boundary conditions for each potential are developed in the time domain. Similar to the expression of the body kinematics in the previous section, the order $\epsilon$ is only related to products of the wave elevation and potential field, and does not imply perturbation of the quantities.

\subsubsection{Incident potential from fully nonlinear waves}
Several nonlinear wave theories employ a pseudo-spectral formulation for the incident velocity potential and the wave elevation, which for unidirectional waves follows

\begin{equation} \label{eq:hos_inc_pot}
\begin{aligned}
    \phi_I(\mathbf{x},t) &= \sum_{n=-N}^N \tilde{B}(k_{n},t) \frac{\text{cosh}~k_n(z+h)}{\text{cosh}~k_n h} \ue^{-\ui k_{n} x} \\ 
\eta(x,y,t) &= \sum_{n=-N}^N \tilde{A}(k_{n},t) \ue^{-\ui k_{n} x}
\end{aligned}
\end{equation} 
where $\tilde{B}$ and $\tilde{A}$ are time-dependent spatial modal amplitudes of complex value. 

An important aspect regarding this expression the similarity in the spatial structure of the nonlinear waves with a linear wavefield at each timestep, from \eqref{eq:linear_waves}, for which it would be $\tilde{B}(k_n,t)= B_n\ue^{\ui \omega_n t}$ and $\tilde{A}(k_n,t)=A_n\ue^{\ui \omega_n t}$, based on the linear dispersion relation. However, the time evolution of the potential field is nonlinearly resolved, and there is no direct relationship between $\tilde{A}(k_n,t)$ and $\tilde{B}(k_n,t)$. In addition, decomposition of the potential and wave elevation into first and second-order contributions is not strictly feasible. Nevertheless, as discussed above, the first- and second-order boundary value problems can be combined, satisfying the boundary value problem of \eqref{eq:nonlinear_laplace}-\eqref{eq:nonlinear_bbc} at second-order accuracy. The velocity potential from \eqref{eq:hos_inc_pot} satisfies exactly the Laplace equation and the impermeability condition at the bottom. The free-surface condition for the incident potential is satisfied with errors that are higher than second-order, and is given as,

\begin{equation} \label{eq:fsbc_inc_hosnwt}
    \partial_{tt}\phi_I + g \partial_z\phi_I = \Theta_{F,II} + O(\epsilon^3) \qquad z=0
\end{equation}
where $\Theta_{F,II}$ corresponds to the terms of $\Theta_{F}$ that only include quadratic interaction of $\phi_I$ with itself. It should be stressed that \eqref{eq:fsbc_inc_hosnwt} was initially derived from a Taylor expansion of the velocity potential around $z=0$ and also using the linear relationship $\eta_I(x,y) = -\partial_t\phi_I(x,y,0)/g$. Therefore, although fully nonlinear incident waves are considered, the formal accuracy of the boundary value problem remains second-order with respect to $\epsilon$.

\subsubsection{Radiation potential in the time domain} \label{sec:radiation_td}
The radiation potential is defined with respect to the total body velocity, 

\begin{equation} \label{eq:radpot_nonlin}
    \begin{aligned}
        \phi_R(\mathbf{x},t) = \boldsymbol{\underline{\phi}}{}_r(\mathbf{x},t)  \circledast  \partial_t \underline{\boldsymbol{\xi}}(t)=\int_{-\infty}^{\infty} \underline{\boldsymbol{\phi}}{}_r(\mathbf{x},t-\tau) \cdot  \partial_{\tau} \underline{\boldsymbol{\xi}}(\tau) ~\ud \tau
    \end{aligned}
\end{equation}
An important element of this expression for the radiation potential is that the radiation impulse response functions are fully defined from the solution of the first-order problem. The solution to the boundary value problem of the individual radiation potentials \citep{bingham_phd} is,

\begin{equation} \label{eq:phir_rad_td}
    \underline{\boldsymbol{\phi}}{}_r(\mathbf{x},t) = \underline{\boldsymbol{\phi}}{}_r^\infty(\mathbf{x}) \delta(t) + \underline{\boldsymbol{\psi}}{}_r(\mathbf{x},t)H(t)
\end{equation}
where $H$ and $\delta$ are Heavyside and Dirac functions. Moreover, $\underline{\boldsymbol{\phi}}{}_r^\infty$ denotes the infinite-frequency impulsive potential in 6DoF, and $ \underline{\boldsymbol{\psi}}{}_r(t)$ denotes the memory radiation kernel. Thus, the total radiation potential can be expressed by combining \eqref{eq:radpot_nonlin} and \eqref{eq:phir_rad_td}, as

\begin{equation}
    \begin{aligned}
        \phi_R(\mathbf{x},t) = \underline{\boldsymbol{\phi}}{}_r^\infty(\mathbf{x}) \cdot \partial_t \underline{\boldsymbol{\xi}}(t) + \int_0 ^t \underline{\boldsymbol{\psi}}{}_r(\mathbf{x},t-\tau) \cdot \partial_t \underline{\boldsymbol{\xi}}(\tau) ~ \ud \tau
    \end{aligned}
\end{equation}
where both elementary radiation components can be evaluated from linear frequency-domain analysis, similarly to the calculation of the added mass and radiation damping kernel, which will be further discussed in Section \ref{sec:rad_dif_output}

\subsubsection{Decomposition of the scattering potential} \label{sec:scattering_decomposition}
The scattering potential satisfies the Laplace equation and the bottom boundary condition, while it also completes the free-surface and the body boundary conditions, 

\begin{equation}
    \begin{aligned}
        \partial_{tt}\phi_S + g \partial_z\phi_S =  \Theta_{F,IB} + \Theta_{F,BB} & \qquad z =0 \\
        \nabla \phi_I \cdot \mathbf{\tilde{n}} + \nabla \phi_S \cdot \mathbf{\tilde{n}} = \Theta_B &\qquad \mathbf{x}=\tilde{\mathbf{x}} \in S_0
    \end{aligned}
\end{equation}
It can be further decomposed into three sub-components,

\begin{equation}
    \phi_S = \phi_{S_W} + \phi_{S_F} + \phi_{S_M}
\end{equation}
where $\phi_{S_W}$ satisfies the homogeneous free-surface boundary condition, and counteracts the effect of $\phi_I$ on the body,

\begin{equation} \label{eq:bc_lin_scattering}
    \begin{aligned}
        \partial_{tt}\phi_{S_W}+g\partial_z\phi_{S_W} = 0 &\qquad z=0 \\
        \nabla \phi_{S_W} \cdot \tilde{\mathbf{n}} = -\nabla \phi_I \cdot \tilde{\mathbf{n}} &\qquad \mathbf{x}=\tilde{\mathbf{x}} \in S_0 \\ 
    \end{aligned}
\end{equation}
The second term, $\phi_{S_F}$, satisfies the free-surface boundary condition, along with the homogeneous body boundary condition,

\begin{equation} \label{eq:bc_surface_potential}
    \begin{aligned}
        \partial_{tt}\phi_{S_F}+g\partial_z\phi_{S_F} = \Theta_{F,IB} + \Theta_{F,BB}  &\qquad z=0 \\
        \nabla \phi_{S_F} \cdot \tilde{\mathbf{n}} = 0 &\qquad \mathbf{x}=\tilde{\mathbf{x}} \in S_0 \\ 
    \end{aligned}
\end{equation} 
Finally, $\phi_{S_M}$ satisfies the homogeneous free-surface condition and the body boundary condition at the displaced position,

\begin{equation} \label{eq:bc_motion_potential}
    \begin{aligned}
        \partial_{tt}\phi_{S_M}+g\partial_z\phi_{S_M} = 0 &\qquad z=0 \\
        \nabla \phi_{S_M}\cdot \tilde{\mathbf{n}} =  \Theta_B  &\qquad \mathbf{x}=\tilde{\mathbf{x}} \in S_0 \\ 
    \end{aligned}
\end{equation}

The strategy behind such a decomposition of the scattering potential lies in the isolation of each scattering contribution, which, as shown in the following sections, facilitates the use of radiation-diffraction output for the expression of the forces.

\subsection{Use of linear radiation-diffraction output} \label{sec:rad_dif_output}
Before formulating the force calculation using the potentials described above, it is demonstrated that the boundary value problem does not need to be solved anew, but the output of a linear frequency-domain analysis can be leveraged to reconstruct the potentials. More specifically, the incident wave potential from \eqref{eq:hos_inc_pot} can be written as,

\begin{equation}
     \phi_I(\mathbf{x},t) =  \sum_{n=-N}^N \tilde{B}(k_{n},t)  ~\hat{\phi}_{I}^{(1)}(k_n,\mathbf{x}) 
\end{equation}
where $\hat{\phi}_{I}^{(1)}(k,\mathbf{x})$ is given by \eqref{eq:linear_inc_tf}. Similarly, the solution of the first-order boundary value problem in the frequency domain provides the linear scattering potential transfer function $\hat{\phi}_{S}^{(1)}(\omega,\mathbf{x})$. Recalling the body boundary condition from \eqref{eq:bc_lin_scattering}, the scattering potential can be reconstructed in the time domain as

\begin{equation}
     \phi_{S_W}(\mathbf{x},t) =  \sum_{n=-N}^N \hat{\phi}_{S}^{(1)}(k\{\omega_n\},\mathbf{x})~\tilde{B}(k_{n},t) 
\end{equation}
where $k\{\omega\}$ indicates that the frequency and wavenumber are related through the linear dispersion relation \eqref{eq:linear_dispersion_relation} and thus $k_n=k\{\omega_n\}$ is implied in the following transfer function expressions. The total diffraction potential $\phi_{W}=\phi_{I}+\phi_{S_W}$ can be reconstructed as,

\begin{equation} \label{eq:phiW_definition}
    \begin{aligned}
        \phi_{W}(\mathbf{x},t) &= \sum_{n=-N}^N\left [ \hat{\phi}^{(1)}_{I}(k_n,\mathbf{x})+\hat{\phi}^{(1)}_{S}(k_n,\mathbf{x})\right]\tilde{B}(k_n,t) \\ &=\sum_{n=-N}^N  \hat{\phi}^{(1)}_{IS}(k_n,\mathbf{x}) \tilde{B}(k_n,t)
    \end{aligned}
\end{equation}
which includes the fully nonlinear incident wavefield and the associated linear scattering effects. The velocity and relative wave elevation can also be evaluated as,

\begin{equation} \label{eq:hos-nwt_coefs}
    \begin{aligned}
        \nabla \phi_{W}(\mathbf{x},t) = \sum_{n=-N}^N \nabla\hat{\phi}^{(1)}_{IS}(k_n,\mathbf{x}) \tilde{B}(k_{n},t)  \qquad 
        \eta_{W}(x,y,t) = \sum_{n=-N}^N \hat{\eta}^{(1)}_{IS}(k_n,x,y) \tilde{A}(k_{n},t)     \end{aligned}
\end{equation}
where $\nabla\hat{\phi}^{(1)}_{IS}$ and $\hat{\eta}^{(1)}_{IS}$ are the respective transfer functions, which are also typically provided by standard radiation-diffraction solvers.

For the radiated wavefield and velocity, the analysis of Section \ref{sec:radiation_td} is employed, and the following expressions are obtained,

\begin{equation} \label{eq:radiated_waves_td}
    \begin{aligned}
     \nabla \phi_R(\mathbf{x},t) 
    &=\underline{\mathbf{u}}{}_r^\infty(\mathbf{x})  \cdot \partial_t \underline{\boldsymbol{\xi}}(t) + \int_0 ^t \boldsymbol{\underline{K}^{\mathbf{u}}}(\mathbf{x},t-\tau) \cdot \partial_{\tau} \underline{\boldsymbol{\xi}}(\tau) ~ \ud \tau \\ 
        \eta_R(x,y,t) 
    &=\underline{\boldsymbol{\eta}}{}_r^\infty(x,y) \cdot \partial_{tt}\underline{\boldsymbol{\xi}}(t) + \int_0 ^t  \boldsymbol{\underline{K}}^{\eta}(x,y,t-\tau) \cdot \partial_{\tau} \underline{\boldsymbol{\xi}}(\tau) ~ \ud \tau 
    \end{aligned}
\end{equation} 
where $ \underline{\boldsymbol{K}}^{\mathbf{u}}$  and $\underline{\boldsymbol{K}}^{\eta}$ denote the impulse response functions for the velocity and wave elevation in all 6 DoFs, while $\underline{\mathbf{u}}{}_r^\infty$ and $\boldsymbol{\underline{\eta}}{}_r^\infty$ denote the respective infinite frequency terms. Letting $\underline{\hat{\mathbf{u}}}{}_r(\omega,\mathbf{x})$ and $\underline{\hat{\boldsymbol{\eta}}}{}_r(\omega,x,y)$ denote the transfer functions for the elementary radiation velocity on the body and wave elevation along the waterline, the impulse response functions are obtained by applying the convolution theorem to \eqref{eq:radiated_waves_td} as,

\begin{equation}
    \begin{aligned}
        \underline{\boldsymbol{K}}^{\mathbf{u}}(\mathbf{x},\tau) &= \frac{2}{\pi}\int_{0}^{\infty}   \Re \Big \{\underline{\hat{\mathbf{u}}}{}_r(\omega,\mathbf{x})  - \underline{\hat{\mathbf{u}}}{}_r(\omega^{\infty},\mathbf{x})  \Big \}\text{cos} ~\omega \tau ~\ud \omega \\
        \underline{\boldsymbol{K}}^{\eta}(x,y,\tau) &= \frac{2}{\pi}\int_{0}^{\infty}  \Re \left \{ \underline{\hat{\boldsymbol{\eta}}}{}_r(\omega,x,y)   \right \} \text{cos} ~\omega \tau ~ \ud \omega
    \end{aligned}
\end{equation}
where $\omega^{\infty}$ denotes the infinite frequency limit. The infinite frequency terms can be obtained in an exactly similar manner as in the case of the infinite frequency added mass \citep{cummins1962impulse}, 

\begin{equation}
    \underline{\mathbf{u}}{}_r^\infty(\mathbf{x}) = \Re \Big \{ \underline{\hat{\mathbf{u}}}{}_r(\omega^{\infty},\mathbf{x}) \Big \} \qquad \qquad \underline{\boldsymbol{\eta}}{}_r^\infty(x,y) = \Im \Big \{  \underline{\hat{\boldsymbol{\eta}}}{}_r(\omega^{\infty},x,y)/\omega^{\infty}  \Big \}
\end{equation} 
where $\Re$ and $\Im$ denote the real and imaginary parts.

\subsection{Hydrodynamic loads} \label{sec:hyd_loads_forcemodel}
Following the same course as the classical second-order approach, and assuming that the total nonlinear motion $\underline{\boldsymbol{\xi}}$ remains small, the pressure is developed in a Taylor series around the mean body position,

\begin{equation} \label{eq:pressure_taylor_nonlin}
    p(\mathbf{x},t) = p(\tilde{\mathbf{x}},t) + \Big ( \boldsymbol{\xi}(t) +\boldsymbol{\alpha}(t)\times \tilde{\mathbf{x}} +  \boldsymbol{\mathcal{H}}(t) \tilde{\mathbf{x}}    \Big ) \cdot \nabla p(\tilde{\mathbf{x}},t) + O(\zeta^2\epsilon)
\end{equation}
Inserting the Bernoulli equation \eqref{eq:bernoulli_x}, the total pressure will follow,

\begin{equation} \label{eq:pressure_nl_formulation}
    \begin{aligned} 
        p(\mathbf{x},t) &= - \rho  \bigg [ g(\tilde{z}+\tilde{z}_0) + \partial_t \phi(\tilde{\mathbf{x}},t) + g \xi_r(\tilde{x}, \tilde{y}, t)  + \frac{1}{2} \nabla \phi(\tilde{\mathbf{x}},t) \cdot \nabla \phi(\tilde{\mathbf{x}},t)   \\ & \qquad\qquad+   \Big(\boldsymbol{\xi}(t) + \boldsymbol{\alpha}(t) \times \tilde{\mathbf{x}} \Big) \cdot \nabla \partial_t\phi(\tilde{\mathbf{x}},t)  + g \boldsymbol{\mathcal{H}}(t) \tilde{\mathbf{x}} \cdot \nabla \tilde{z}  \bigg ]
    \end{aligned}
\end{equation}
where $\xi_r=\xi_3 + \alpha_1\tilde{y}- \alpha_2 \tilde{x}$. Inserting expression \eqref{eq:pressure_nl_formulation} into \eqref{eq:force_exact} the total force follows,

\begin{equation} \label{eq:force_moving_body}
\begin{aligned}
        \underline{\mathbf{F}}(t) &= -\rho \int_{S_0} \bigg[ g(\tilde{z}+\tilde{z}_0) + \partial_t \phi(\tilde{\mathbf{x}},t)  + g \xi_r(\tilde{x}, \tilde{y}, t)  + \frac{1}{2} \nabla \phi(\tilde{\mathbf{x}},t) \cdot \nabla \phi(\tilde{\mathbf{x}},t) + g \boldsymbol{\mathcal{H}}(t) \tilde{\mathbf{x}} \cdot \nabla \tilde{z}\\ 
        &\qquad \quad + \Big ( \boldsymbol{\xi}(t) + \boldsymbol{\alpha}(t) \times \tilde{\mathbf{x}} \Big) \cdot \nabla \partial_t \phi(\tilde{\mathbf{x}},t) \bigg] ~ \Big (\underline{\tilde{\mathbf{n}}} +  \tilde{\mathbf{N}}_{\underline{\boldsymbol{\xi}}}(t) + \tilde{\mathbf{N}}_{\underline{\boldsymbol{\xi}}\underline{\boldsymbol{\xi}}}(t) \Big ) ~ \ud S \\ 
        & \qquad \quad +\frac{1}{2} \rho g\int_{\mathrm{WL}}  \Big (\eta(\tilde{x},\tilde{y},t)-g\xi_r(\tilde{x}, \tilde{y}, t) \Big )^2 \left(1-\tilde{n}_z^2\right)^{-1/2} \underline{\tilde{\mathbf{n}}} ~\ud l \\
    &= \underline{\mathbf{F}}_B(t) + \underline{\mathbf{F}}_P(t) + \underline{\mathbf{F}}_H(t) + \underline{\mathbf{F}}_Q (t)
\end{aligned}
\end{equation}
It shall be noted that the force thus obtained becomes equivalent to the loads of Section \ref{sec:ordered_wave_loads} if \eqref{eq:series_motion} and \eqref{eq:series_potential} are inserted, along with neglecting terms of $O(\epsilon^3)$. This proves that the general expression \eqref{eq:force_moving_body} is second-order-consistent, since all terms included are of at least second order. However, the present formulation allows for the incorporation of some higher-order contributions through products between the nonlinear velocity potential and body motions. Proceeding with the development of each term, the hydrostatic loads $ \underline{\mathbf{F}}_B$ remain unchanged from Section \ref{sec:ordered_wave_loads}. Similarly, the restoring loads are defined proportional to the nonlinear motion $\underline{\boldsymbol{\xi}}$,

\begin{equation}\label{eq:restoring_loads_nonlin_mot}
            \underline{\mathbf{F}}_H(t) =-\rho g\int_{S_0} \left( \xi_r(\tilde{x}, \tilde{y}, t) \underline{\tilde{\mathbf{n}}}  + (\tilde{z} + \tilde{z}_0)\mathbf{N} _{\underline{\boldsymbol{\xi}}}(t) \right )\ud S  = - \boldsymbol{\mathcal{C}}_H  \underline{\boldsymbol{\xi}}(t)
\end{equation} 

\subsubsection{Potential force}

Integrating the time derivative of the total nonlinear potential on the mean body surface provides the so-called potential force, 

\begin{equation}
\begin{aligned}
\underline{\mathbf{F}}_P(t) &=-\rho \int_{S_0} \partial_t\phi(\tilde{\mathbf{x}},t) ~\underline{\tilde{\mathbf{n}}} ~\ud S= -\rho \int_{S_0} \partial_t \Big ( \phi_I(\tilde{\mathbf{x}},t) +\phi_S(\tilde{\mathbf{x}},t) + \phi_R(\tilde{\mathbf{x}},t) \Big ) ~ \underline{\tilde{\mathbf{n}}} ~\ud S  \\ &= \underline{\mathbf{F}}_{P_{IS}}(t) + \underline{\mathbf{F}}_{P_R}(t)
\end{aligned}
\end{equation}
Due to the definition of the radiation potential in \eqref{eq:radpot_nonlin}, the corresponding force can be formulated in accordance with the standard approach \citep{cummins1962impulse}. More precisely, using the infinite frequency added mass $\boldsymbol{\mathcal{M}}^{\infty}_A$ and the radiation convolution kernel $\boldsymbol{\mathcal{K}}$, the respective force will be proportional to the total body velocity and acceleration,

\begin{equation} \label{eq:radiation_force_nonlinear}
   \underline{\mathbf{F}}_{P_R}(t) = -\boldsymbol{\mathcal{M}}^{\infty}_A  \partial_{tt}\underline{\boldsymbol{\xi}}(t) - \int_{0}^t \boldsymbol{\mathcal{K}}(t-\tau) \partial_{\tau}\underline{\boldsymbol{\xi}}(\tau) ~\ud \tau 
\end{equation}
Then, the following decomposition can be made for the remaining terms,

\begin{equation} \label{eq:FPFully}
\begin{aligned}
\underline{\mathbf{F}}_{P_{IS}}(t) 
& = -\rho \int_{S_0} \partial_t\Big (\phi_W(\tilde{\mathbf{x}},t) +\phi_{S_F}(\tilde{\mathbf{x}},t) +\phi_{S_M}(\tilde{\mathbf{x}},t) \Big ) \underline{\tilde{\mathbf{n}}} ~\ud S = \underline{\mathbf{F}}_{P_W}(t) + \underline{\mathbf{F}}_{P_F}(t) + \underline{\mathbf{F}}_{P_M}(t) 
\end{aligned}
\end{equation}
Starting with the first component, use of \eqref{eq:phiW_definition} provides,

\begin{equation} \label{eq:force_potential_nonlin}
    \begin{aligned}
        \underline{\mathbf{F}}_{P_W}(t) & =-\rho \int_{S_0} \partial_t \phi_W(\tilde{\mathbf{x}},t)  ~ \underline{\tilde{\mathbf{n}}} ~\ud S = -\rho \sum_{n=-N}^N \partial_t\tilde{B}(k_n,t)\int_{S_0} \hat{\phi}^{(1)}_{IS}(k_n,\tilde{\mathbf{x}})\underline{\tilde{\mathbf{n}}} ~\ud S \\
        &=  -\frac{1}{g}\sum_{n=-N}^N\hat{\underline{\mathbf{F}}}_{P_{IS}}^{(1)}(k_n)\tilde{B}_t(k_n,t)
    \end{aligned}
\end{equation} 
where $\tilde{B}_t(k,t)$ corresponds to the modal amplitudes of the velocity potential time derivative, and the linear excitation force transfer function $\hat{\underline{\mathbf{F}}}_{P_{IS}}$ is defined by \eqref{eq:linear_exc_force_tf}.

Such implementation of the time derivative in \eqref{eq:force_potential_nonlin} retains the nonlinearity of the force obtained, and incorporates bound wave effects, since no assumption is made regarding the link between the frequency and wavenumber of each wave component. Nevertheless, it is noted that applying a linear force transfer function, and considering the same wavenumber for $\phi_{S_W}$ and $\phi_I$ constitutes an approximation. In particular, $\phi_{S_W}$ satisfies the linear dispersion relation, as suggested by \eqref{eq:bc_lin_scattering}, in contrast with $\phi_I$, which follows the inhomogeneous free-surface boundary condition \eqref{eq:fsbc_inc_hosnwt}. As demonstrated also numerically in Section \ref{sec:forcemodel_validation}, for a second-order incident wavefield, this expression is fully equivalent to the Pinkster approximation \citep{pinkster1980}.

Moreover, the potential $\phi_{S_F}$ is related to the free-surface boundary condition $\Theta_F$, and therefore, the evaluation of the associated potential force $\underline{\mathbf{F}}_{P_F}$ requires solution of the second-order boundary value problem of Section \ref{sec:bvp_ordered}. This procedure is cumbersome and computationally intensive, and for low-frequency responses, it is a common approach to omit its contribution \citep{de_hauteclocque_review_2012}, through the use of the indirect method \citep{second_order_molin_1979}. The contribution of $\phi_{S_F}$ can be evaluated at the additional expense of solving the second-order radiation-diffraction problem with an adequately refined free-surface panel mesh. More specifically, appropriate manipulation of the potential force QTF provides,

\begin{equation} \label{eq:pot_force_pf}
    \begin{aligned}
         \underline{\mathbf{F}}_{P_F}= \sum_{m=-N}^N \sum_{n=-N}^N \left ( \hat{\underline{\mathbf{F}}}_{P_{IS}}^{(2)}(k_m, k_n) - \hat{\underline{\mathbf{F}}}_{P_{IS},Pinkster}^{(2)}(k_m, k_n) \right ) \tilde{B}(k_m,t) \tilde{B}(k_n,t)
    \end{aligned}
\end{equation}
where $\hat{\underline{\mathbf{F}}}_{P_{IS}, Pinkster}^{(2)}$ denotes the potential force QTF, obtained under the Pinskter approximation. This correction considers the quadratic free-surface forcing function \eqref{eq:fs_forcing_functions} in the frequency domain. Thus, the radiation potential included on the right-hand side of the forcing function is based on the first-order body motions and is not treated as in \eqref{eq:radpot_nonlin}. Nevertheless, low-frequency loads are primarily of interest in this study, and therefore, this practical correction is considered sufficient. 

The evaluation of $\underline{\mathbf{F}}_{P_M}$ does not require explicit knowledge of $\phi_{S_M}$, but according to the analysis of Appendix \ref{sec:F_P_M} can be expressed as,

\begin{equation} \label{eq:fpm_nonlin}
    \begin{aligned}
        \underline{\mathbf{F}}_{P_M}(t) = -\rho \int_{S_0}  \partial_{t} \Theta_B(\tilde{\mathbf{x}}, t)~\underline{\boldsymbol{\phi}}{}^{\infty}_r(\tilde{\mathbf{x}})~\ud S -\rho \int_0^t\int_{S_0}  \partial_{\tau} \Theta_B(\tilde{\mathbf{x}}, \tau)~\underline{\boldsymbol{\psi}}{}_r(\tilde{\mathbf{x}},t-\tau)~\ud S ~\ud \tau
    \end{aligned}
\end{equation}
and thus can be evaluated through the time-variant body forcing function $\Theta_B$ and the elementary radiation quantities.

\subsubsection{Quadratic force}
The quadratic force contains all terms arising from the quadratic interaction of wave and motion quantities and is given by,

\begin{equation} \label{eq:force_quadratic_nonlin}
    \begin{aligned}
        \underline{\mathbf{F}}_Q(t) &= -\rho \int_{S_0} \left [ \frac{1}{2} \nabla \phi(\tilde{\mathbf{x}},t) \cdot \nabla \phi(\tilde{\mathbf{x}},t) ~\underline{\tilde{\mathbf{n}}}+ \Big (\boldsymbol{\xi}(t) + \boldsymbol{\alpha}(t) \times \tilde{\mathbf{x}} \Big ) \cdot \nabla \partial_t \phi(\tilde{\mathbf{x}},t) ~\underline{\tilde{\mathbf{n}}}   \right ] ~\ud S \\
     & \qquad  - \rho \int_{S_0}\tilde{\mathbf{N}}_{\underline{\boldsymbol{\xi}}} \partial_t \phi(\tilde{\mathbf{x}},t)~\ud S  +\frac{1}{2} \rho g\int_{\mathrm{WL}}  [\eta(\tilde{x},\tilde{y},t)-g\xi_r(\tilde{x}, \tilde{y}, t)]^2 \left(1-\tilde{n}_z^2\right)^{-1/2} \underline{\tilde{\mathbf{n}}} ~\ud l \\ 
     &\qquad \qquad   \rho g \int_{S_0} \left [  \xi_r(\tilde{x}, \tilde{y}, t)  \tilde{\mathbf{N}}_{\underline{\boldsymbol{\xi}}}(t) + (\tilde{z}+\tilde{z}_0) \tilde{\mathbf{N}}_{\underline{\boldsymbol{\xi}}\underline{\boldsymbol{\xi}}}(t) + (\boldsymbol{\mathcal{H}}(t) \tilde{\mathbf{x}} \cdot \nabla \tilde{z})~\underline{\tilde{\mathbf{n}}} \right ] \ud S  \\ 
     & = \underline{\mathbf{F}}^{QP}_{Q}(t) + \underline{\mathbf{F}}_{Q}^D(t) + \underline{\mathbf{F}}_{Q}^A(t) + \underline{\mathbf{F}}_{Q}^{WL}(t) + \underline{\mathbf{F}}^H_{Q}(t)
    \end{aligned}
\end{equation} 
At this point, it should be stressed that the scattering potentials $\phi_{S_F}$ and $\phi_{S_M}$, as defined from the boundary conditions in \eqref{eq:bc_surface_potential} and \eqref{eq:bc_motion_potential}, involve $O(\epsilon^2,\epsilon \zeta,\zeta^2)$ terms and thus their products will involve tertiary or higher-order interactions between $\phi$ and $\underline{\boldsymbol{\xi}}$. Therefore, in the evaluation of the quadratic force $\mathbf{F}_Q$, the scattering potential contribution can be limited to the $\phi_{S_W}$ component, which considerably simplifies the analysis.

Starting with the so-called quadratic pressure force, it is decomposed into the following components,

\begin{equation} \label{eq:quadratic_pressure_nonlinear} 
    \begin{aligned} 
       \underline{\mathbf{F}}_{Q}^{QP}(t) &= -\frac{1}{2}\rho \int_{S_0} \nabla \Big( \phi_{W}(\tilde{\mathbf{x}},t)+\phi_R(\tilde{\mathbf{x}},t)\Big)\cdot \nabla \Big(\phi_{W}(\tilde{\mathbf{x}},t)+\phi_R(\tilde{\mathbf{x}},t)\Big)  ~\underline{\tilde{\mathbf{n}}}~ \ud S\\
        &=  -\frac{1}{2}\rho \int_{S_0} \nabla \phi_{W}(\tilde{\mathbf{x}},t)\cdot \nabla \phi_{W} (\tilde{\mathbf{x}},t) ~\underline{\tilde{\mathbf{n}}}~ \ud S -\rho \int_{S_0} \nabla \phi_{W}(\tilde{\mathbf{x}},t)\cdot \nabla \phi_R(\tilde{\mathbf{x}},t)  ~\underline{\tilde{\mathbf{n}}}~ \ud S \\
        &\qquad \qquad- \frac{1}{2}\rho \int_{S_0} \nabla \phi_R(\tilde{\mathbf{x}},t)\cdot \nabla \phi_R(\tilde{\mathbf{x}},t)  ~\underline{\tilde{\mathbf{n}}}~ \ud S \\ 
        &=\underline{\mathbf{F}}_{Q_{WW}}^{QP}(t) + \underline{\mathbf{F}}_{Q_{WR}}^{QP}(t) + \underline{\mathbf{F}}_{Q_{RR}}^{QP}(t)
    \end{aligned}
\end{equation} 
Upon insertion of \eqref{eq:hos-nwt_coefs}, the following relation is obtained for $\underline{\mathbf{F}}_{Q_{WW}}^{QP}$, 

\begin{equation}  \label{eq:quadratic_pressure_ww}
    \begin{aligned}
        \underline{\mathbf{F}}_{Q_{WW}}^{QP}(t) &=  -\frac{1}{2}\rho \int_{S_0} \left ( \sum_{m=-N}^N \nabla \hat{\phi}^{(1)}_{IS}(k_m,\tilde{\mathbf{x}}) \tilde{B}(k_m,t) \right ) \cdot \left ( \sum_{n=-N}^N \nabla\hat{\phi}^{(1)}_{IS}(k_n,\tilde{\mathbf{x}}) \tilde{B}(k_n,t)  \right )  ~\underline{\tilde{\mathbf{n}}} ~\ud S\\
         &= \sum_{m=-N}^N\sum_{n=-N}^N \underline{\boldsymbol{\mathcal{Q}}}{}^{QP}_{WW}(k_m, k_n) ~\tilde{B}(k_m,t)\tilde{B}(k_n,t)
    \end{aligned}
\end{equation}
where $\underline{\boldsymbol{\mathcal{Q}}}{}^{QP}_{WW}$ denotes a quadratic transfer function in all 6 DoFs, which can be reconstructed using the precomputed linear diffraction velocity transfer functions as,

\begin{equation}
\underline{\boldsymbol{\mathcal{Q}}}^{QP}_{WW}(k_m, k_n) = -\frac{1}{2}\rho\int_{S_0}\nabla\hat{\phi}^{(1)}_{IS}(k_m,\tilde{\mathbf{x}})\cdot \nabla \hat{\phi}^{(1)}_{IS}(k_n,\tilde{\mathbf{x}}) ~ \underline{\tilde{\mathbf{n}}} ~\ud S \quad 
\end{equation} 
Performing the double summation in \eqref{eq:quadratic_pressure_ww} at each time step may be computationally intensive. However, this procedure can be greatly accelerated using the approach of \cite{bredmose_second-order_2021}, which reduces the number of frequency-wavenumber components in the double summation, based on an eigenvalue decomposition of the QTF. 

The second quadratic pressure component, upon substituting \eqref{eq:radiated_waves_td} for the radiated velocity in the time domain, is developed as,

\begin{equation}
    \begin{aligned}
\underline{\mathbf{F}}_{Q_{WR}}^{QP}(t) &= \sum_{n=-N}^N \tilde{B}(k_n,t) \left ( \underline{\boldsymbol{\mathcal{L}}}{}^{QP}_{WR}(k_n) \partial_t\underline{\pmb{\xi}}(t) +\int_{0}^t~ \underline{\boldsymbol{\mathcal{Q}}}{}^{QP}_{WR}(k_n,t-\tau)   \partial_{\tau}\underline{\pmb{\xi}}(\tau) \ud \tau \right )
    \end{aligned}
\end{equation}
where $\underline{\boldsymbol{\mathcal{L}}}{}^{QP}_{WR}$ and $\underline{\boldsymbol{\mathcal{Q}}}{}^{QP}_{WR}$ are defined as,

\begin{equation}
    \begin{aligned}
        \underline{\boldsymbol{\mathcal{L}}}{}^{QP}_{WR}(k_n) & = -\rho\int_{S_0} \left ( \nabla \hat{\phi}^{(1)}_{IS}(k_n,\tilde{\mathbf{x}}) \underline{\mathbf{u}}{}_r^\infty(\tilde{\mathbf{x}}) \right )~\underline{\tilde{\mathbf{n}}} ~\ud S\\
        \underline{\boldsymbol{\mathcal{Q}}}{}^{QP}_{WR}(k_n, t-\tau) &= -\rho\int_{S_0} \left ( \nabla\hat{\phi}^{(1)}_{IS}(k_n,\tilde{\mathbf{x}}) \underline{\boldsymbol{K}}^{\mathbf{u}}(\tilde{\mathbf{x}},t-\tau) \right ) ~ \underline{\tilde{\mathbf{n}}} ~\ud S  
    \end{aligned}
\end{equation}
The last term is treated in a similar manner and yields,

\begin{equation} \label{eq:Fqp_RR}
    \begin{aligned} 
        \underline{\mathbf{F}}_{Q_{RR}}^{QP}(t)
        & = \partial_t\underline{\pmb{\xi}} (t)  ~\underline{\boldsymbol{\mathcal{N}}}{}_{RR}^{QP} ~ \partial_t\underline{\pmb{\xi}}(t) + \partial_t\underline{\pmb{\xi}} (t) \int_{0}^t \underline{\boldsymbol{\mathcal{K}}}{}^{QP}_{RR}(t-\tau)  \partial_{\tau}\underline{\pmb{\xi}}(\tau) ~\ud \tau  \\ 
        & \qquad \quad + \int_{0}^t \int_0^t \partial_{\tau_1}\underline{\pmb{\xi}}(\tau_1) \underline{\boldsymbol{\mathcal{Q}}}{}^{QP}_{RR}(t-\tau_{1}, t-\tau_{2})  \partial_{\tau_2}\underline{\pmb{\xi}}(\tau_2) \ud \tau_1 \ud \tau_2
    \end{aligned}
\end{equation}
where, letting $\otimes$ denote the tensor product, the different operators are defined as, 

\begin{equation}
    \begin{aligned}
        \underline{\boldsymbol{\mathcal{N}}}{}^{QP}_{RR} &= - \frac{1}{2}\rho \int_{S_0} \Big ( \underline{\mathbf{u}}{}_r^{\infty}(\tilde{\mathbf{x}}) \otimes \underline{\mathbf{u}}{}_r^{\infty}(\tilde{\mathbf{x}}) \Big ) ~ \underline{\tilde{\mathbf{n}}} ~ \ud S \\
        \underline{\boldsymbol{\mathcal{K}}}{}^{QP}_{RR}(t-\tau) &= - \rho \int_{S_0} \Big ( \underline{\mathbf{u}}{}_r^{\infty}(\tilde{\mathbf{x}}) \otimes \underline{\boldsymbol{K}}^{\mathbf{u}}(\tilde{\mathbf{x}},t-\tau) \Big ) ~ \underline{\tilde{\mathbf{n}}} ~ \ud S \\
        \underline{\boldsymbol{\mathcal{Q}}}{}^{QP}_{RR}(t-\tau_1, t-\tau_2) &= -\frac{1}{2}\rho\int_{S_0} \Big ( \underline{\boldsymbol{K}}^{\mathbf{u}}(\tilde{\mathbf{x}},t-\tau_1)\otimes  \underline{\boldsymbol{K}}^{\mathbf{u}}(\tilde{\mathbf{x}},t-\tau_2) \Big ) ~ \underline{\tilde{\mathbf{n}}} ~\ud S  
    \end{aligned}
\end{equation}

It is shown that, decomposition of the total potential into the individual diffraction and radiation contributions, replaces the application of frequency-domain QTFs with a combination of QTFs ($\underline{\boldsymbol{\mathcal{Q}}}{}^{QP}_{WW}$), two-dimensional convolution kernels ($\underline{\boldsymbol{\mathcal{Q}}}{}^{QP}_{RR}$) and hybrid transfer-function/kernel structures ($\underline{\boldsymbol{\mathcal{Q}}}{}^{QP}_{WR}$). Similar analysis can be applied for the waterline integral term $\underline{\mathbf{F}}_Q^{WL}$ and therefore its explicit derivation is omitted here but can be found in \cite{dermatis_phd}. 

Proceeding with the term that arises from the interaction of the potential time derivative $\partial_t\phi$ with the normal vector operator $\tilde{\mathbf{N}}_{\underline{\boldsymbol{\xi}}}$,

\begin{equation}
\begin{aligned}
    \underline{\mathbf{F}}_{Q}^A(t) &= - \rho \int_{S_0} \tilde{\mathbf{N}}_{\underline{\boldsymbol{\xi}}}(t)\partial_t\phi(\tilde{\mathbf{x}},t) ~ \ud S  =- \rho \int_{S_0} \tilde{\mathbf{N}}_{\underline{\boldsymbol{\xi}}}(t)\partial_t \Big ( \phi_{W}(\tilde{\mathbf{x}},t)+\phi_{R}(\tilde{\mathbf{x}},t) \Big ) ~ \ud S \\& =\underline{\mathbf{F}}_{Q_W}^A(t) +  \underline{\mathbf{F}}_{Q_R}^A(t)
\end{aligned}
\end{equation}
Recalling the skew-symmetric matrix of the rotation vector in \eqref{eq:rot_skew_symmetric}, the equivalent matrix for nonlinear rotational motions is denoted $\boldsymbol{\mathcal{A}}_{\times}$ and for translational motions $\boldsymbol{\mathcal{\Xi}}_{\times}$. Therefore, employing the expression of the transformation operator $\tilde{\mathbf{N}}_{\underline{\boldsymbol{\xi}}}$ from \eqref{eq:transform_operators}, both terms can be expressed in a concise form as a function of the respective potential force terms,

\begin{equation}
    \begin{aligned}
        \underline{\mathbf{F}}^A_{Q_W}(t) = \begin{bmatrix}
            \boldsymbol{\mathcal{A}}_{\times}(t) & [0]_{3\times3} \\
            \boldsymbol{\mathcal{\Xi}}_{\times}(t) & \boldsymbol{\mathcal{A}}_{\times}(t) 
        \end{bmatrix}  \underline{\mathbf{F}}_{P_W}(t) \qquad  \qquad  \underline{\mathbf{F}}_{Q_R}^A(t) = \begin{bmatrix}
            \boldsymbol{\mathcal{A}}_{\times}(t) & [0]_{3\times3} \\
            \boldsymbol{\mathcal{\Xi}}_{\times}(t) & \boldsymbol{\mathcal{A}}_{\times}(t) 
        \end{bmatrix}  \underline{\mathbf{F}}_{P_R}(t)
    \end{aligned}
\end{equation}

Furthermore, the Taylor expansion of the pressure around the mean body position gives rise to the following term,

\begin{equation}
    \begin{aligned} 
            \underline{\mathbf{F}}_{Q}^D(t) &=  -\rho \int_{S_0}  \Big(\pmb{\xi}(t)+\pmb{\alpha}(t)\times \tilde{\mathbf{x}} \Big)\cdot \nabla \partial_t\phi(\tilde{\mathbf{x}},t) ~\underline{\tilde{\mathbf{n}}} ~\ud S  \\ 
            &=  -\rho \int_{S_0}  \Big(\pmb{\xi}(t)+\pmb{\alpha}(t)\times \tilde{\mathbf{x}}  \Big)\cdot \nabla \partial_t \Big (\phi_{W} (\tilde{\mathbf{x}},t) +\phi_{R}(\tilde{\mathbf{x}},t) \Big )~\underline{\tilde{\mathbf{n}}} ~\ud S \\
            &= \underline{\mathbf{F}}_{Q_W}^D(t) + \underline{\mathbf{F}}_{Q_R}^D(t)
    \end{aligned}
\end{equation}
The incident and scattering wave contribution is developed using \eqref{eq:hos-nwt_coefs} as,

\begin{equation}
    \begin{aligned}
        \underline{\mathbf{F}}^D_{Q_W}(t) & = -\rho \int_{S_0} \boldsymbol\xi(t) \cdot \nabla \partial_t\phi_{W}(\tilde{\mathbf{x}},t)~\underline{\tilde{\mathbf{n}}} ~\ud S - \rho \int_{S_0} \Big (\boldsymbol\alpha(t) \times \tilde{\mathbf{x}}\Big ) \cdot \nabla \partial_t\phi_{W}(\tilde{\mathbf{x}},t) ~\underline{\tilde{\mathbf{n}}} ~\ud S \\
        & = \underline{\boldsymbol{\xi}}(t) \left (\sum_{n=-N}^N \underline{\boldsymbol{\mathcal{L}}}{}_{W}^{D}(k_n) \tilde{B}_t(k_n,t) \right )
    \end{aligned}
\end{equation}
with the linear transfer function $\underline{\boldsymbol{\mathcal{L}}}{}_{W}^{D}$ being

\begin{equation}
    \underline{\boldsymbol{\mathcal{L}}}{}_{W}^{D}(k_n) = -\rho \int_{S_0}\begin{bmatrix} \nabla \hat{\phi}^{(1)}_{IS}(k_n,\tilde{\mathbf{x}}) \\ \tilde{\mathbf{x}} \times \nabla \hat{\phi}^{(1)}_{IS}(k_n,\tilde{\mathbf{x}}) \end{bmatrix}^T ~\underline{\tilde{\mathbf{n}}} ~\ud S
\end{equation}
and for the radiation term, replacing \eqref{eq:radiated_waves_td} provides, 

\begin{equation} \label{eq:disp_pot_force_rad}
    \begin{aligned}
        \underline{\mathbf{F}}^D_{Q_R}(t) & = -\rho \int_{S_0} \boldsymbol\xi(t) \cdot \nabla \partial_t\phi_{R} (\tilde{\mathbf{x}},t)~\underline{\tilde{\mathbf{n}}} ~\ud S - \rho \int_{S_0} \Big(\boldsymbol\alpha(t) \times \tilde{\mathbf{x}}\Big) \cdot \nabla \partial_t\phi_{R}(\tilde{\mathbf{x}},t) ~\underline{\tilde{\mathbf{n}}} ~\ud S \\
        & = \underline{\boldsymbol{\xi}}(t)  \left (\underline{\boldsymbol{\mathcal{N}}}{}_{R}^{D} \partial_{tt}\underline{\boldsymbol{\xi}}(t) +  \int_0^t \underline{\boldsymbol{\mathcal{K}}}{}_{R}^{D}(t-\tau) \partial_{\tau}\underline{\boldsymbol{\xi}}(\tau) ~\ud \tau \right )
    \end{aligned}
\end{equation}
with $\underline{\boldsymbol{\mathcal{N}}}{}_{R}^{D}$ and $\underline{\boldsymbol{\mathcal{K}}}{}_{R}^{D}$ being

\begin{equation}
\begin{aligned}
    \underline{\boldsymbol{\mathcal{N}}}_{R}^{D} & =-\rho \int_{S_0}\begin{bmatrix}  \underline{\mathbf{u}}{}_r^\infty(\tilde{\mathbf{x}}) \\ \tilde{\mathbf{x}} \times  \underline{\mathbf{u}}{}_r^\infty(\tilde{\mathbf{x}}) \end{bmatrix}^T ~\underline{\tilde{\mathbf{n}}} ~\ud S \\
    \underline{\boldsymbol{\mathcal{K}}}{}_{R}^{D}(t-\tau) &= -\rho \int_{S_0}\begin{bmatrix}  \partial_t\underline{\boldsymbol{K}}{}^{\mathbf{u}}(\tilde{\mathbf{x}},t-\tau) \\ \tilde{\mathbf{x}} \times  \partial_t\underline{\boldsymbol{K}}{}^{\mathbf{u}}(\tilde{\mathbf{x}},t-\tau) \end{bmatrix}^T ~\underline{\tilde{\mathbf{n}}} ~\ud S \\
\end{aligned}
\end{equation}

Finally, the quadratic hydrostatic term is obtained as, 

\begin{equation}
    \begin{aligned} 
        \underline{\mathbf{F}}^H_{Q}(t) =  -\rho g \int_{S_0} \Big( (\boldsymbol{\mathcal{H}}(t) \tilde{\mathbf{x}}\cdot \nabla \tilde{z})~\underline{\tilde{\mathbf{n}}} +  \tilde{\mathbf{N}}_{\underline{\boldsymbol{\xi}}}(t)\xi_r(\tilde{x}, \tilde{y}, t)  + (\tilde{z} +\tilde{z}_0)\tilde{\mathbf{N}}_{\underline{\boldsymbol{\xi}}\underline{\boldsymbol{\xi}}}(t) \Big) ~\ud S 
    \end{aligned}
\end{equation}
and can be explicitly evaluated in the time domain using the instantaneous body motions at each timestep. This force can also be expressed in parametric form using the body volume and waterplane properties, as shown by \cite{lee1995wamit}.

\subsection{Equations of motion} \label{sec:dtumotionsimulator}
The definition of the restoring and radiation loads in \eqref{eq:restoring_loads_nonlin_mot} and \eqref{eq:radiation_force_nonlinear} permit the use of the Cummins equation \citep{cummins1962impulse} directly for the nonlinear motions as,

\begin{equation} \label{eq:cummins_eq_nonlin}
    \begin{aligned}
        (\boldsymbol{\mathcal{M}} + \boldsymbol{\mathcal{M}}^{\infty}_{A}) \partial_{tt}\underline{\boldsymbol{\xi}}(t) + \int_{0}^t \boldsymbol{\mathcal{K}}(\tau) \partial_t\underline{\boldsymbol{\xi}}(t-\tau) ~\ud \tau + \boldsymbol{\mathcal{C}}_{H}  \underline{\boldsymbol{\xi}}(t) = \underline{\mathbf{F}}(t) + \underline{\mathbf{F}}_E(t) + \underline{\mathbf{F}}_{O}(t)
    \end{aligned}
\end{equation}
where $\boldsymbol{\mathcal{M}}$ is the $6\times6$ mass and inertia matrix, $\underline{\mathbf{F}}_E$ corresponds to externally imposed loads and $\underline{\mathbf{F}} = \underline{\mathbf{F}}_{P_{IS}} + \underline{\mathbf{F}}_Q$. In addition, $\underline{\mathbf{F}}_{O}$ is given by, 

\begin{equation} \label{eq:weird_loads_forcemodel} 
    \begin{aligned}
        \underline{\mathbf{F}}_{O}(t) &= \begin{bmatrix}
            - M\partial_{tt}\boldsymbol{\mathcal{H}}(t)  \tilde{\mathbf{x}}_g \\ \\
             - \boldsymbol{\xi}(t) \times \mathbf{F}(t)  - \boldsymbol{\alpha}(t) \times \mathbf{Q}(t) + \tilde{\mathbf{x}}_g\times \big( \boldsymbol{\alpha}(t)\times\mathbf{F}(t) \big) -M\tilde{\mathbf{x}}_g \times \partial_t \boldsymbol{\mathcal{H}}(t)  \tilde{\mathbf{x}}_g\\   - \partial_t\boldsymbol{\alpha}(t) \times\boldsymbol{\mathcal{I}}_g  \partial_t\boldsymbol{\alpha}(t)   - \boldsymbol{\mathcal{I}}_g  \partial_t\boldsymbol{\omega}_{\boldsymbol{\alpha}}(t)
        \end{bmatrix}
    \end{aligned}    
\end{equation}
where $M$ is the body mass, $\tilde{\mathbf{x}}_g$ is the vector of the COG coordinates in the body-fixed frame, $\boldsymbol{\mathcal{I}}_g$ is the $3\times 3$ inertia matrix around the COG and $\boldsymbol{\omega}_{\boldsymbol{\alpha}} = [ \partial_t\alpha_2 ~\alpha_3, \partial_t\alpha_1~\alpha_3, \partial_t\alpha_1 ~\alpha_2]^T$. It is worth noting that most of the terms in $\underline{\mathbf{F}}_{O}$ disappear when $\tilde{z}_0=0$ and thus $\mathbf{O}_b$ corresponds to the centre of gravity (COG) of the body \citep{lee1995wamit}, in which case, only the second line of the moments remain.

The open-source time-domain solver \texttt{DTUMotionSimulator} \citep{bingham2000hybrid} was used to solve this set of motion equations. The solver uses a $4^{th}$-order Runge-Kutta scheme for time marching, and at each timestep, the instantaneous body motion and velocity are explicitly provided in the force calculation. Hence, the force evaluation is performed considering the instantaneous nonlinear body kinematics, rather than the precomputed linear counterparts. 
It is noted that the body acceleration is considered within several force terms, such as \eqref{eq:disp_pot_force_rad}, which are nonlinear and $O(\epsilon \zeta, \zeta^2)$. Transferring these nonlinear terms to the left-hand side of \eqref{eq:cummins_eq_nonlin} would rather complicate the numerical procedure, as a nonlinear equation would need to be solved. For this reason, a 3-point backwards finite difference scheme is employed to estimate the body acceleration at each time step and evaluate the respective force terms, which are preserved on the right-hand side. 

Another issue concerns the consistent representation of the forces within the equations of motion. As previously mentioned, \cite{teng2018decomposition} proposed a similar decomposition for the second-order quadratic force terms of \eqref{eq:force_quadratic_nonlin} into three contributions arising from products between the linear potentials $\phi^{(1)}_{IS}$ and $\phi^{(1)}_R$. The convolution kernels associated with the radiation contribution were obtained by applying an IFFT to the frequency-domain QTFs, while the infinite-frequency terms were not treated explicitly. As pointed out by \cite{malenica2025practical}, this leads to a practical limitation, since the resulting convolution integral extends to infinity rather than being truncated at the current timestep. Consequently, both past and future motion histories are required in advance, which is only feasible if the force evaluation assumes purely first-order motions \citep{teng2018nonlinear}. However, in the RK4 scheme of the present formulation, the motion history is only known up to the current time instant. Hence, when nonlinear body motions are included in the force computation, it becomes necessary to account for the infinite-frequency contributions.

\section{Experiments}\label{sec:experiments}

\begin{figure}
\centering
\includegraphics[width=0.44\linewidth]{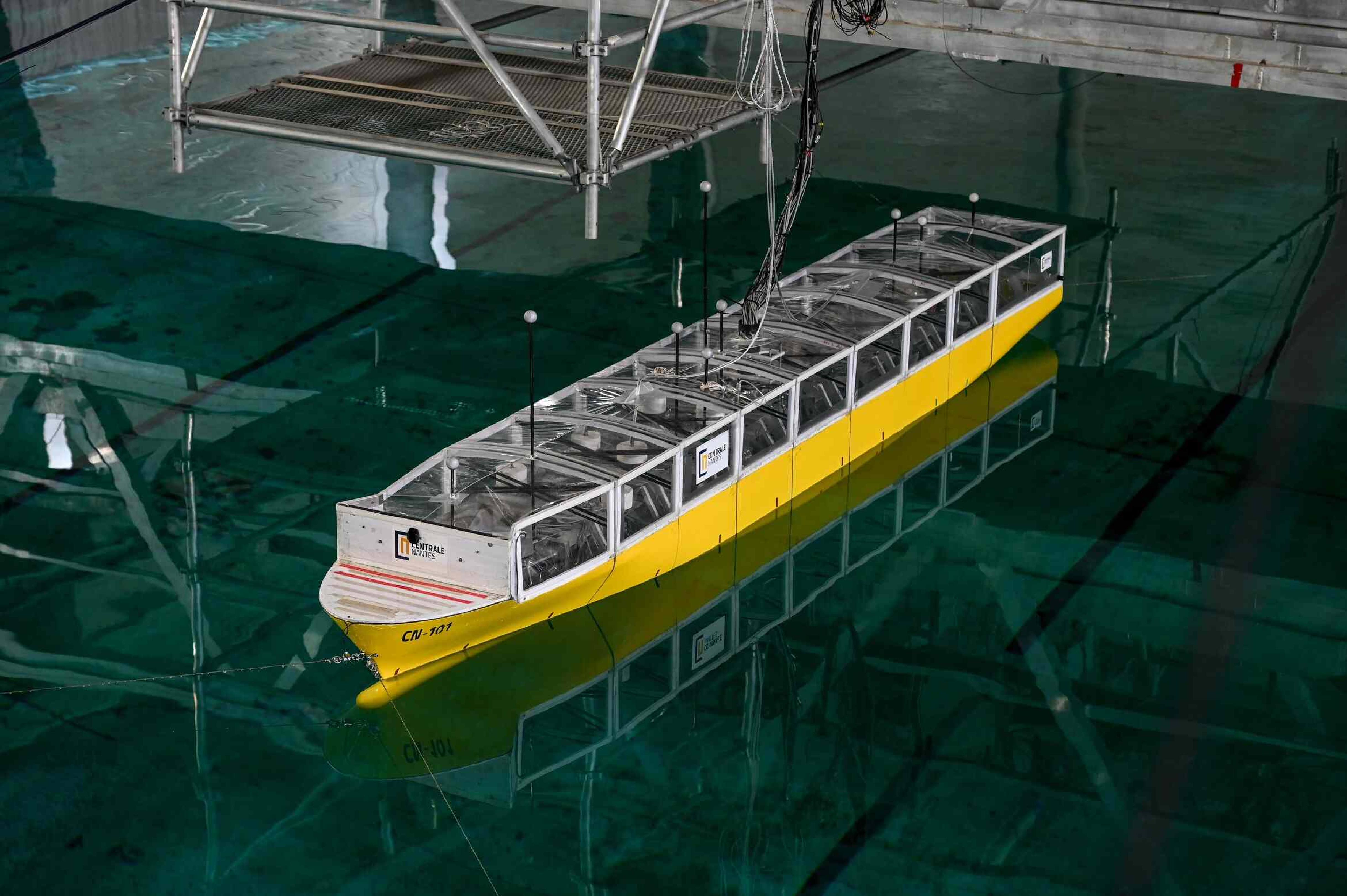}
\includegraphics[width=0.54\linewidth]{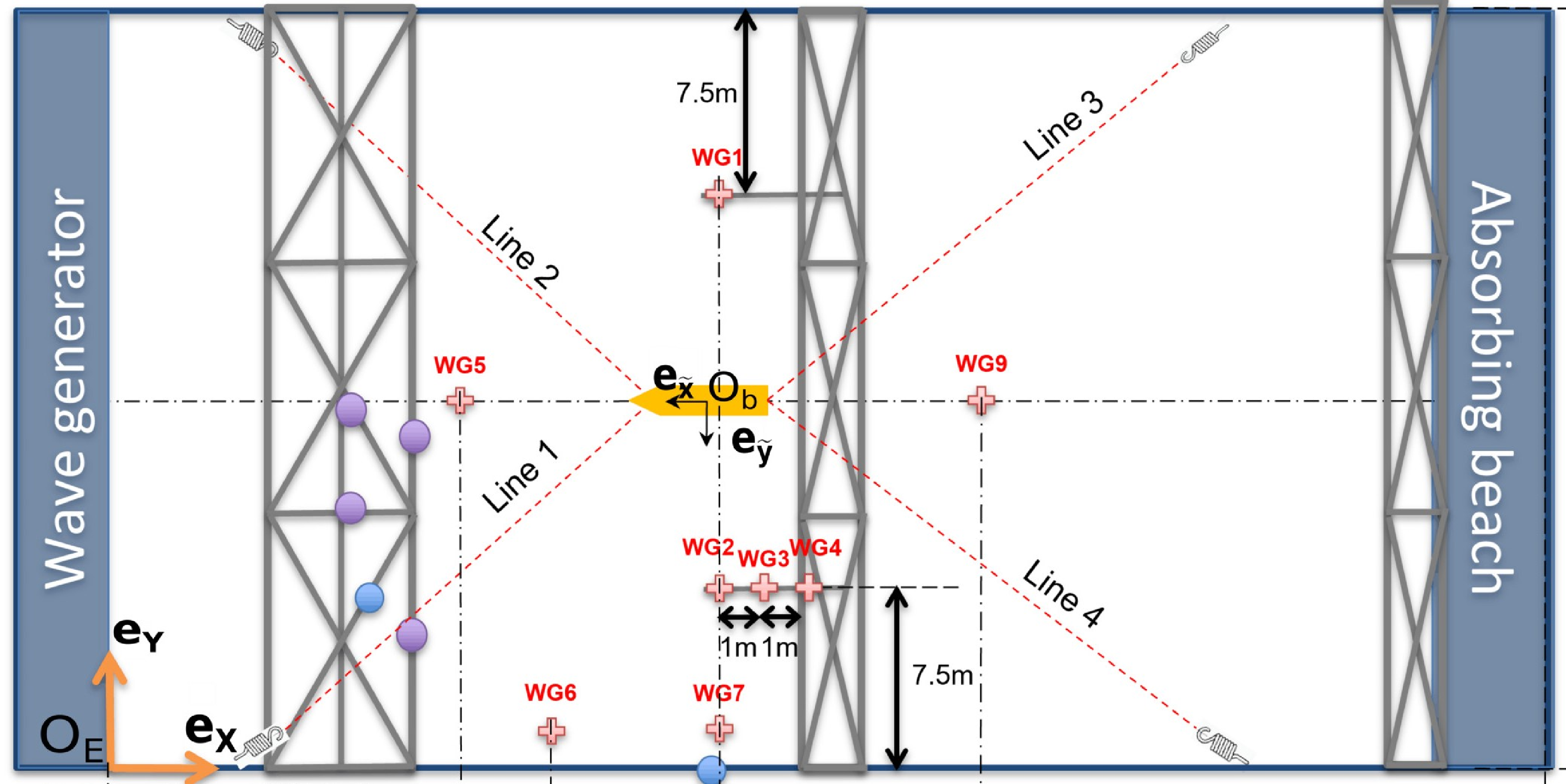}
\caption{6750-TEU container ship model and experimental setup.}
\label{fig:model_exp}
\end{figure}

Experimental tests were carried out in the Ocean Engineering wave tank of the Ecole Centrale de Nantes (ECN). It consists of a rectangular basin of length $L_x=50$ m, width $L_y=30$ m and depth $h=5$ m. On one side, it is equipped with a single-hinge wavemaker of 48 individual flaps, while the opposite side is equipped with an absorbing beach.

A rigid 6750-TEU container ship under a scale of $1/65$ was investigated, the dimensions of which are identical to those used in the ITTC-ISSC benchmark study of \cite{kim_benchmark_2016} and are presented in Table \ref{tab:containership_dimensions}. The ship hull was framed by 9 segments, which ensure the rigidity of the vessel, and \cite{bouscasse2022experimental} provides a detailed description of the structural properties. The model is illustrated in Figure \ref{fig:model_exp}, along with a layout of its installation in the tank. A global reference frame $R_E = (\mathbf{O}_E, \mathbf{e}_{X}, \mathbf{e}_Y, \mathbf{e}_Z)$ is defined at the location of the wavemaker, and the relative position between the two global reference frames is $\mathbf{X}_g=\mathbf{O}_E\mathbf{O}=(x_0,y_0,z_0)=(18.2~m,~15.0~m,~0.073~m)$. A Qualisys optical tracking system was used to measure the 6-DoF motions of the vessel, consisting of 4 cameras that tracked several reflective markers on the model. The Qualisys body-fixed reference frame $R_b$ was considered at the COG of the vessel.

The mooring arrangement consisted of four lines, which were equipped with a spring of $k = 58$ N/m stiffness and a pretension of $T_0=17$ N, while they had an angle of $45^\circ$ with the horizontal axis. The lines were maintained horizontal at a height of $0.1$ m above the still water line and they were attached to the sidewalls of the tank, at longitudinal positions of $2.2$ m and $37.8$ m from the wavemaker, as shown in Figure \ref{fig:model_exp}. In addition, they were connected to the vessel through fairleads, at points along the $\mathbf{e}_x$ axis, of distance $15.9$ m and $20.4$ m in $R_E$.

\begin{table}
\centering
\begin{tabular}{cc}
\textbf{Scale} & 1/65 \\ 
\textbf{Length $L_{PP}$ } (m)    & 4.409   \\ 
\textbf{Breadth $B$} (m)  & 0.615   \\ 
\textbf{Draft $T$ } (m)   & 0.188   \\ 
\textbf{Displacement $\Delta$}  (kg)    & 311.93 \\ 
\textbf{Vertical COG} $KG$ (m)  & 0.257   \\ 
\textbf{Longitudinal COG from aft end} $LCG$ (m)    & 2.147   \\ 
\textbf{Radius of gyration} $k_{yy}$ (m)  & 1.109   \\ 
\bottomrule
\end{tabular}
\caption{Principal dimensions of the 6750-TEU container ship.}\label{tab:containership_dimensions}
\end{table}

\begin{table}
\centering
\begin{tabular}{c|c|cc|cc}
\multicolumn{2}{c}{} & \multicolumn{2}{c}{\textbf{Full scale}} & \multicolumn{2}{c}{\textbf{Model scale}}  \\
\textbf{Case} & $\gamma$ & \textbf{$H_s$ (m)} & \textbf{$T_p$ (s)}  & \textbf{$H_s$ (m)} & \textbf{$T_p$ (s)} \\
\hline
SS6  & 1 & 6  & 12.3 & 0.092  & 1.52 \\ 
SS10 & 1.5 & 10 & 14 & 0.154  & 1.74  \\ 
SS17 & 2.6 & 17 & 15.5 & 0.262 & 1.92 \\ 
\end{tabular}
\caption{Irregular wave conditions.}
\label{tab:testCondition}
\end{table}

Extensive experimental tests have been performed for the present configuration, for regular \citep{bouscasse2022experimental}, irregular \citep{kim_experimental_2024} and design waves \citep{dermatis_2025_prediction}. In the present study, 3 unidirectional irregular sea states are considered, which are described by the JONSWAP spectrum and their properties are shown in Table \ref{tab:testCondition} in both model and full scale. It is worth noting that SS17 corresponds to an extremely severe sea state, of a 1000-year return period in the Gulf of Mexico. Thus, considerable nonlinear phenomena occurred during this sea state, such as wave breaking and green water on deck. 


\section{Results and discussion} \label{sec:results}
\subsection{Radiation-diffraction analysis} \label{sec:raddif_hydrostar}

The 3D-BEM solver \texttt{Hydrostar} \citep{chen_middle-field_2007} was employed to solve the radiation-diffraction problem in the frequency domain. The half hull was considered for the panel discretisation, and 2,500 panels were distributed over its submerged surface. Given the peak frequency of the sea states under investigation, a frequency range of $[0.1$-$10]$ rad/s with an interval of $0.1$ rad/s was considered. The effect of the mooring system was taken into account through an additional stiffness matrix, which, upon linearisation of the mooring arrangement, was determined in model scale as,

\begin{equation} \label{eq:mooring_matrix}
    \boldsymbol{\mathcal{C}}_M =  \begin{bmatrix}
        121.8 & 0 & 0 & 0 & -12.2 & 0 \\
        0 & 113.3 & 0 & 11.3 & 0 & 28.6 \\
        0 & 0 & 3.2 &  0 & 2.4 & 0 \\
        0 & 11.3 & 0 &  1.1 & 0 & 2.9 \\
        -12.2 & 0 & - 0.5 & 0 &  130.4 & 0  \\
        0 & 31.4 & 0 & 3.1 & 0 & 712.0
    \end{bmatrix} \quad kg/s^2
\end{equation}
Finally, for the QTF calculation, the near-field approach was used and the indirect approach \citep{second_order_molin_1979} for the potential component. To treat the free-surface integral, a semi-circular mesh of $12$ m radius and $3200$ panels was considered on the free-surface.

As discussed in Section \ref{sec:freqdom_solution}, the radiation-diffraction analysis provides the linear transfer functions of the combined incident and scattering velocity $\nabla\hat{\phi}{}^{(1)}_{IS}(\omega,\tilde{\mathbf{x}})$ for all panels on the body mesh, and wave elevation $\eta{}^{(1)}_{IS}(\omega,\tilde{x},\tilde{y})$ along the waterline. In addition, the elementary radiation 6DoF vectors $\hat{\boldsymbol{\underline{\eta}}}{}_r(\omega, \tilde{x},\tilde{y})$ and $\hat{\underline{\mathbf{u}}}{}_r(\omega,\tilde{\mathbf{x}})$ are obtained, based on which all quantities regarding the time-domain expression of the radiation potential can be explicitly evaluated according to Section \ref{sec:radiation_td}. Moreover, the linear excitation force transfer function $\hat{\mathbf{F}}_{P_{IS}}^{(1)}$ from \eqref{eq:linear_exc_force_tf} is provided, and using those elements, the hydrodynamic loads of Section \ref{sec:hyd_loads_forcemodel} can be evaluated. Finally, the analysis also provides the force QTFs, $\hat{\mathbf{F}}_Q^{(2)}$ and $\hat{\mathbf{F}}_{P_{IS}}^{(2)}$, which will be employed for the verification and validation of the proposed approach.

\subsection{Force model verification} \label{sec:forcemodel_validation}

To verify the proposed approach, the force model must be reduced to a purely second-order regime, enabling direct comparison with the results of standard second-order theory. To this end, the second-order incident velocity potential from \eqref{eq:quadratic_incident_potential} can be substituted in \eqref{eq:force_potential_nonlin} to provide,

\begin{equation} \label{eq:sharma_dean_force}
\underline{\mathbf{F}}_{P_W}^{(2)}(t) =  -\frac{\ui}{g} \sum_{m=-N}^N\sum_{n=-N}^N (\omega_m + \omega_n) B_m B_n \hat{\mathbb{T}}^{(2)}_{\phi}(\omega_m, \omega_n) ~\hat{\underline{\mathbf{F}}}^{(1)}_{P_{IS}}(k_m+k_n) \ue^{\ui(\omega_m+\omega_n)t}
\end{equation}
which constitutes the pure second-order potential force contribution. Regarding the quadratic force, it is sufficient to consider the interaction of linear waves and linear body motions. To this end, the time-dependent modal wave amplitudes can be replaced by their linear equivalent,

\begin{equation} \label{eq:linear_waves_modal_amplitudes}
\begin{aligned}
    \tilde{A}_{lin}(k_n,t) = A_n e^{\ui \omega_n t} \qquad 
    \tilde{B}_{lin}(k_n,t) = \frac{\ui g}{\omega_n} \tilde{A}_{lin}(k_n,t)
\end{aligned}
\end{equation} 
Under these terms, the wavefield obtained from \eqref{eq:hos_inc_pot} is fully equivalent to \eqref{eq:linear_waves}. The linear motions are reconstructed based on their RAOs, 

\begin{equation} \label{eq:linear_motions_td}
    \underline{\boldsymbol{\xi}}(t) = \underline{\boldsymbol{\xi}}^{(1)}(t) = \sum_{n=-N}^N \hat{\underline{\boldsymbol{\xi}}}{}^{(1)}(\omega_n) \tilde{B}_{lin}(k_n,t)
\end{equation}
Finally, the linear radiation velocity can be reconstructed as

\begin{equation}
    \begin{aligned} \label{eq:linear_radiation_tf}
       \nabla \phi_R^{(1)}(\mathbf{x},t) =\sum_{n=-N}^N \nabla \hat{\phi}^{(1)}_R(\omega_n,\mathbf{x}) \tilde{B}_{lin}(k_n\{\omega_n\},t) 
    \end{aligned}
\end{equation}
where $\hat{\phi}^{(1)}_R(\omega_n,\mathbf{x})$ is given by \eqref{eq:rad_pot_freq} and is based on the first-order body motions. In a similar manner, the linear radiated wave elevation $\eta_R^{(1)}(x,y,t)$ can be obtained.

\begin{figure}
    \centering
    \includegraphics[width=\linewidth]{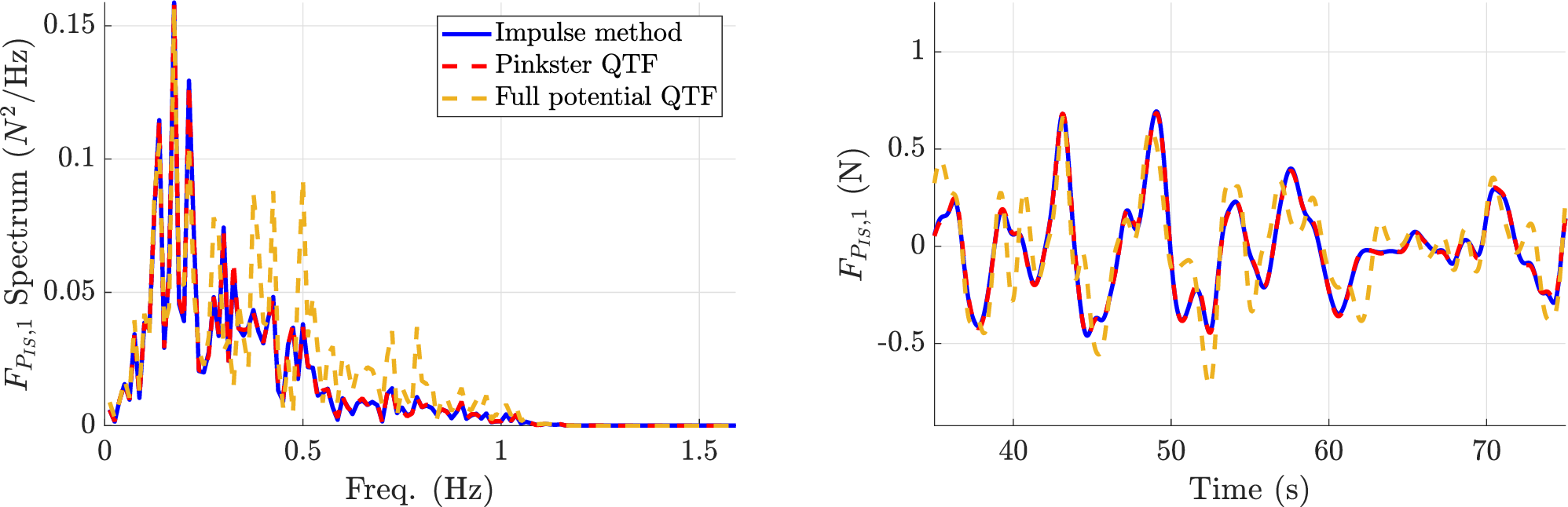}
    \caption{Potential component of the second-order horizontal force on the container ship in terms of spectrum (left) and time series (right).}
    \label{fig:ship_pot_force}
\end{figure}
\begin{figure}
    \centering
    \includegraphics[width=\linewidth]{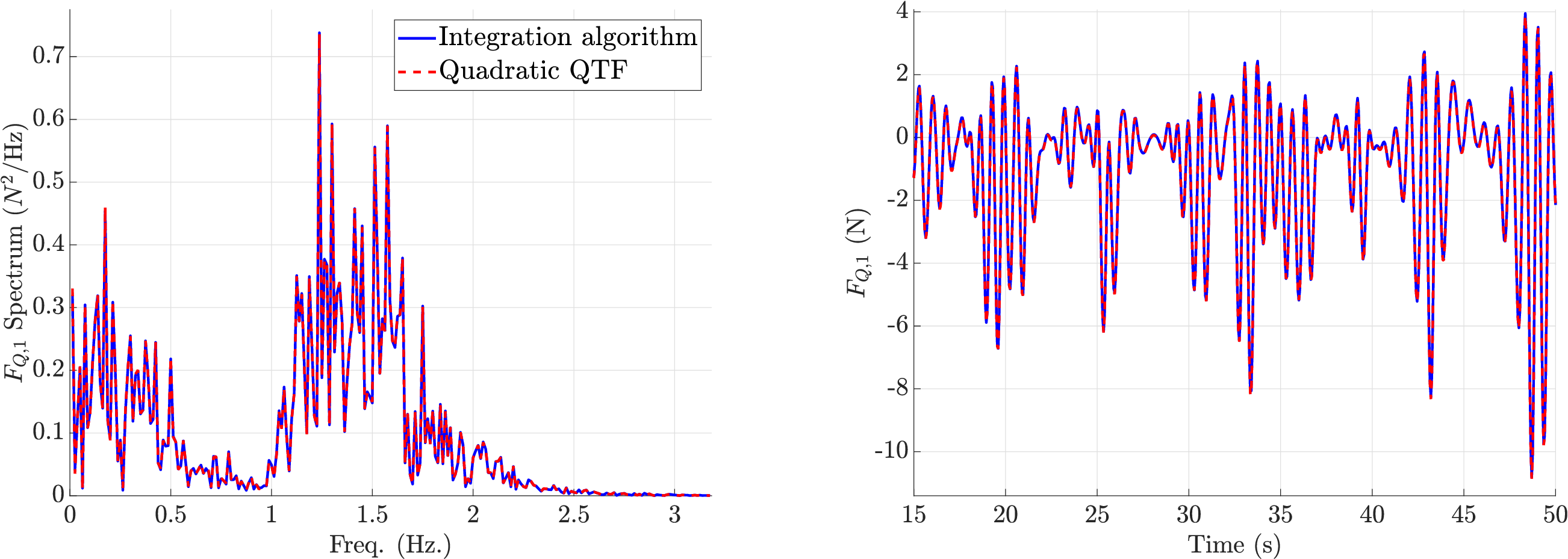}
    \caption{Quadratic component of the second-order horizontal force on the container ship in terms of spectrum (left) and time series (right).}
    \label{fig:motion_terms_validation}
\end{figure}

The irregular sea state SS6 from Table \ref{tab:testCondition} was considered, and a linear wavefield was constructed using a random phase model, with the first- and second-order potentials given by \eqref{eq:linear_waves}-\eqref{eq:quadratic_incident_potential}. Figure \ref{fig:ship_pot_force} illustrates the pure second-order force in the $\mathbf{e}_x$ direction using \eqref{eq:sharma_dean_force}, which is hereafter referred to as the 'Impulse method', following the terminology introduced by \cite{bredmose2024forcemodel}. An excerpt of the force timeseries, and the associated power spectrum, are compared against the force reconstructed by the potential component of the difference-frequency QTF and the Pinkster QTF. It is noted that the sum-frequency potential force results are not included in the figure, since they are mostly related to the free-surface integral, which is not treated directly by \eqref{eq:sharma_dean_force}. Excellent agreement is found between the impulse method and the Pinkster approach, verifying the equivalence of the two methods at second order. In contrast, comparison with the full potential-force solution shows some discrepancies attributed to second-order scattering effects that are not fully taken into account by \eqref{eq:sharma_dean_force}. Moreover, the results regarding the quadratic force component $\mathbf{F}_Q$ in the $\mathbf{e}_x$ direction, from \eqref{eq:force_quadratic_nonlin}, are presented in Figure \ref{fig:motion_terms_validation}. The close agreement between the force obtained in the time domain through the proposed approach and the reconstruction of the quadratic component force QTF verifies the implementation of the numerical schemes for each term of \eqref{eq:force_quadratic_nonlin}. A similar verification study for the case of a monopile can be found in \cite{bredmose2024forcemodel}.

\subsection{Comparison with experiments and second-order theory} 

Upon verifying the formulation and implementation of the proposed approach against classical second-order theory, it is now utilised at its full capacity, employing nonlinear wave input and the total instantaneous body motions. To this end, the sea states of Table \ref{tab:testCondition} were generated using the open-source \texttt{HOS-NWT} solver \citep{ducrozet2012modified}, which accounts for the nonlinear wave propagation inside a numerical wave tank based on the High-Order Spectral (HOS) method. The iterative calibration procedure of \cite{canard_varying_2022} was employed to match the generated spectrum at the location of the vessel COG with the theoretical wave spectra. The output of \texttt{HOS-NWT} is the modal amplitudes $\tilde{A}(k,t)$, $\tilde{B}(k,t)$ and $\tilde{B}_t(k,t)$, based on which the nonlinear incident potential and wave elevation are reconstructed through \eqref{eq:hos_inc_pot}, and therefore the wave loads can be evaluated. 

In the following, the responses obtained from the implementation of the time-domain solver, described in Section \ref{sec:dtumotionsimulator}, coupled with the hydrodynamic loads evaluation procedure from Section \ref{sec:hyd_loads_forcemodel}, are compared against standard second-order theory, and experiments for the case of the container ship vessel of Section \ref{sec:experiments}. Comparison between the two numerical approaches is feasible by preserving the left-hand side of \eqref{eq:cummins_eq_nonlin} the same, and changing the forces on the right-hand side. Within the proposed formulation, hereby referred to as the 'QME' approach, a two-way coupling is formed between the forces and the resulting body kinematics, which are interdependent during each timestep. On the contrary, the second-order approach employs the linear and quadratic force transfer functions to reconstruct the total force in the time domain as,

\begin{equation} \label{eq:exact-hydrostar-force}
    \underline{\mathbf{F}}(t)=\sum_{n=-N}^N \hat{\underline{\mathbf{F}}}{}^{(1)}_{P_{IS}}(\omega_n) B_n \ue^{\ui \omega_n t } + \sum_{m=-N}^N\sum_{n=-N}^N \hat{\underline{\mathbf{F}}}{}^{(2)}(\omega_m,\omega_n) B_m B_n \ue^{\ui (\omega_m + \omega_n) t }
\end{equation}
where $\hat{\underline{\mathbf{F}}}{}^{(2)}=\hat{\underline{\mathbf{F}}}{}^{(2)}_{P_{IS}}+\hat{\underline{\mathbf{F}}}{}^{(2)}_Q$ is the total force QTF, as obtained by the procedure described in Section \ref{sec:freqdom_solution}. Moreover, the wave amplitudes $A_n$ were obtained through a Fast Fourier Transform of the wave elevation from $\texttt{HOS-NWT}$ at the in-tank location of the structure. The associated linear potential amplitudes were then obtained as $B_n=\ui g A_n/\omega_n$. This force, when inserted to the right-hand side of \eqref{eq:cummins_eq_nonlin}, provides the total first- and second-order body motions, and the respective results are hereby denoted as 'Second-order'. It is emphasised that the force from \eqref{eq:exact-hydrostar-force} is not directly dependent on the instantaneous total motion and velocity of the body. Instead, the motion-related contribution to the second-order force stems from the inclusion of the first-order motion RAOs in the evaluation of the frequency-domain QTF at the preprocessing stage.

\subsubsection{Time-domain solver setup}
The effect of the mooring system is taken into account in the time-domain simulations through the additional linear stiffness matrix $\boldsymbol{\mathcal{C}}_M$ from \eqref{eq:mooring_matrix}, so that an external force is imposed as,

\begin{equation}
    \mathbf{F}_{E,moor}(t) = -\boldsymbol{\mathcal{C}}_M \boldsymbol{\xi}(t)
\end{equation}
Moreover, due to the limited radiation damping at low frequencies, additional viscous damping terms must be considered to prevent unphysically large resonant oscillations,

\begin{equation}
    \mathbf{F}_{E,damp}(t) = -\boldsymbol{\mathcal{B}}_L  \partial_t \underline{\boldsymbol{\xi}}(t)- \boldsymbol{\mathcal{B}}_Q \partial_t \underline{\boldsymbol{\xi}}(t)| \partial_t \underline{\boldsymbol{\xi}}(t)|
\end{equation}
where $\boldsymbol{\mathcal{B}}_L$ and $\boldsymbol{\mathcal{B}}_Q$ are $6\times6$ matrices containing the linear and quadratic damping coefficients, respectively. Different approaches exist for the calibration of those coefficients, such as free-decay tests, or more refined techniques that account for the wave conditions \citep{pegalajar-jurado_reproduction_2019}. In the present work, the decay test approach of \cite{faltinsen1990} was employed, from which the coefficients only in pure surge motion were extracted as $\boldsymbol{\mathcal{B}}_{L,11}=7.63$ $kg/s$ and $\boldsymbol{\mathcal{B}}_{Q,11}=232.74$ $kg/m$. Calibrating these parameters in such a manner, rather than for each model separately, permits an unbiased comparison between the two numerical approaches.

\begin{figure}
    \centering
    \includegraphics[width=0.45\linewidth]{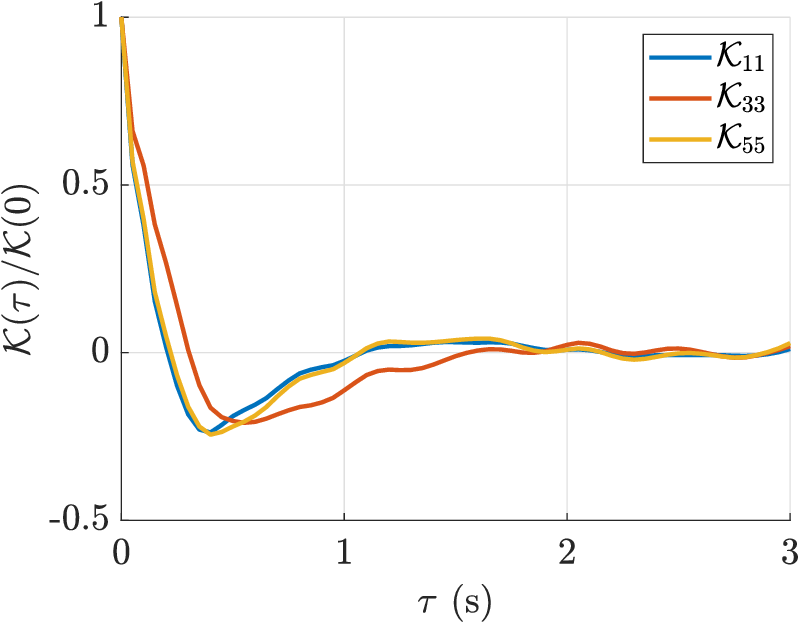}
    \includegraphics[width=0.45\linewidth]{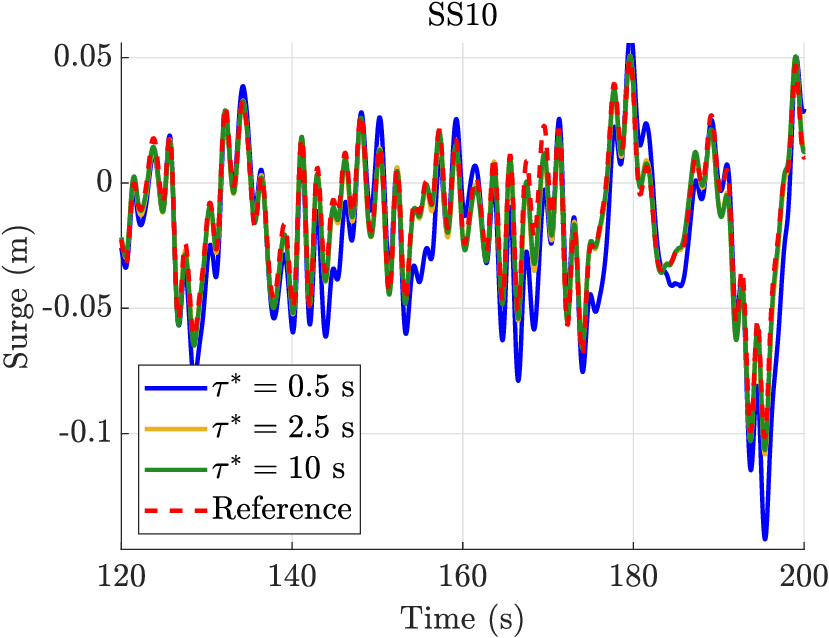}
    \caption{Convolution kernels for the radiation damping force (left) and effect of lower cut-off limit of convolution integrals within the hydrodynamic force evaluation to surge motion (right).}
    \label{fig:radiation_kernels}
\end{figure}

The timestep for the time-domain simulations was set as $\ud t = 0.05~s$, which presented adequate convergence for all sea states. In addition, the upper limit for the convolutional integrals in the force calculation was investigated due to their significance in the computational performance of the method. The radiation damping convolution kernels in pure surge $\boldsymbol{\mathcal{K}}_{11}$, heave $\boldsymbol{\mathcal{K}}_{33}$ and pitch $\boldsymbol{\mathcal{K}}_{55}$ are shown in Figure \ref{fig:radiation_kernels}, where after approximately $\tau=2.5~s$ the memory effects are negligible. It is noted that all single and double convolution kernels used for the evaluation of the radiation contributions in \eqref{eq:force_quadratic_nonlin} are constructed based on the elementary potentials $\underline{\boldsymbol{\phi}}{}_r$. Therefore, it is expected that their behaviour will be similar to $\boldsymbol{\mathcal{K}}(\tau)$. To verify this assumption, different lower cut-off limits $t-\tau^\ast$ were used for all convolution integrals in the hydrodynamic loads evaluation, and the resulting surge motion is illustrated in Figure \ref{fig:radiation_kernels}. The reference solution consists of the second-order response obtained by using the precomputed QTFs from \texttt{Hydrostar} in the right-hand side of \eqref{eq:cummins_eq_nonlin}, where radiation effects are accounted for in the frequency domain. For the force model to be directly comparable to the reference solution, its pure second-order regime was employed, as used for the verification study in Section \ref{sec:forcemodel_validation} along with the first-order force from \eqref{eq:linear_exc_force_tf}. As suggested by Figure \ref{fig:radiation_kernels}, the results for $\tau^\ast \geq 2.5~s$ are almost indistinguishable, hence this value was chosen.

\subsubsection{Irregular waves}
\begin{figure}
    \centering
    \includegraphics[width=0.63\linewidth]{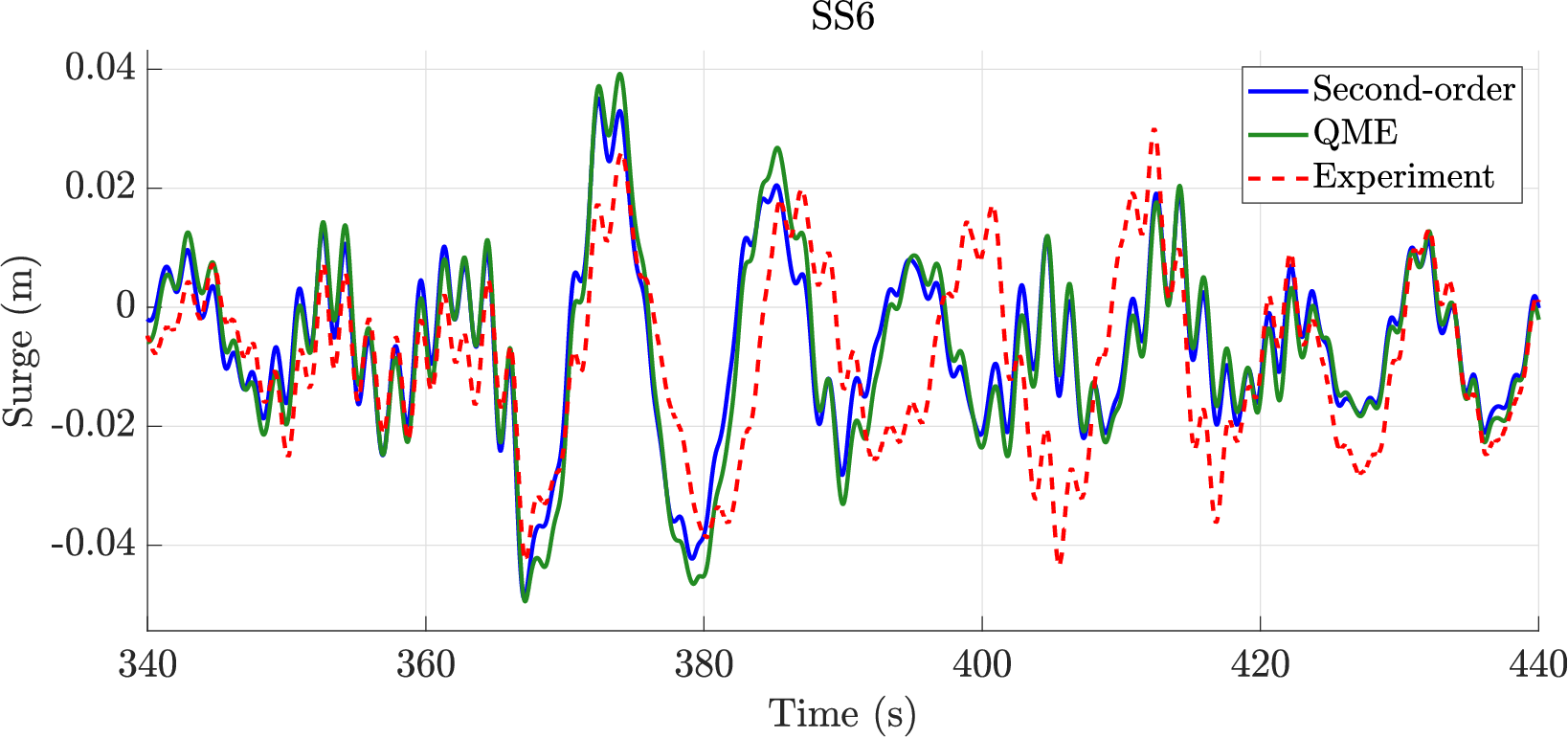}
    \includegraphics[width=0.36\linewidth]{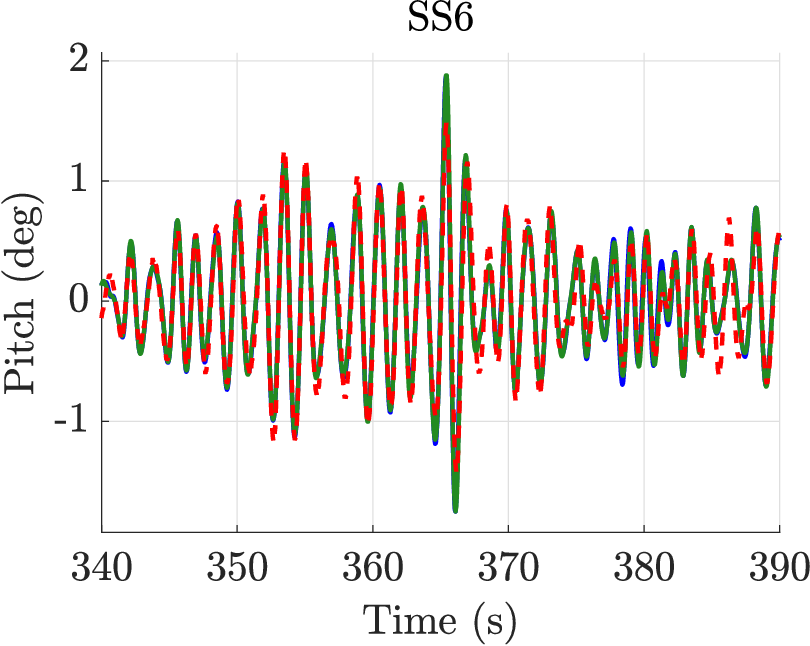}
    \includegraphics[width=0.63\linewidth]{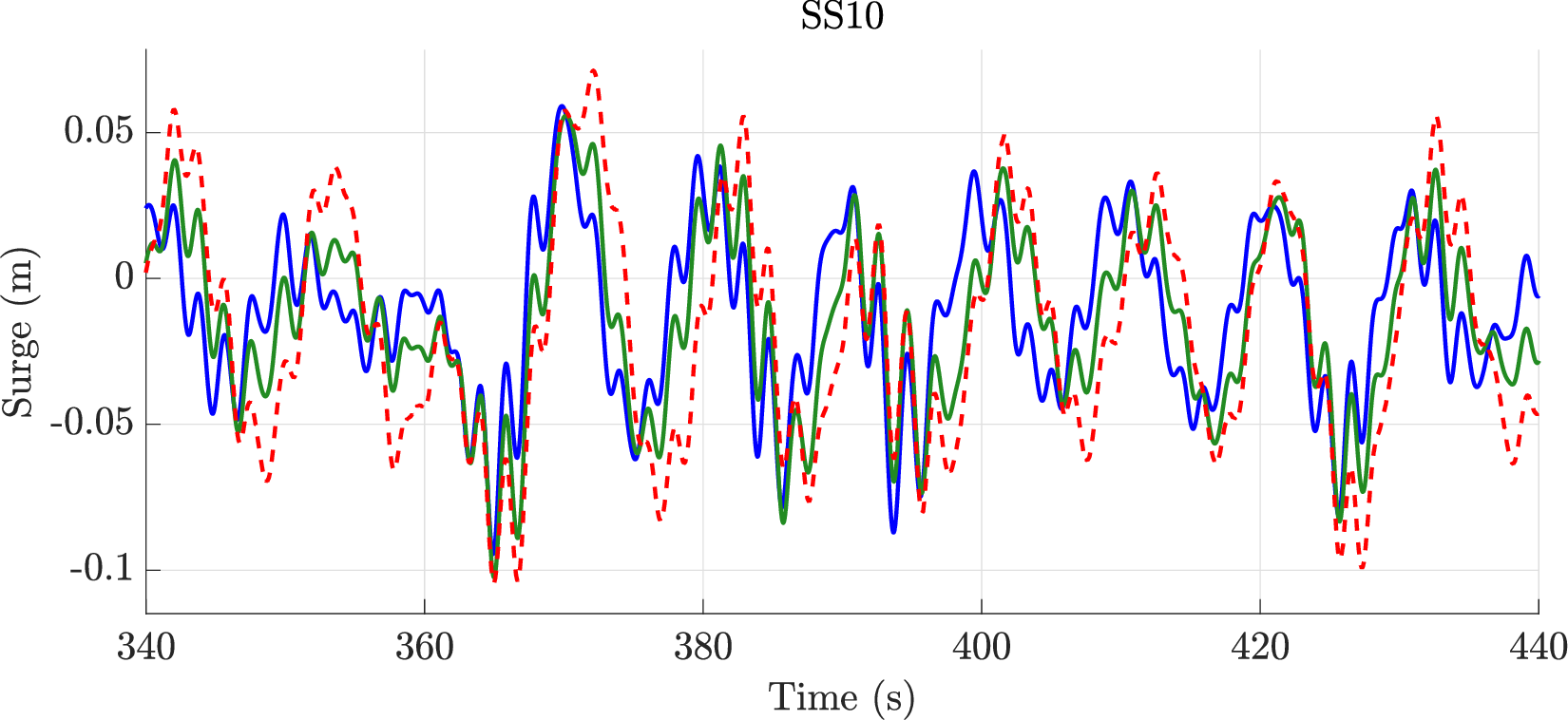}
    \includegraphics[width=0.36\linewidth]{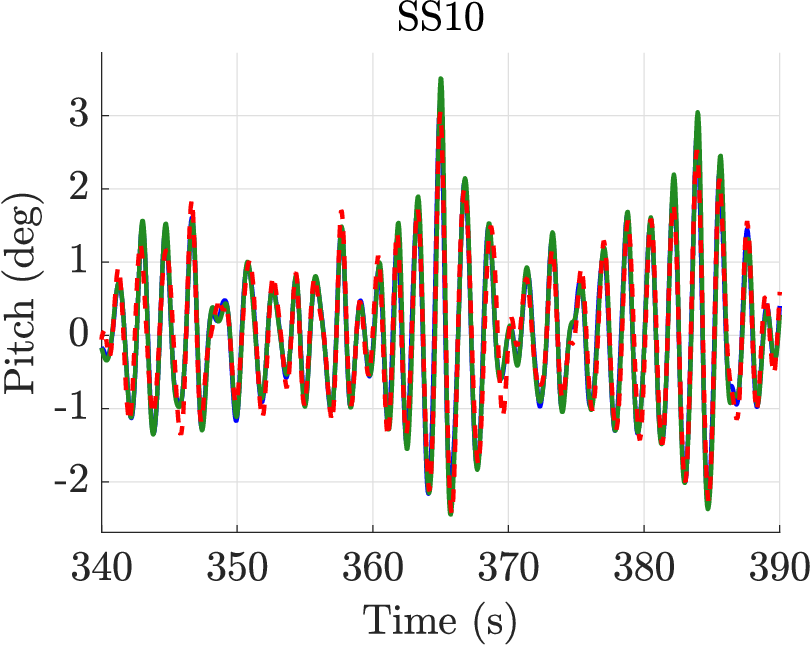}
    \includegraphics[width=0.63\linewidth]{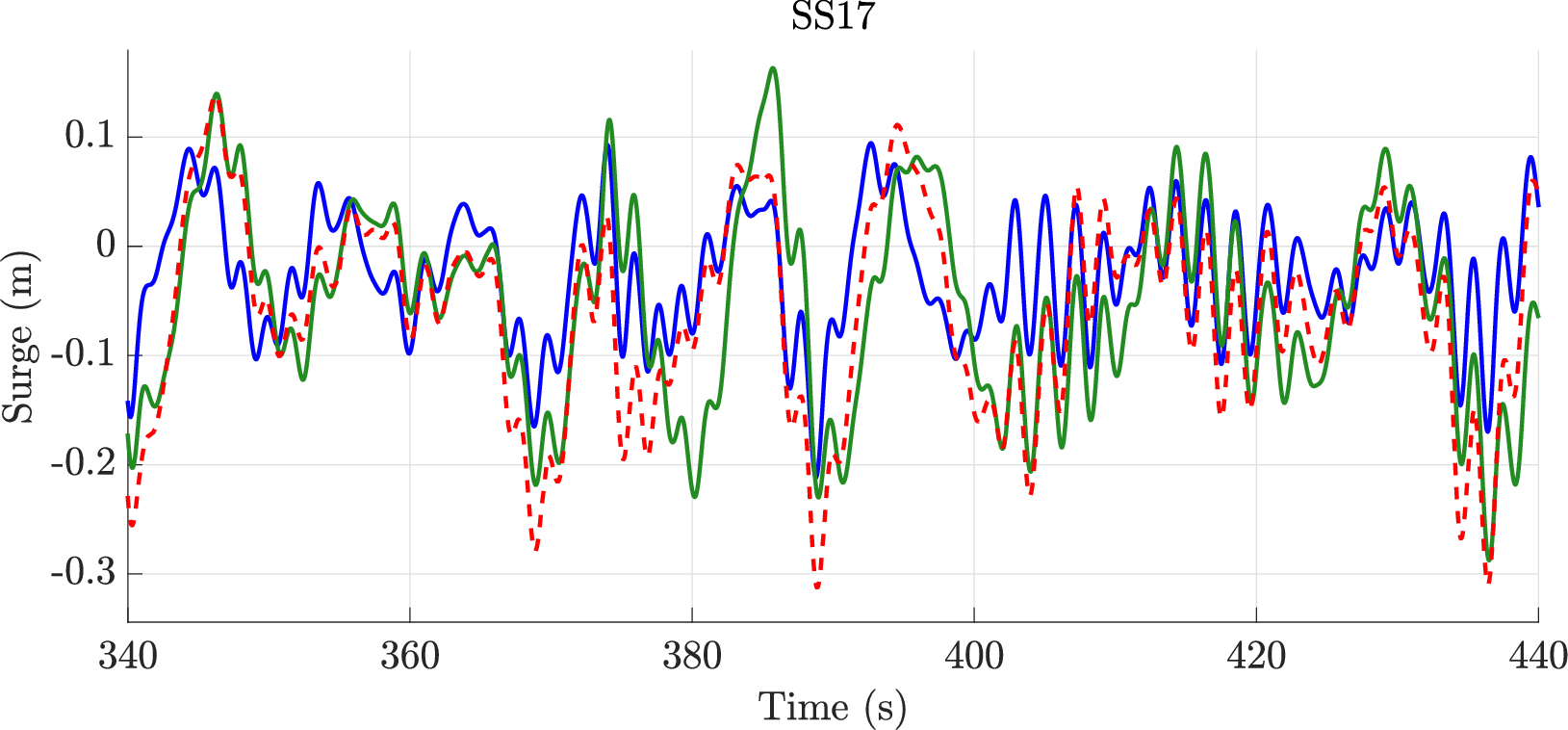}
    \includegraphics[width=0.36\linewidth]{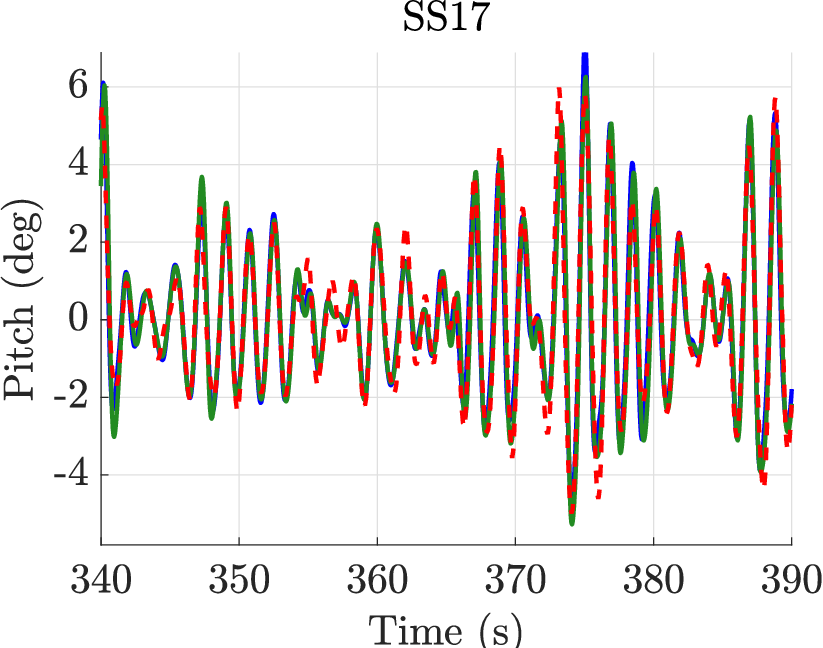}
    \caption{Comparison of numerical and experimental surge (left column) and pitch motion (right column) time series in irregular waves for all sea states.}
    \label{fig:force_model_iw_results_surge}
\end{figure}

\begin{figure}
    \centering
    \includegraphics[width=0.328\linewidth]{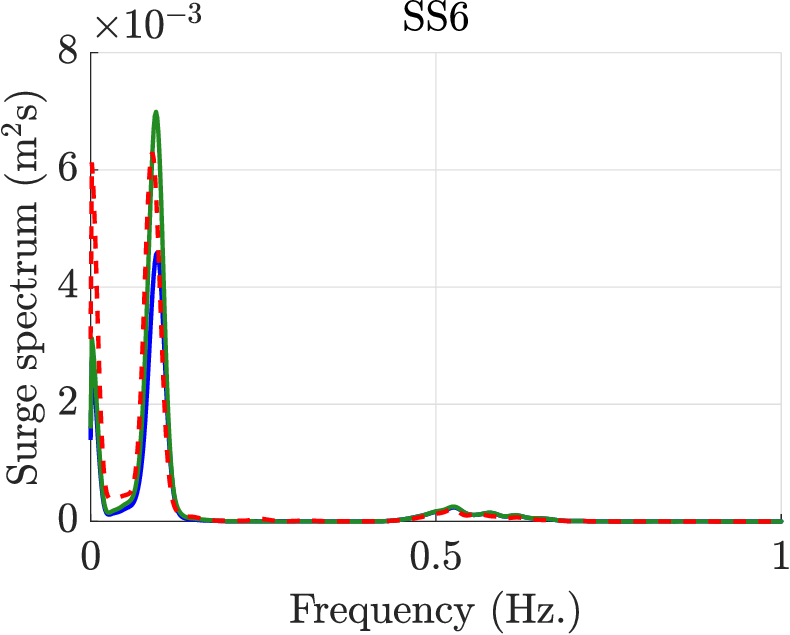}
    \includegraphics[width=0.328\linewidth]{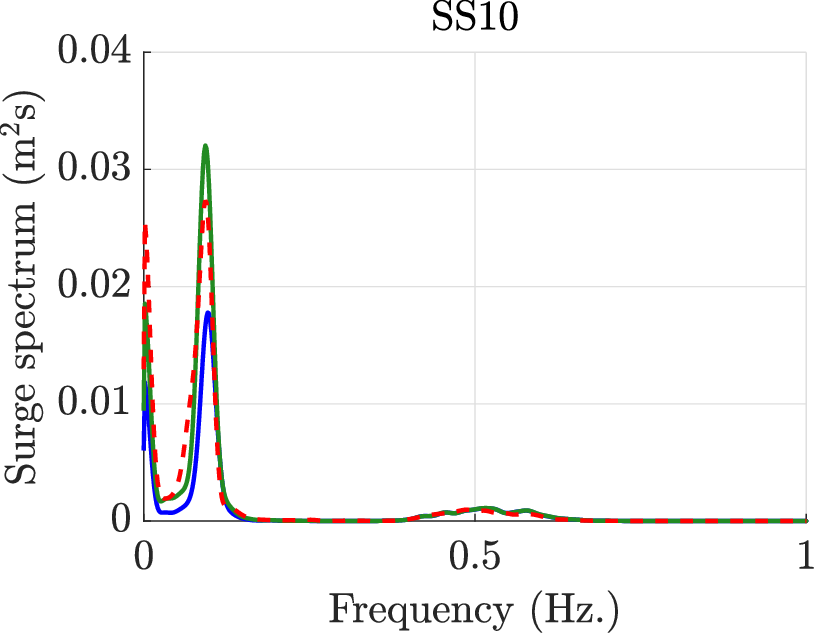}
    \includegraphics[width=0.328\linewidth]{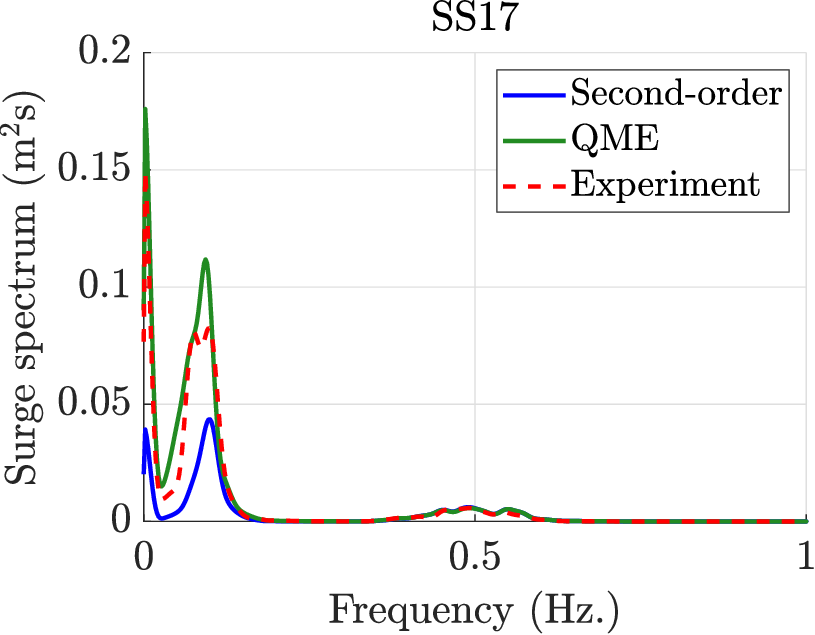}
    \includegraphics[width=0.328\linewidth]{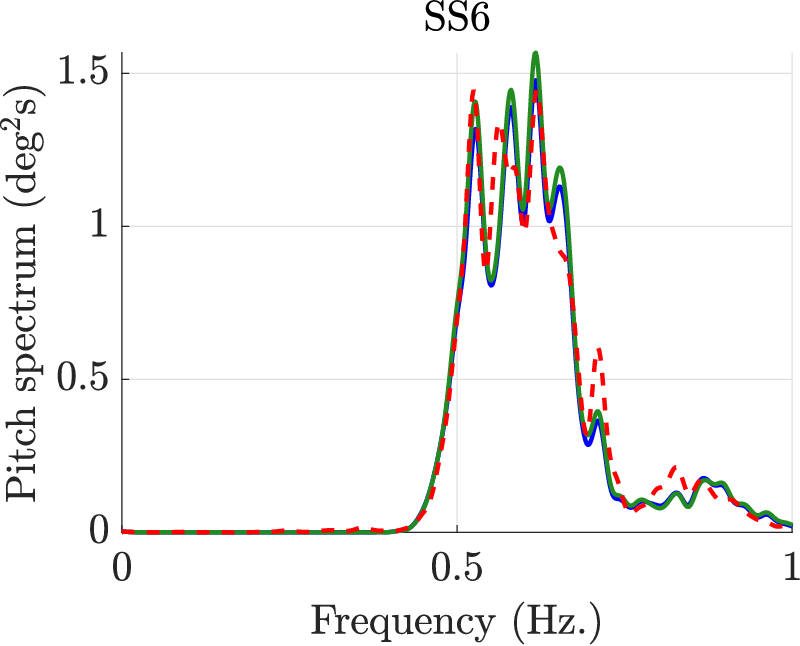}
    \includegraphics[width=0.328\linewidth]{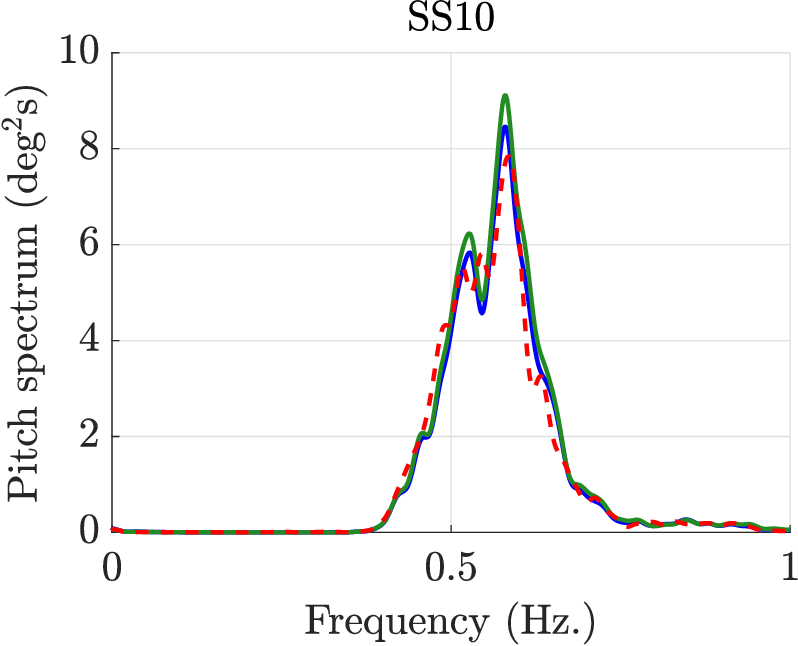}
    \includegraphics[width=0.328\linewidth]{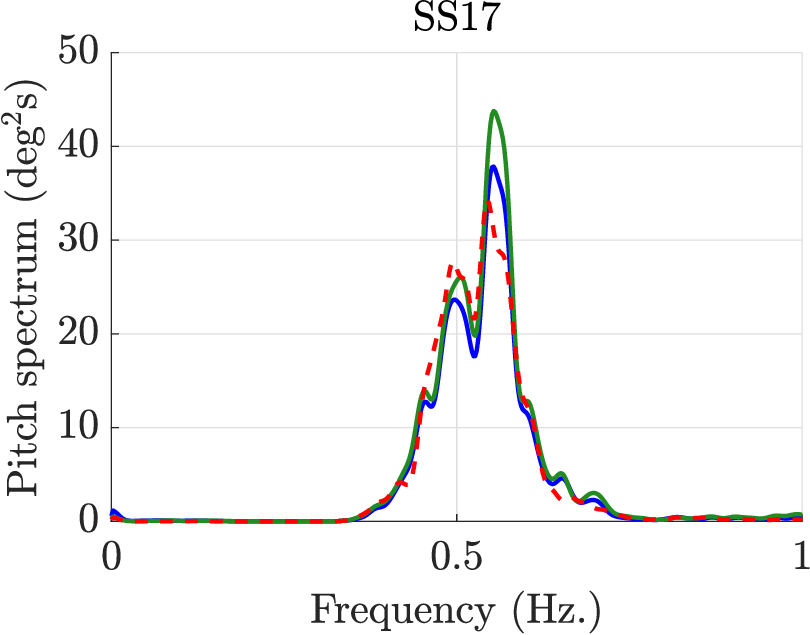}
    \caption{Comparison of numerical and experimental surge (top row) and pitch motion (bottom row) spectra in irregular waves for all sea states.}
    \label{fig:force_model_iw_spectra}
\end{figure}

\begin{figure}
    \centering
    \includegraphics[width=0.328\linewidth]{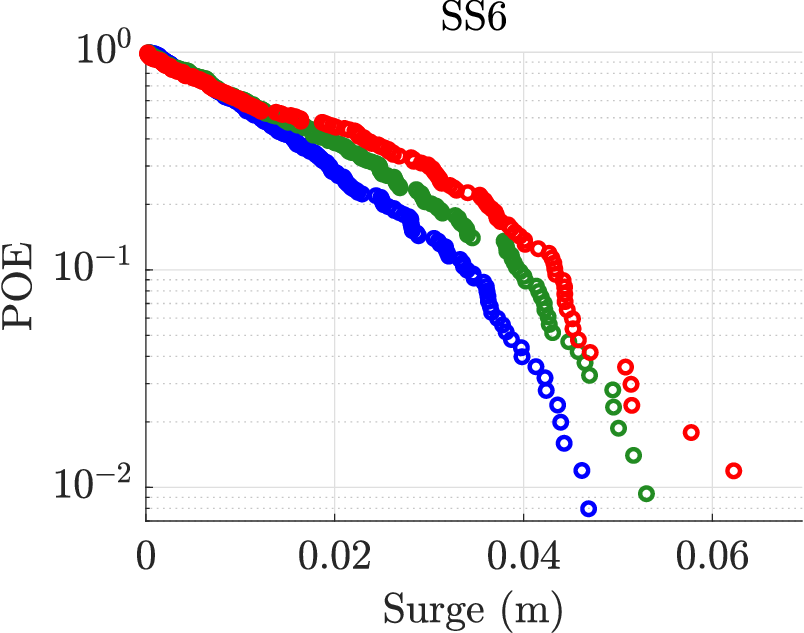}
    \includegraphics[width=0.328\linewidth]{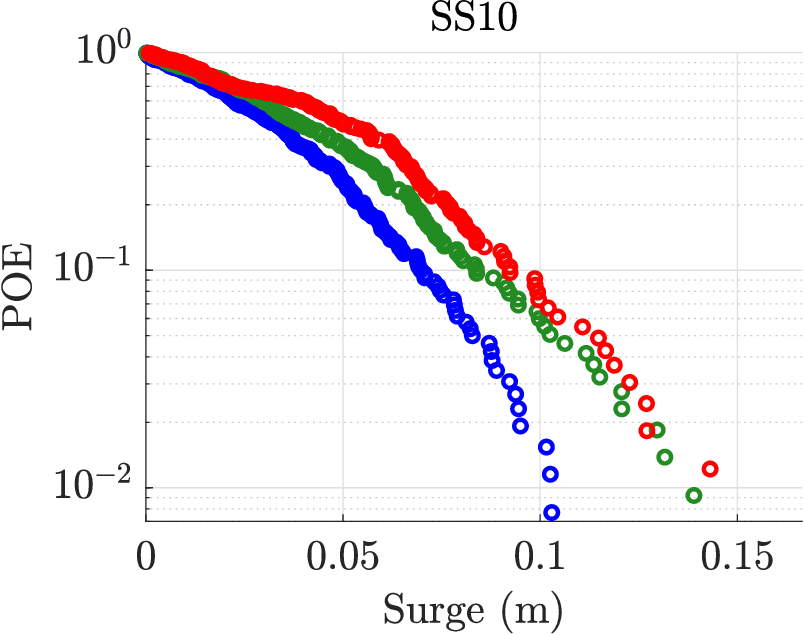}
    \includegraphics[width=0.328\linewidth]{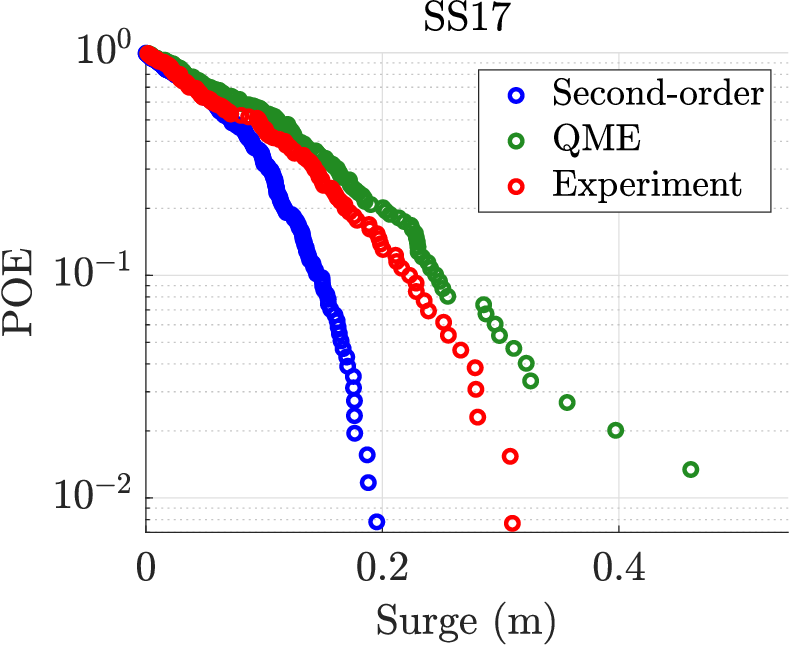}
    \includegraphics[width=0.328\linewidth]{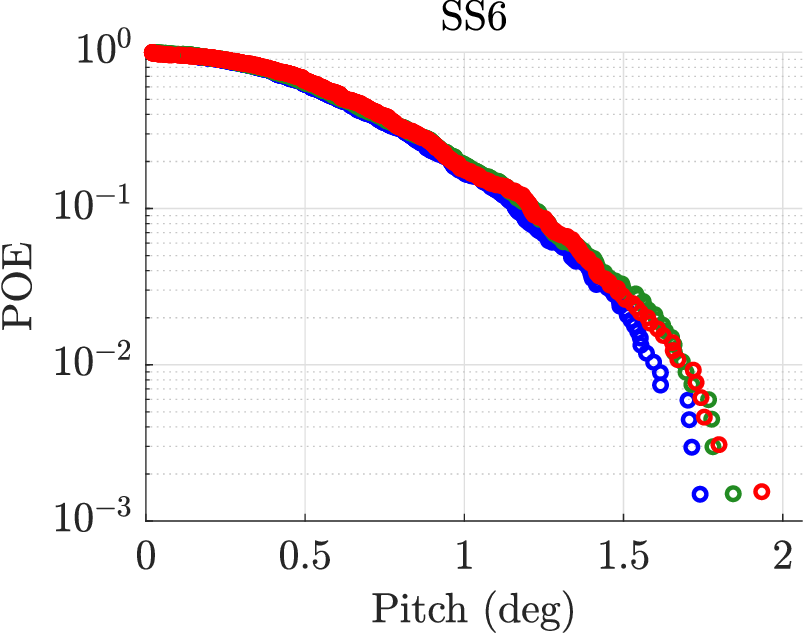}
    \includegraphics[width=0.328\linewidth]{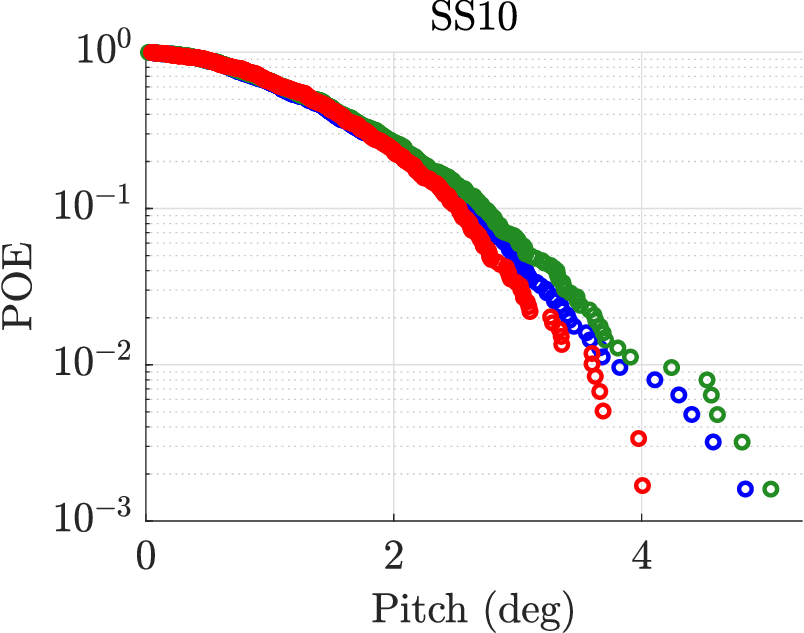}
    \includegraphics[width=0.328\linewidth]{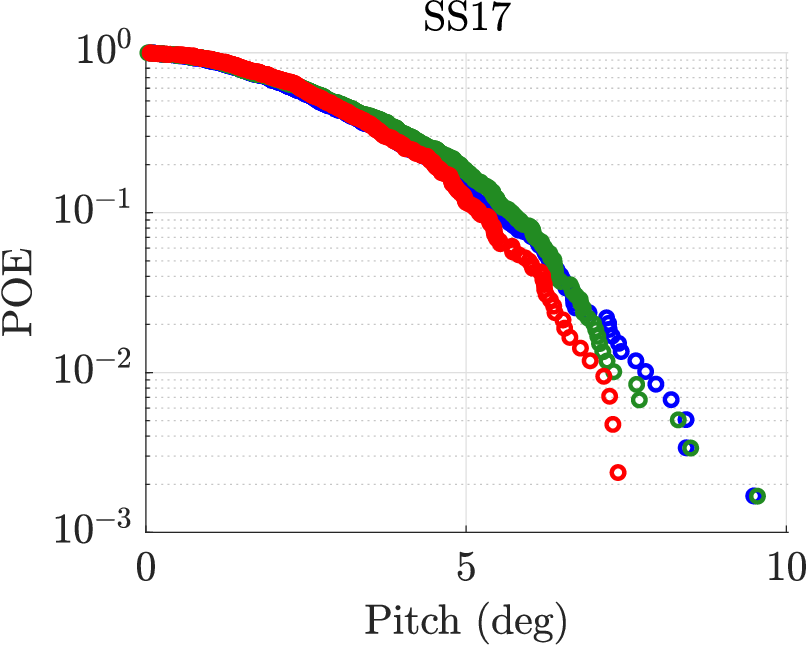}
    \caption{Comparison of numerical and experimental surge (top row) and pitch motion (bottom row) exceedance probability in irregular waves for all sea states.}
    \label{fig:force_model_iw_poe}
\end{figure}

The analysis continues with the validation of the approach for a 20-minute irregular wave realisation for each sea state listed in Table \ref{tab:testCondition}, through comparison against experimental measurements and classical second-order theory. Figure \ref{fig:force_model_iw_results_surge} shows an excerpt of the surge and pitch motion time series for each sea state. In addition, the corresponding motion spectra and exceedance probability curves are shown in Figures \ref{fig:force_model_iw_spectra} and \ref{fig:force_model_iw_poe}, respectively. Regarding the probability distributions, they were empirically obtained by extracting the response peaks from the time series through zero-crossing analysis, and sorting them in ascending order of magnitude. Moreover, the response spectra were estimated using the Welch method with $30T_p$ time windows and $50\%$ overlap. To remove transient effects, the first minute of each time history was excluded from the statistical and spectral analyses. 

For the surge motion, the time-series comparison shows that the QME model reproduces the response phase and overall profile more accurately for all sea states. This is likely due to accounting for the instantaneous body position rather than relying on a linear approximation. This is also reflected in the surge spectra, where the QME model exhibits a resonant peak at a frequency closer to the experimental. Moreover, the overall profile of the spectrum in the low-frequency region resembles the experimental throughout the whole range of wave conditions. In addition, the proposed approach captures the response amplitudes more accurately than the classical second-order approach, as suggested by the trend of the surge exceedance probability curves. For the most extreme sea state of SS17, the QME approach overestimates the experimental distribution, although being more consistent than the second-order results. This can potentially be attributed to the inevitable violation of the small body motion assumption in \eqref{eq:pressure_taylor_nonlin}. Considering that the additional damping level was obtained from the decay tests, model-specific calibration might potentially treat this issue. Nevertheless, it should be stressed that this sea state is particularly extreme, corresponding to a 1000-year return period in the Gulf of Mexico, and its purpose was to test the robustness of the method.

Differences between the two numerical models are less pronounced in pitch, which is a predominantly linear response, with less significant second-order effects. This is also verified by the pitch response spectra, where all the energy is concentrated at the wave frequency range. Some slight differences are observed in the tail of the exceedance probability curves, while for SS10 and SS17, both models overestimate the experimental results. It is worth noting that the heave response was also investigated and presented similar trends with the pitch response regarding the performance of the QME approach. Therefore, the results are omitted here, but the relevant analysis can be found in \cite{dermatis_phd}. 

An important aspect of the proposed method is that it does not require solving the boundary-value problem in the time domain. Instead, it is based on the output of a frequency-domain radiation-diffraction analysis. Therefore, the required computational time remains low and is mostly related to the computation of the double-summation and convolutional schemes described in Section \ref{sec:hyd_loads_forcemodel}. For the 20-minute irregular sea states and the numerical setup of this study, the QME approach required around 30 minutes of simulation, while the respective time for the standard second-order approach was around 5 minutes.

\begin{figure}
    \centering
    \includegraphics[width=0.328\linewidth]{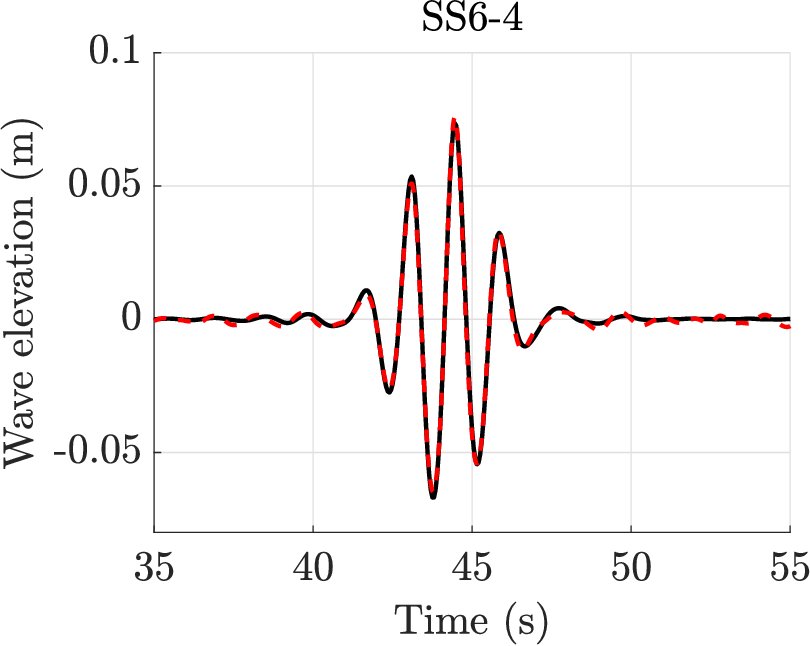} 
    \includegraphics[width=0.328\linewidth]{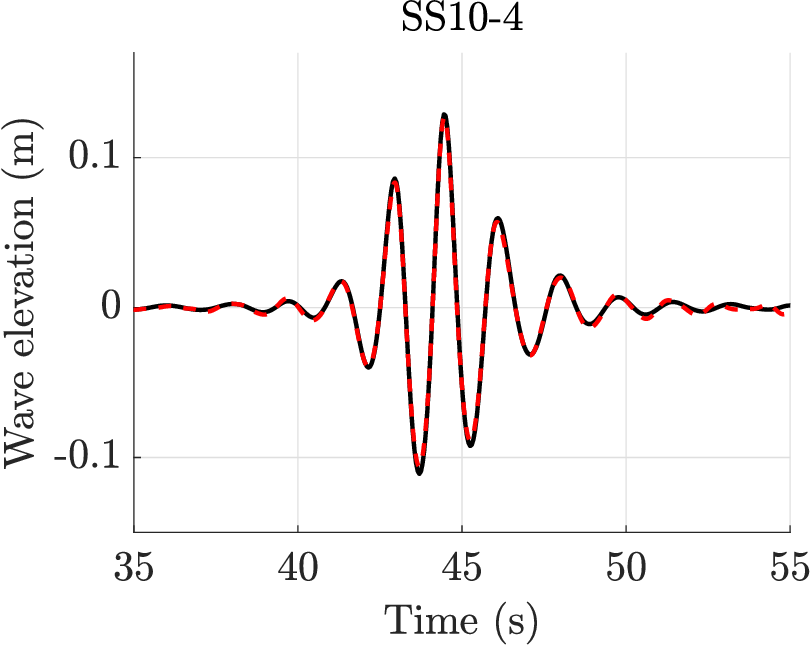}
    \includegraphics[width=0.328\linewidth]{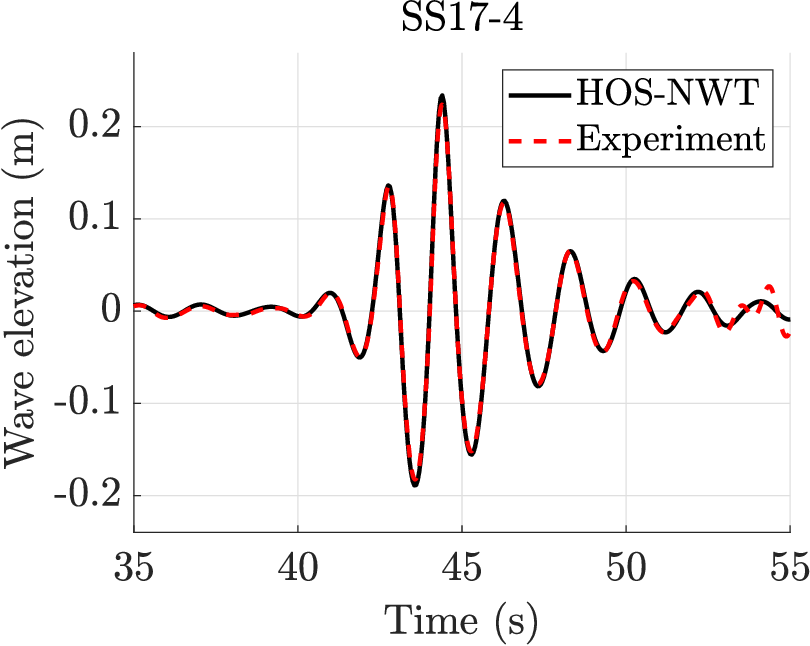}
    \includegraphics[width=0.328\linewidth]{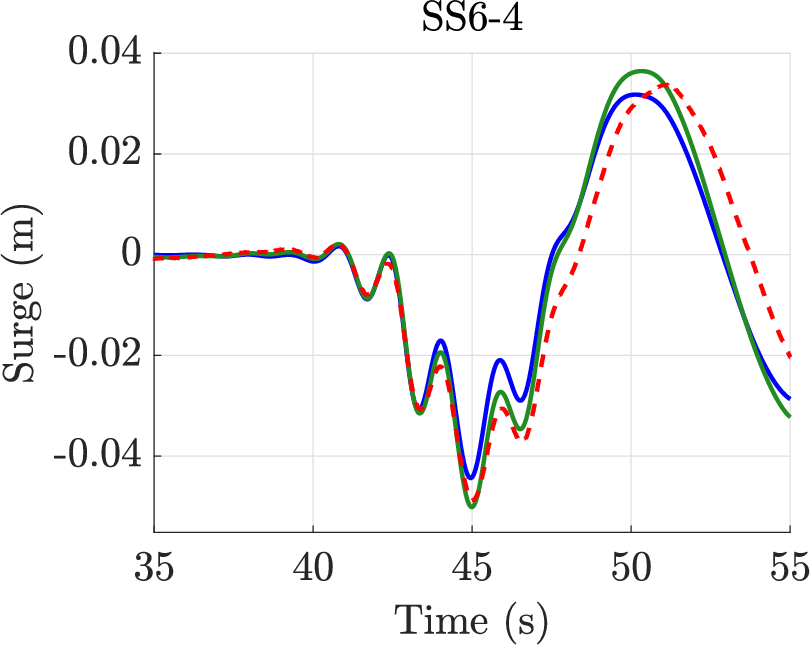} 
    \includegraphics[width=0.328\linewidth]{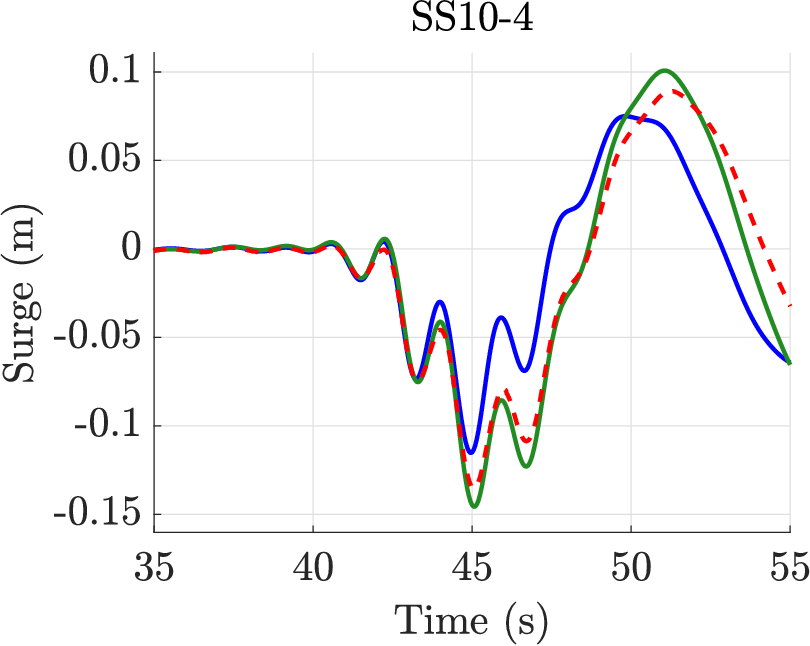}
    \includegraphics[width=0.328\linewidth]{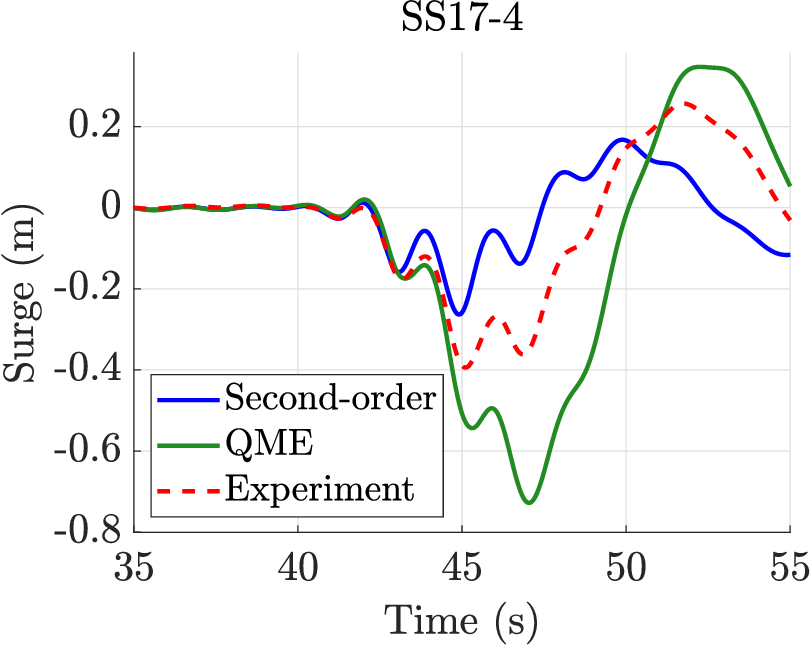}
    \includegraphics[width=0.328\linewidth]{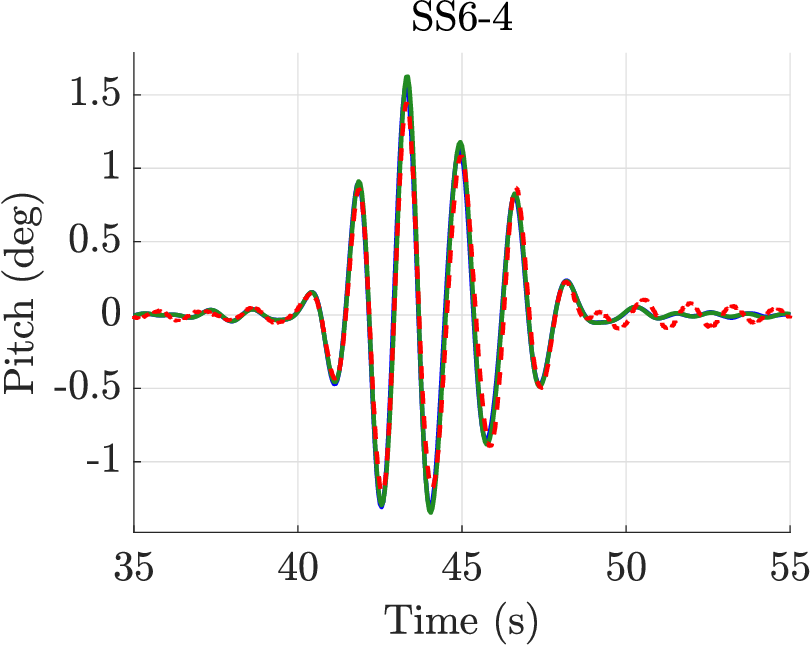}
    \includegraphics[width=0.328\linewidth]{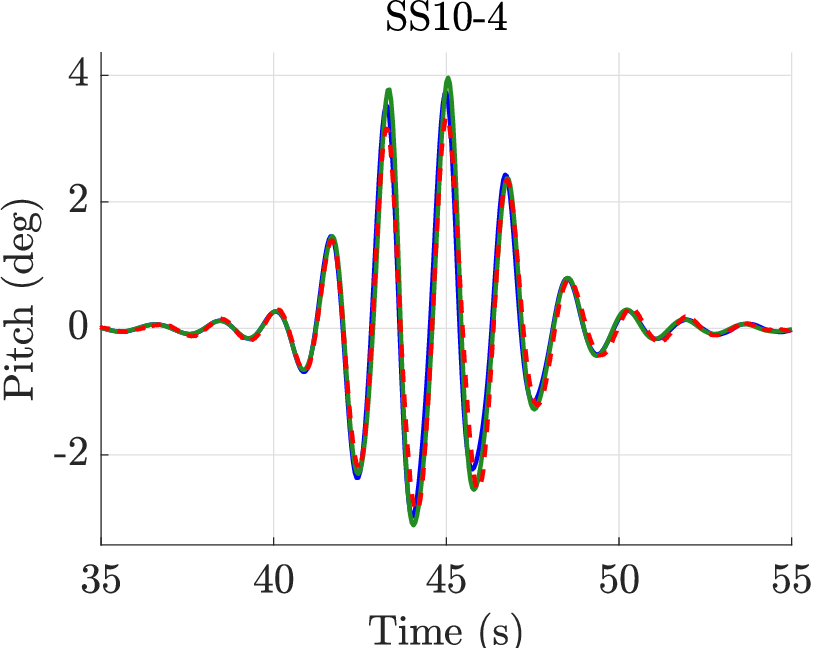}
    \includegraphics[width=0.328\linewidth]{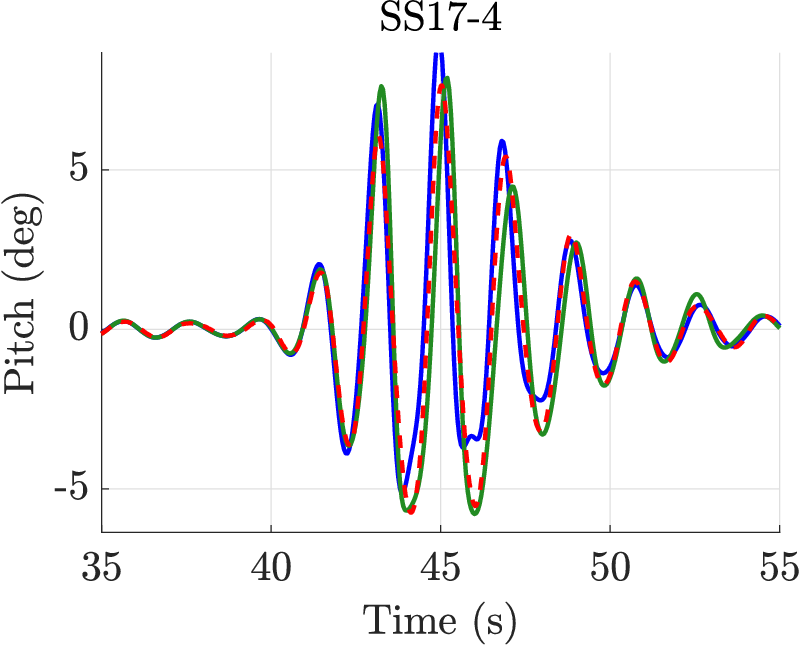}
    \caption{Comparison of numerical and experimental wave elevation (top row), surge motion (middle row), and pitch motion (bottom row) in design wave episodes.}
    \label{fig:force_model_rcw_results}
\end{figure} 

\subsubsection{Design wave episodes}

The investigation proceeds with the design wave episodes, which are short-duration wave packets, designed to excite a target surge response level of 4 standard deviations of the total first- and second-order response spectra \citep{dermatis_2025_prediction}. These waves are specifically constructed to trigger large surge excursions, and therefore, they are representative of the extreme events the vessel may experience within each sea state. The free-surface elevation of those wave packets, as well as the resulting surge and pitch motions of the vessel, are presented in Figure \ref{fig:force_model_rcw_results}. The overall surge response profiles obtained by the QME approach are consistently closer to the experimental results than the standard second-order approach. The wave episode of SS17 is an exception to this trend, since the surge motion is considerably overestimated after the main wave impact at $t=45$ s, which agrees with the behaviour of the model under irregular waves in the same sea state. Regarding the pitch response, the two numerical approaches yield comparable results and remain relatively close to the measured responses for the first two sea states. Nevertheless, in SS17, the standard second-order approach diverges from the experimental results after $t=42$ s, demonstrating a significant phase offset, while the proposed approach remains in closer agreement with experiments.

\begin{figure}
    \centering
    \includegraphics[width=0.328\linewidth]{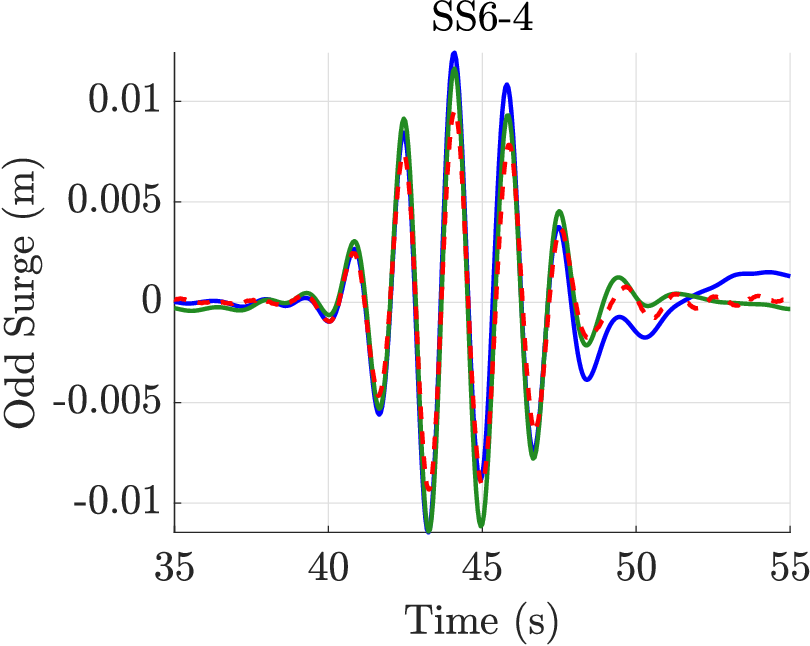}
    \includegraphics[width=0.328\linewidth]{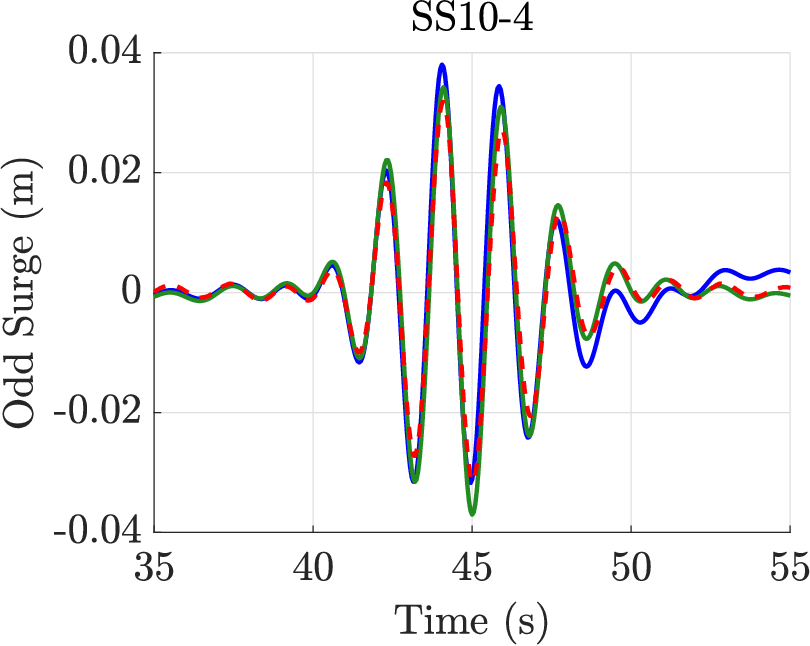}
    \includegraphics[width=0.328\linewidth]{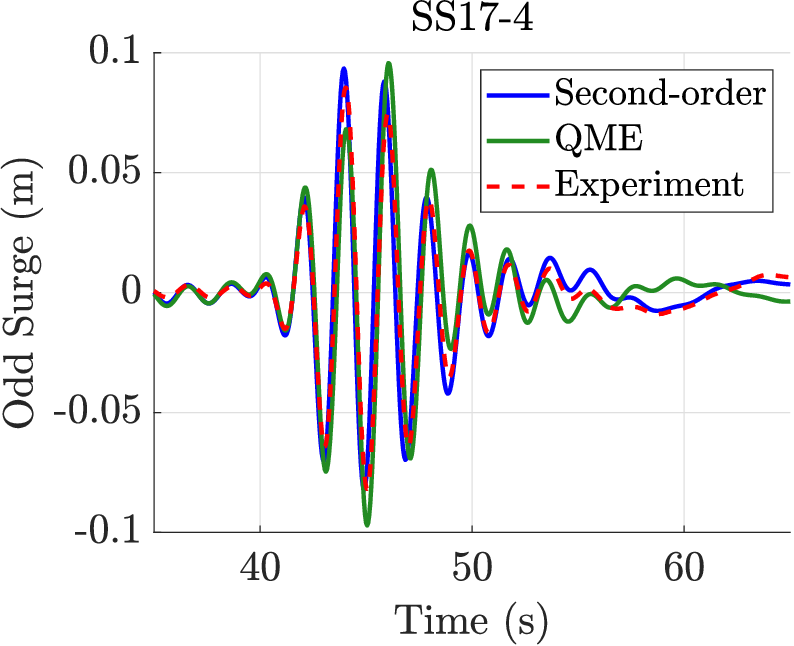}
    \includegraphics[width=0.328\linewidth]{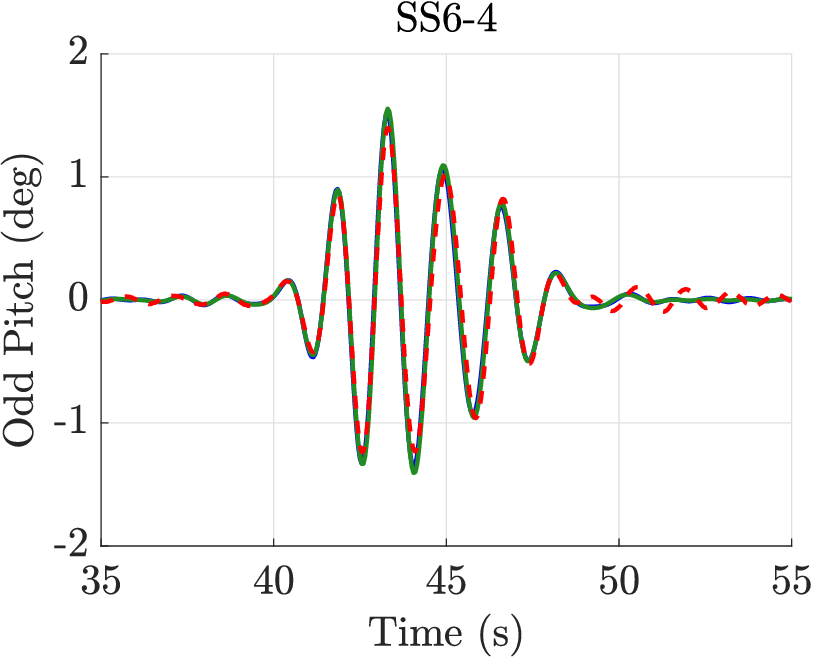}
    \includegraphics[width=0.328\linewidth]{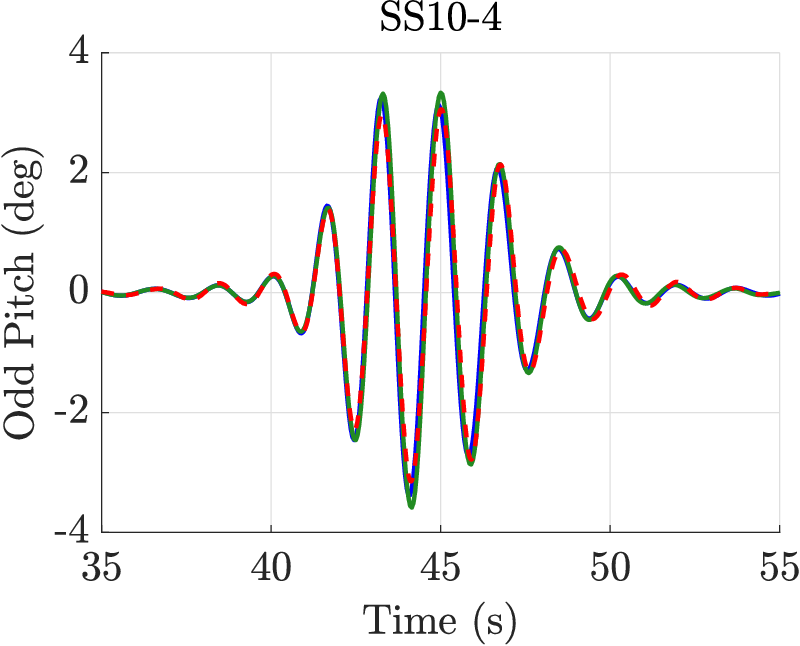}
    \includegraphics[width=0.328\linewidth]{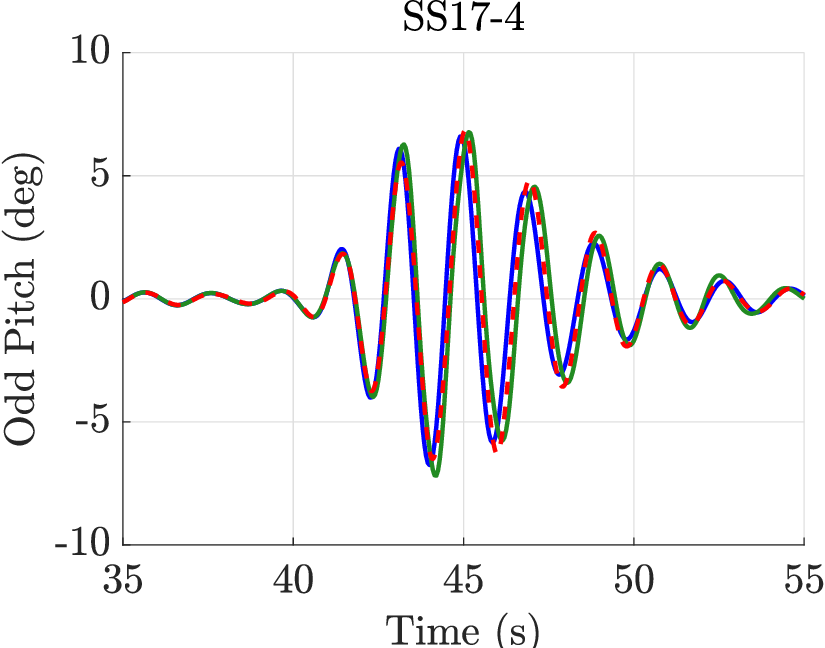}
    \caption{Comparison between the numerical and experimental odd harmonics for the surge (top row) and pitch motion (bottom row).}
    \label{fig:forcemodel_odd_harmonics}
\end{figure} 
\begin{figure}
    \centering
    \includegraphics[width=0.328\linewidth]{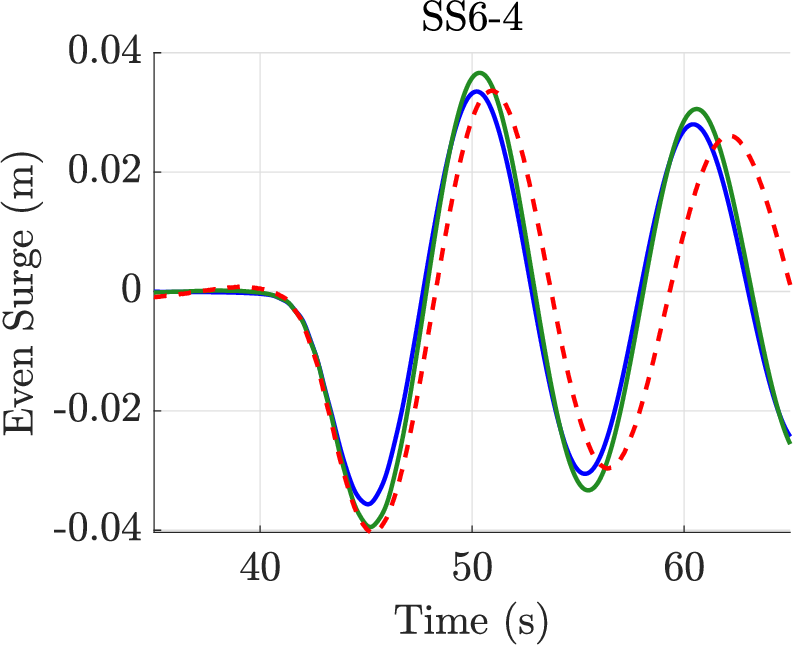}
    \includegraphics[width=0.328\linewidth]{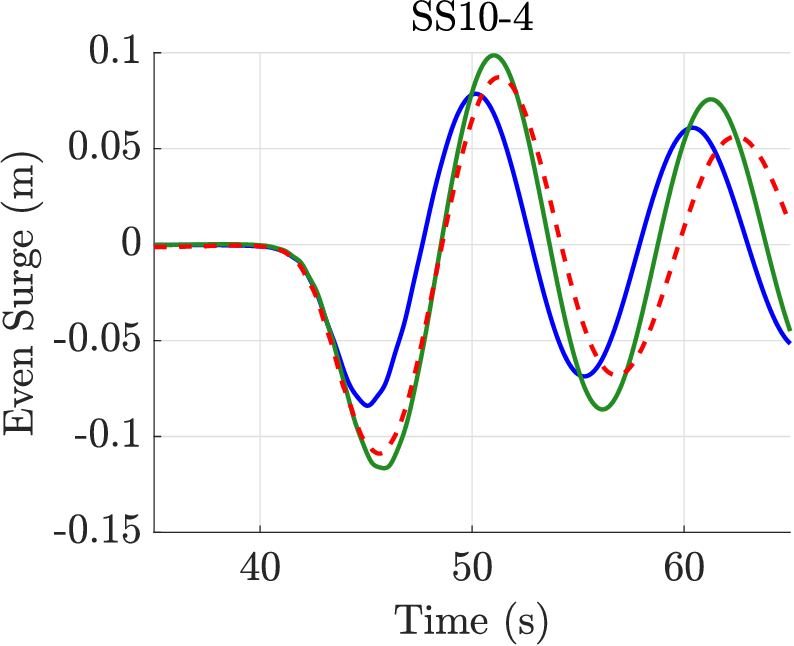}
    \includegraphics[width=0.328\linewidth]{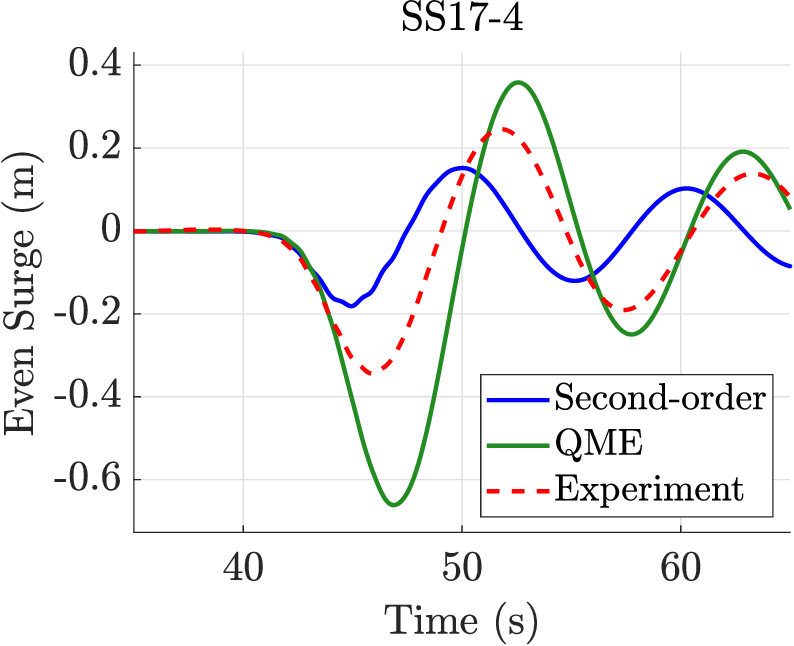}
    \includegraphics[width=0.328\linewidth]{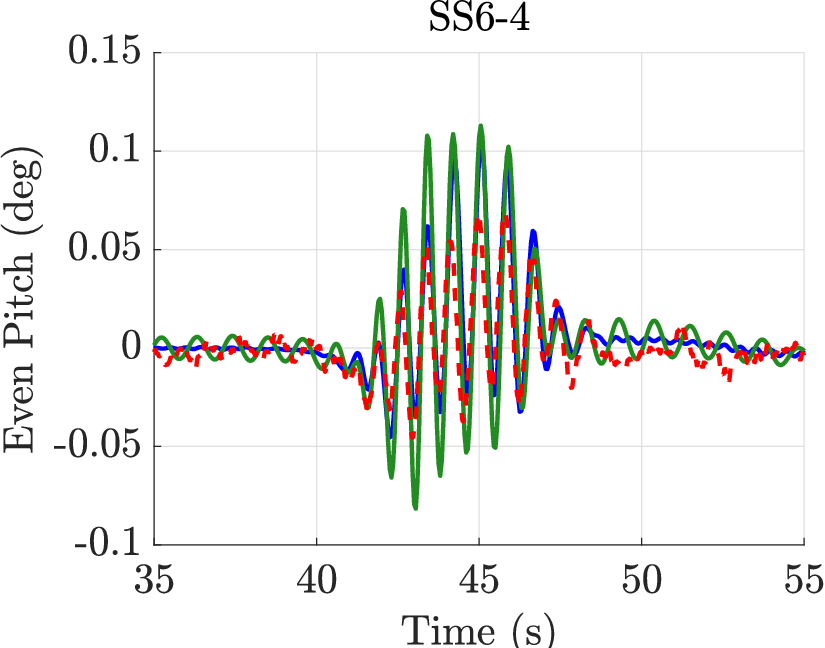}
    \includegraphics[width=0.328\linewidth]{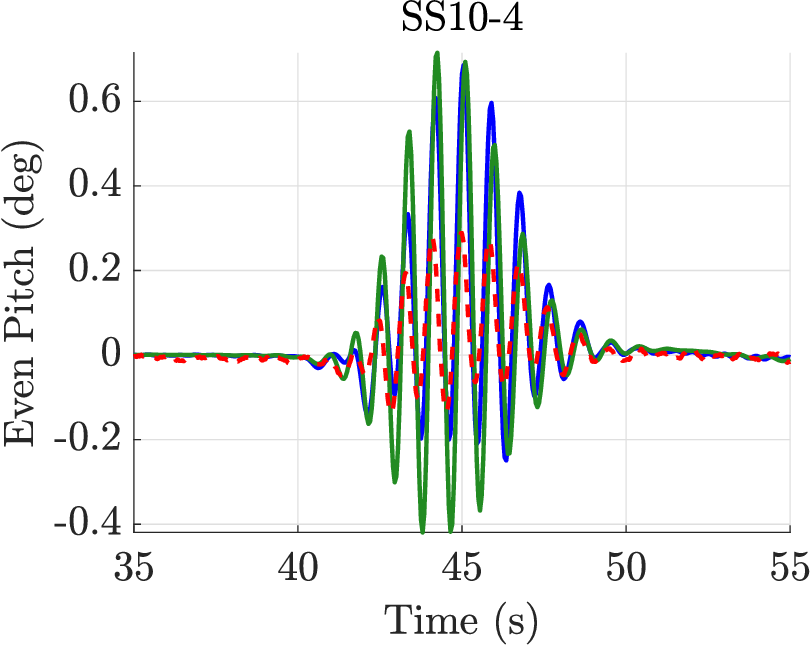}
    \includegraphics[width=0.328\linewidth]{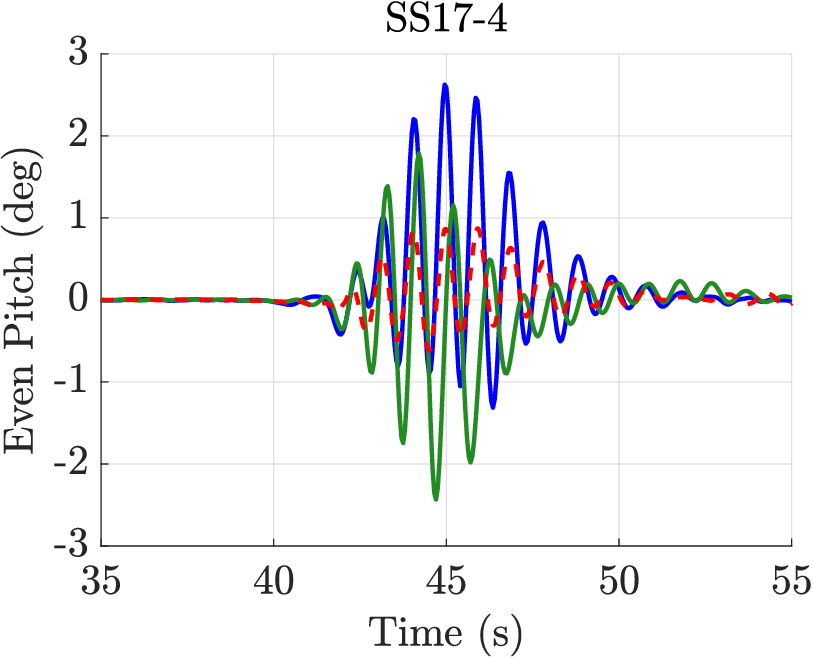}
    \caption{Comparison between the numerical and experimental even harmonics for the surge (top row) and pitch motion (bottom row).}
    \label{fig:forcemodel_even_harmonics}
\end{figure}

Towards a more in-depth investigation, decomposition of the numerical and experimental responses into odd and even harmonic contributions can be achieved through appropriately manipulating the phase of the wave signal \citep{walker2004shape,fitzgerald_phase_2014}. More precisely, upon phase-shifting the wave elevation, or the wavemaker motion history in case of experiments, by $180^\circ$, subtracting or adding the two response signals provides, 

\begin{equation} \label{eq:harmonic_decomposition_equation}
\begin{aligned}
    \text{odd:} & \quad \frac{1}{2}\left ( \underline{\boldsymbol{\xi}} - \underline{\boldsymbol{\xi}}^\prime \right ) = \underline{\boldsymbol{\xi}}^{(1)} + O(\epsilon^3) \\ 
    \text{even:} & \quad \frac{1}{2}\left ( \underline{\boldsymbol{\xi}} + \underline{\boldsymbol{\xi}}^\prime \right ) = \underline{\boldsymbol{\xi}}^{(0)}+ \underline{\boldsymbol{\xi}}^{(2)} + O(\epsilon^4). \\
\end{aligned}
\end{equation}
where $\underline{\boldsymbol{\xi}}$ and $\underline{\boldsymbol{\xi}}^\prime$ denote the body motions at the $0^\circ$ and $180^\circ$ tests respectively.

The results obtained for both vessel responses through this procedure are shown in Figures \ref{fig:forcemodel_odd_harmonics} and \ref{fig:forcemodel_even_harmonics}, for the odd and even harmonics, respectively. Regarding the former, both numerical models demonstrate good agreement with the experimental results. Especially for SS17, the QME result matches the experimental motions slightly better, indicating the existence of third-order contributions in the developed force model. More precisely, the quadratic force $\underline{\mathbf{F}}_Q$ within the QME approach entails the interaction of the total wavefield with the induced motion. These variables entail second-order contributions, and thus their products are higher-order compared to the pure second-order expression of \eqref{eq:quadratic_force_wamit}. Regarding the even harmonics, the low-frequency surge motion follows the experimental measurements closely, both in amplitude and phase. Once again, the response amplitude is considerably overestimated in the limiting sea state SS17, as discussed previously for Figure \ref{fig:force_model_rcw_results}. Finally, both numerical models overestimate the even pitch harmonic contributions, which is expected since no additional damping was applied in this degree of freedom.

\subsubsection{Investigation of individual nonlinear contribution}
\begin{figure}
    \centering
    \includegraphics[width=0.45\linewidth]{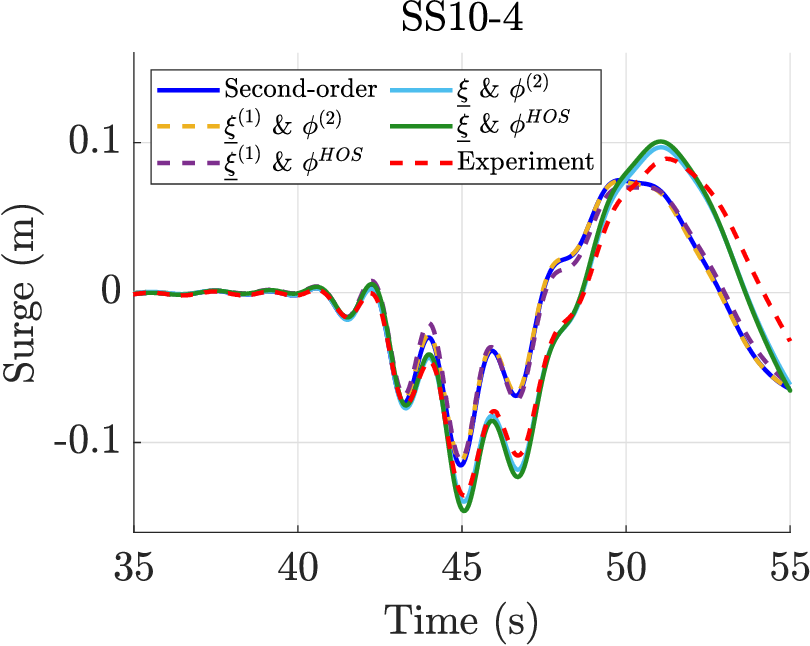}
    \includegraphics[width=0.45\linewidth]{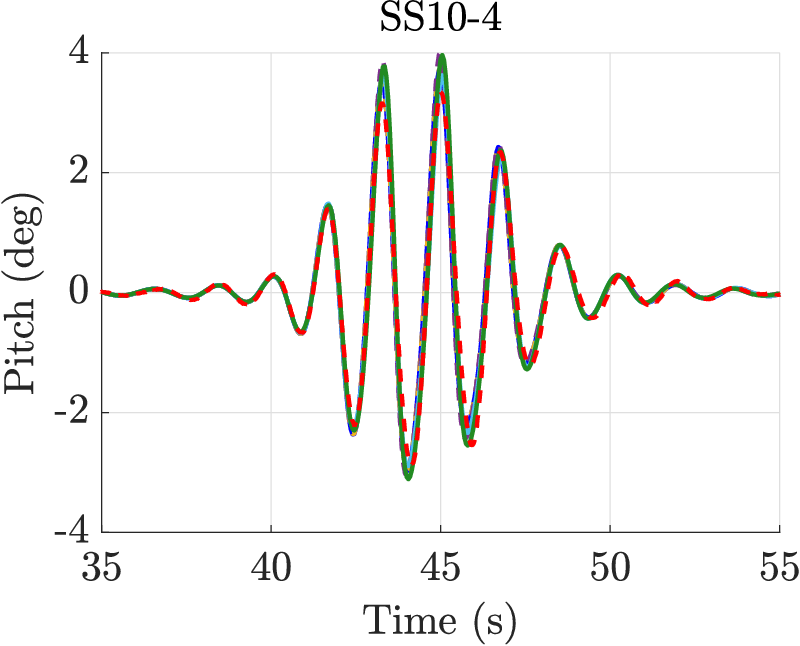}    
    \caption{Investigation of the nonlinear motions and waves contribution to the surge (left), and pitch motion (right) in design wave episode SS10-4.}
    \label{fig:wavemotion_contribution_rcw}
\end{figure}

\begin{figure}
    \centering
    \includegraphics[width=0.8\linewidth]{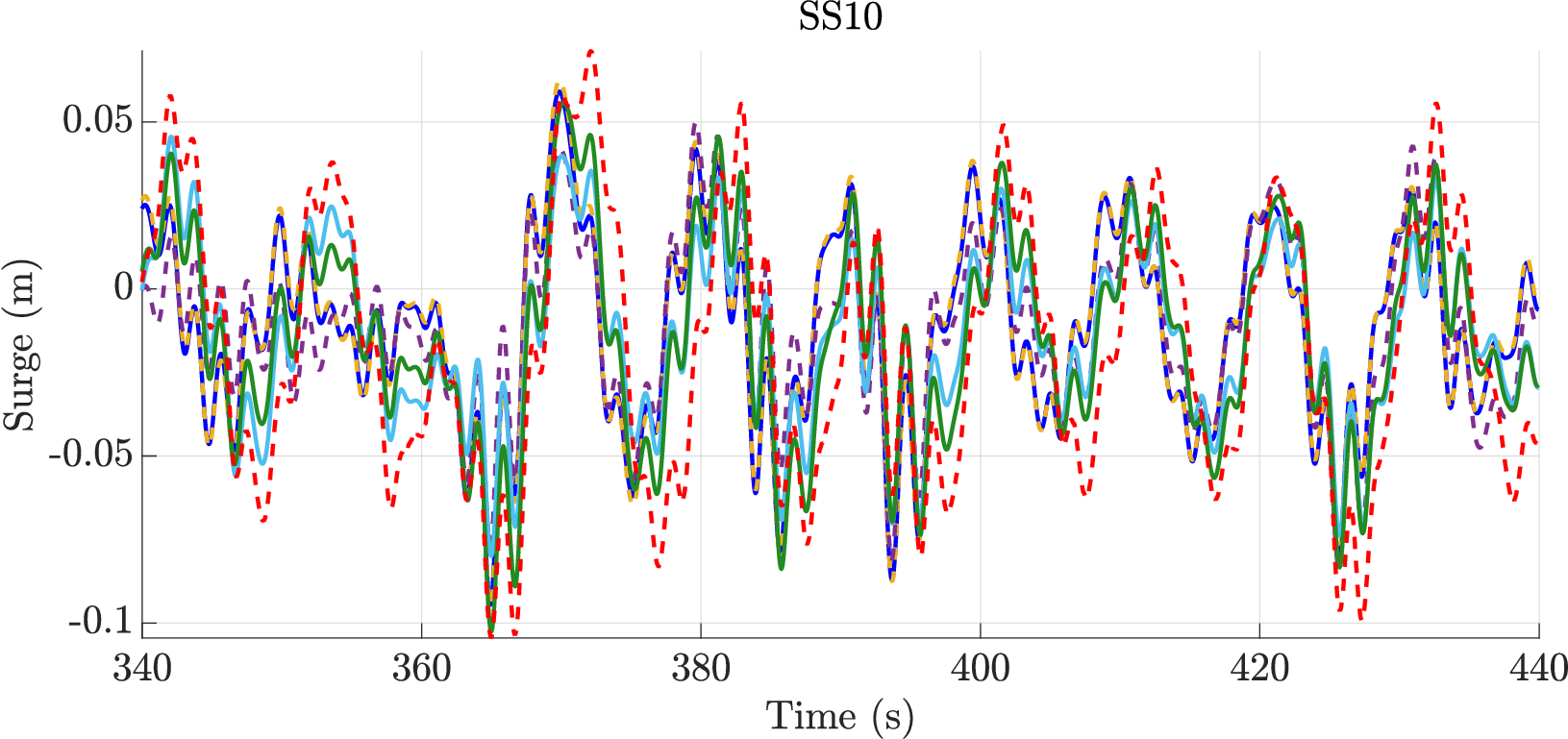}
    \includegraphics[width=0.4\linewidth]{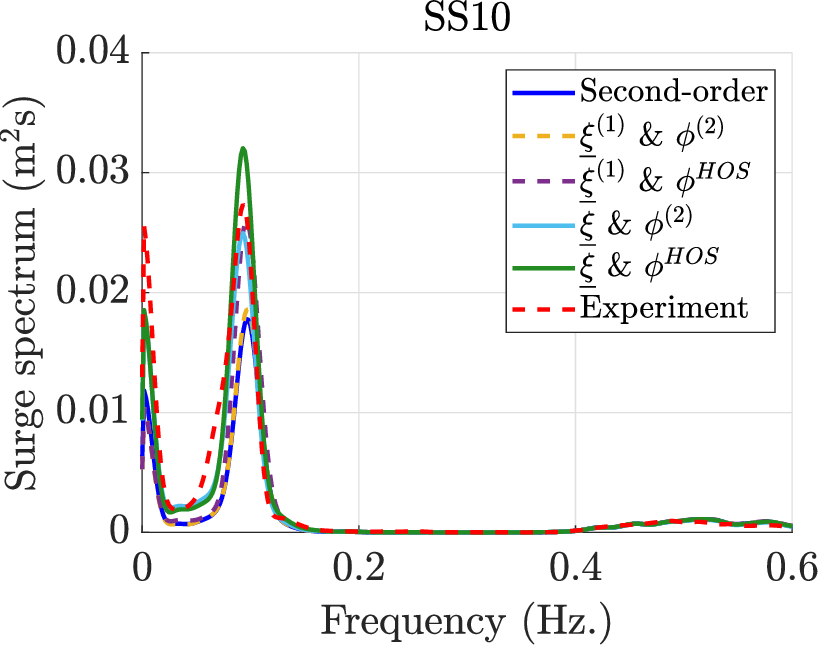}
    \includegraphics[width=0.4\linewidth]{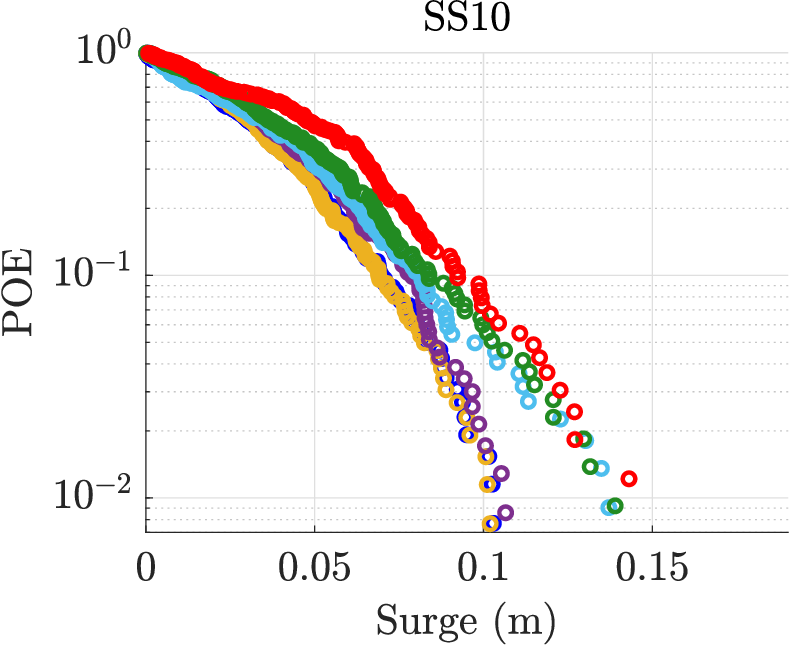} 
    \caption{Investigation of the nonlinear motions and waves contribution to the surge motion timeseries (top row), spectrum and exceedance probability (bottom row) in irregular waves of SS10.}
    \label{fig:wavemotion_contribution_iw}
\end{figure}

The force model developed in this work incorporates two distinct sources of additional nonlinearity in the evaluation of the hydrodynamic loads, namely the inclusion of: (i) a nonlinear incident wave field and (ii) the instantaneous, nonlinear body motions. It is important to consider that these contributions can also be employed independently, and their relative importance in the improved performance of the QME approach is hereby investigated. Figure \ref{fig:wavemotion_contribution_rcw} presents the vessel motions under the design wave episode SS10-4 for different combinations of wave and motion nonlinearity in the force model. The respective results for the surge motion under irregular waves of SS10 are presented in Figure \ref{fig:wavemotion_contribution_iw}. 

Regarding the wave kinematics,  $\phi^{(2)}$ corresponds to a second-order wavefield, where \eqref{eq:sharma_dean_force} is used for the potential force $\underline{\mathbf{F}}_{P_W}$. Moreover, the linear modal amplitudes from \eqref{eq:linear_waves_modal_amplitudes} are used in the evaluation of the quadratic force $\underline{\mathbf{F}}_{Q}$ and the potential terms $\underline{\mathbf{F}}_{P_F}$ and $\underline{\mathbf{F}}_{P_M}$. On the other hand, $\phi^{HOS}$ corresponds to the use of a nonlinear incident wavefield from \texttt{HOS-NWT} and the corresponding nonlinear modal amplitudes. Regarding the body motions, $\underline{\boldsymbol{\xi}}^{(1)}$ denotes the use of the precomputed linear motions from \eqref{eq:linear_motions_td}, whereas $\underline{\boldsymbol{\xi}}$ means the use of the motion-explicit schemes of Section \ref{sec:hyd_loads_forcemodel} to take into account the instantaneous body motions. Overall, the responses obtained with the present force model under the purely second-order regime ($\phi^{(2)}$ and $\underline{\boldsymbol{\xi}}^{(1)}$) are essentially equivalent to those derived from the precomputed QTF-based forces, within the classical second-order approach. This also serves as verification regarding the second-order consistency of the responses obtained. Incorporating nonlinear body motions in the force calculation leads to a substantial improvement in the accuracy of the surge response, whereas the influence of nonlinear waves is relatively minor. The pitch response is relatively insensitive to these modelling choices, as suggested by the design wave results in Figure \ref{fig:wavemotion_contribution_rcw}, and therefore the respective irregular wave results are omitted.

These findings suggest that the primary source of the improved accuracy demonstrated by the proposed method is the use of instantaneous total motions in the load evaluation. It should be stressed, however, that this conclusion is case-specific and should be assessed further, as it might be different for other configurations. Notably, the use of modal amplitudes from \texttt{HOS-NWT} inherently accounts for bound waves and nonlinear wave–wave interactions. Consequently, the impact of nonlinear waves could become significant if transfer of wave energy occurred into frequency regions where the linear transfer functions $\hat{\underline{\mathbf{F}}}^{(1)}_{P_{IS}}$, $\hat{\eta}^{(1)}_{IS}$, and $\nabla \hat{\phi}^{(1)}_{IS}$ exhibited large values. This effect is not observed in the present study due to the sea-state calibration procedure of \cite{canard_varying_2022}. More precisely, this technique matches the experimental wave spectrum at the in-tank position of the structure with the theoretical spectrum, and thus eliminates the low-frequency content of the wave spectrum.

\section{Conclusions} \label{sec:conclusions}

In the present paper, a novel approach is presented that employs nonlinear wave and body kinematics to evaluate hydrodynamic loads and motion responses in the time domain. The method is consistent with the classical second-order radiation-diffraction theory, but permits inclusion of some higher-order contributions. The principal advantage of the proposed force model is the formulation based on the results of linear frequency-domain radiation-diffraction analysis, which is widely accessible and efficient. The pressure integration approach is adopted, and diffraction effects are accounted for by applying linear transfer functions in the wavenumber domain to the time-dependent nonlinear wave modal amplitudes. These are obtained through the spatial Fourier decomposition of the incident wave field, and are a standard output of modern nonlinear wave solvers. At the same time, radiation effects are treated in the time domain, permitting their expression based on the instantaneous total body kinematics. It was shown that in a purely second-order wavefield, and considering the first-order body motions solely, the present force model is fully consistent with standard second-order theory, but subject to the Pinkster approximation \citep{pinkster1980}.

Furthermore, the developed force model was coupled with a time-domain motion solver. This two-way coupling allows the use of the instantaneous body kinematics during each time step to evaluate the wave-induced loads. The resulting motions were compared against the classical hybrid frequency-time domain approach and experiments for the configuration of a moored container ship under irregular waves and design wave episodes. For the majority of the test cases, the proposed approach yields highly accurate results for the surge motion compared to the classical second-order approach. Nevertheless, for an extremely severe sea state of $H_s=17$ m in full scale, a tendency to overestimate the magnitude of surge excursions was observed. This is likely linked to the violation of the small motion assumption in the approximation of the instantaneous body surface through Taylor expansion of the mean body surface position from \eqref{eq:pressure_taylor_nonlin}. However, the same additional damping in the surge DoF was employed for both types of numerical simulations, which was derived from model-free calibration according to experimental decay tests. Therefore, this limitation might potentially be treated through more refined damping calibration methods \citep{pegalajar-jurado_reproduction_2019}. Alternatively, higher-order terms in the Taylor expansion of the instantaneous body surface can be considered. Regarding the pitch response, second-order contributions were relatively minimal, and thus the differences between results of the proposed force model and the classical second-order approach were less pronounced and mainly observed in the extreme response range. 

Finally, the relative contribution of the nonlinear wave and body kinematics to the performance of the force model was investigated. For the present configuration, the use of the instantaneous body motions in the force calculation was the primary source of higher accuracy, while the effect of nonlinear waves was secondary. However, this finding is considered case-specific, and the ability of the model to accommodate combinations of free and bound waves is promising for other applications, which will be investigated in future work. The  approach takes explicitly into account the body motions and partially decouples them from the incident wavefield. This feature renders the QME approach well-suited for applications such as floating wind turbines, where the forcing of the system is not only wave-driven.

\begin{bmhead}[Funding.]
This research was supported by the WASANO project (French National Research Agency, Grant No. ANR-16-IDEX-0007) and the FloatLab project (Innovation Fund Denmark, Grant No. 2079-00082B).
\end{bmhead}

\begin{bmhead}[Declaration of interests.]
The authors report no conflict of interest.
\end{bmhead}



\appendix
\begin{appen}

\section{Development of $\underline{\mathbf{F}}_{P_M}$ term} \label{sec:F_P_M}
The potential force contribution of $\phi_{S_M}$ can be developed using the body boundary condition for the elementary radiation potentials from \eqref{eq:bc-phirj}, 

\begin{equation}
    \begin{aligned}
        \underline{\mathbf{F}}_{P_M}(t) &= -\rho \int_{S_0} \partial_t\phi_{S_M}(\tilde{\mathbf{x}},t) ~\underline{\tilde{\mathbf{n}}}~\ud S=-\rho \int_0^t\int_{S_0} \partial_{\tau}\phi_{S_M}(\tilde{\mathbf{x}},\tau) ~\underline{\tilde{\mathbf{n}}} ~\delta(t-\tau)~\ud S ~\ud \tau \\ 
        &=-\rho\int_0^t\int_{S_0} \partial_{\tau}\phi_{S_M}(\tilde{\mathbf{x}},\tau) ~\partial_{\tilde{\mathbf{n}}}\underline{\boldsymbol{\phi}}{}_r(\tilde{\mathbf{x}},t-\tau)~\ud S ~\ud \tau
    \end{aligned}
\end{equation}
Application of Green's theorem to $\phi_{S_M}$ and the components of $\underline{\boldsymbol{\phi}}{}_r$ and further exploiting \eqref{eq:bc_motion_potential}, the force is transformed into,

\begin{equation}
    \begin{aligned}
        \underline{\mathbf{F}}_{P_M}(t) &=-\rho \int_0^t\int_{S_0}  \left ( \nabla \partial_{\tau}\phi_{S_M}(\tilde{\mathbf{x}},\tau) \cdot \tilde{\mathbf{n}}\right )~\underline{\boldsymbol{\phi}}{}_r(\tilde{\mathbf{x}},t-\tau)~\ud S ~\ud \tau =-\rho \int_0^t\int_{S_0}  \partial_{\tau} \Theta_B(\tilde{\mathbf{x}}, \tau)~\underline{\boldsymbol{\phi}}{}_r(\tilde{\mathbf{x}},t-\tau)~\ud S ~\ud \tau \\ 
        &=-\rho \int_{S_0}  \partial_{t} \Theta_B(\tilde{\mathbf{x}}, t)~\underline{\boldsymbol{\phi}}{}^{\infty}_r(\tilde{\mathbf{x}})~\ud S -\rho \int_0^t\int_{S_0}  \partial_{\tau} \Theta_B(\tilde{\mathbf{x}}, \tau)~\underline{\boldsymbol{\psi}}{}_r(\tilde{\mathbf{x}},t-\tau)~\ud S ~\ud \tau
    \end{aligned}
\end{equation}
where it is noted that the free-surface integral from Green's theorem vanishes due to \eqref{eq:phiR_fsbc} and \eqref{eq:bc_motion_potential}. In addition, the time derivative of the body boundary condition is given as,

\begin{equation} \label{eq:thetaB_derivative}
\begin{aligned}
    \partial_{\tau} \Theta_B(\tilde{\mathbf{x}}, \tau) = &
(\partial_{\tau}\pmb{\alpha}(\tau)\times \tilde{\mathbf{n}})\cdot \Big(\partial_{\tau}\pmb{\xi}(\tau) +\partial_{\tau}\pmb{\alpha}(\tau)\times \tilde{\mathbf{x}} - \nabla \phi(\tilde{\mathbf{x}}, \tau) \Big ) - \tilde{\mathbf{n}}  \Big [ \left ( \partial_{\tau}\boldsymbol{\xi}(\tau) + \partial_{\tau}\boldsymbol{\alpha}(\tau) \times \tilde{\mathbf{x}}  \right) \cdot \nabla \Big ] \nabla \phi(\tilde{\mathbf{x}}, \tau)\\  
& +(\pmb{\alpha}(\tau)\times\tilde{\mathbf{n}})\cdot \Big(\partial_{\tau \tau}{\pmb{\xi}}(\tau)+\partial_{\tau \tau}\pmb{\alpha}(\tau)\times \tilde{\mathbf{x}} - \nabla \partial_{\tau}\phi(\tilde{\mathbf{x}}, \tau)\Big) - \tilde{\mathbf{n}}  \Big [ \left ( \boldsymbol{\xi}(\tau) + \boldsymbol{\alpha}(\tau) \times 
    \tilde{\mathbf{x}}  \right) \cdot \nabla \Big ] \nabla \partial_{\tau}\phi (\tilde{\mathbf{x}}, \tau)\\  
& + \tilde{\mathbf{n}} \cdot \partial_{\tau \tau}\boldsymbol{\mathcal{H}}(\tau)\tilde{\mathbf{x}}
\end{aligned}
\end{equation}
It is also noted that the double spatial derivative in \eqref{eq:thetaB_derivative} is transformed into a single derivative using Stokes' theorem \citep{lee1995wamit}.

\end{appen}

\bibliographystyle{1_jfm}
\bibliography{1_jfm}

\end{document}